\documentclass[11pt]{article} %
\usepackage[a4paper, margin=1.25in]{geometry} %
\AtBeginDocument{%
  }

\usepackage{comment}
\usepackage{tabularx}
\usepackage[table]{xcolor} 
\usepackage{makecell}

\usepackage{amssymb}
\usepackage[dvipsnames,svgnames,x11names]{xcolor}

\colorlet{RED}{red} %

\colorlet{WHITE}{white}
\colorlet{FORESTGREEN}{ForestGreen}
\colorlet{MAROON}{Maroon}
\newcommand{\pc}[2]{
 {\colorbox{#2}{\bfseries\sffamily\scriptsize\textcolor{white}{#1}}}
}
\newcommand{\pgs}[1]{%
\pc{#1\,pgs}{ForestGreen}%
}

\renewcommand{\pgs}[1]{}

\usepackage[normalem]{ulem} %

\usepackage{balance}

\usepackage{rotating}
\usepackage{adjustbox}

\usepackage{fontawesome}
\newcommand{\xmark}{\textcolor{Mahogany!50}{\faTimesCircle}}%
\newcommand{\cmark}{\textcolor{LimeGreen}{\faCheckCircle}}%

\usepackage{listings}

\lstdefinelanguage{Agent}
{
  basicstyle=\ttfamily\footnotesize,
  columns=fullflexible,
  moredelim=[s][\color{blue}\bfseries]{[}{]},
  moredelim=[s][\color{Maroon}\bfseries]{<}{>},
  alsoletter=:-,
  morekeywords=[1]{Plan:,Tool:,Result:,Parameter:,-Overall:,-Step},
  keywordstyle=[1]\bfseries,  %
  frame=single,            %
  frameround=tttt,         %
  rulesep=.4pt,            %
  framesep=6pt,            %
  rulecolor=\color{gray},%
  linewidth=\linewidth,    %
  backgroundcolor=\color{gray!5}, %
  showstringspaces=false,
  breaklines=true,
}

\usepackage{moreverb}

\usepackage[nounderscore]{syntax}

\usepackage{graphicx}
\graphicspath{{figures/}{Txx/figures/}}

\usepackage[caption=false,font=footnotesize]{subfig}

\usepackage{mathtools}
\usepackage{amsthm}
\usepackage{amsfonts}

\usepackage{algorithmicx}
\usepackage{algpseudocode}
\usepackage[ruled]{algorithm}
\definecolor{light-gray}{gray}{0.75}
\algrenewcommand{\algorithmiccomment}[1]{\hskip3em{{\footnotesize \textcolor{light-gray}{$\blacktriangleright$}}} #1}

\usepackage{multirow} %
\usepackage{rotating} %
\usepackage{booktabs} %
\usepackage{colortbl} %
\usepackage{tablefootnote} %

\usepackage{array}
\newcolumntype{L}[1]{>{\raggedright\let\newline\\\arraybackslash\hspace{0pt}}m{#1}}
\newcolumntype{C}[1]{>{\centering\let\newline\\\arraybackslash\hspace{0pt}}m{#1}}
\newcolumntype{R}[1]{>{\raggedleft\let\newline\\\arraybackslash\hspace{0pt}}m{#1}}

\usepackage{xspace}

\usepackage{enumitem}
\newcommand{\para}[1]{\noindent\textbf{#1.~}}

\usepackage{soul}
\sethlcolor{blue!10}
\newcommand{\takeaway}[1]{
\par\vspace{0.3\baselineskip}
\noindent\hl{\textsf{#1}}}

\newcommand{\fw}{FAME\xspace}

\newcommand{\react}{ReAct\xspace}
\newcommand{\reflect}{Reflexion\xspace}
\newcommand{\llmc}{LLM Compiler\xspace}
\newcommand{\mlzero}{MLZero\xspace}

\usepackage[english]{babel}
\usepackage{blindtext}

\usepackage[colorlinks,bookmarksopen,bookmarksnumbered,citecolor=blue,urlcolor=blue]{hyperref}

\begin{document}

\title{\huge \fw: QoS-aware Async Service Orchestration for Agentic Workflows}

\author{
Haseeb Abdullah Kollorath$^{1}$, Vaibhav Jha$^{1}$, Varad Kulkarni$^{1}$, Vinay$^{1}$,\\
Nikhil Reddy$^{1}$, Anand Eswaran$^{2}$, Praveen Jayachandran$^{2}$ and \\
Yogesh Simmhan$^{1}$
\\~\\
\textit{$^{1}$~Indian Institute of Science, Bangalore 560012 India}\\
\textit{$^{2}$~IBM Research Lab, Bangalore, India}
\\~\\
\texttt{haseebabdul1@iisc.ac.in, vaibhavjha@iisc.ac.in}\\
\texttt{varadk@iisc.ac.in, simmhan@iisc.ac.in}
}
\date{}

\maketitle

\begin{abstract}

Agentic workflows built on Large Language Models (LLMs) increasingly rely on external tools, with Model Context Protocol (MCP) services emerging as a common interface for tool discovery and invocation. These workflows are stateful, bursty and can block for minutes to hours while external tools execute. Always-on virtual machines and managed agent runtimes simplify deployment, but can over-provision capacity, bill idle wait time or encounter execution limits for long-running tool calls. 
We present \fw, a service-oriented middleware that optimizes MCP-enabled agentic workflows through modular service orchestration, explicit state management, MCP service input/output reduction and timeout-safe async invocation for long-running external services.
\fw realizes these composite services using FaaS-based agent roles, persists workflow state through agent memory injection, and reduces MCP service overhead using S3 handle passing and tool-output caching.
For long-running external MCP tools, \fw checkpoints agent and orchestrator state, suspends execution and resumes from callbacks without billing idle waits.
Across 26 short-running service tasks over 10 MCP servers, \fw improves service-level QoS by reducing infrastructure cost by $8$--$12\times$ relative to virtual machines and $43$--$106\times$ relative to managed runtimes, while reducing latency by up to $17\times$, input tokens by up to $88\%$ and total cost by up to $66\%$.
On two long-running \mlzero{} workloads from MLE-Bench, \fw completes all iterations while synchronous AWS Step Functions time out, and is $3.64$--$4.24\times$ cheaper than AWS Durable Functions at the agent level and $2.67$--$8.23\times$ cheaper at the orchestrator level.
\end{abstract}

\section{Introduction \pgs{1.25}}

\begin{figure}[t!]
\centering%
    \includegraphics[width=0.7\columnwidth]{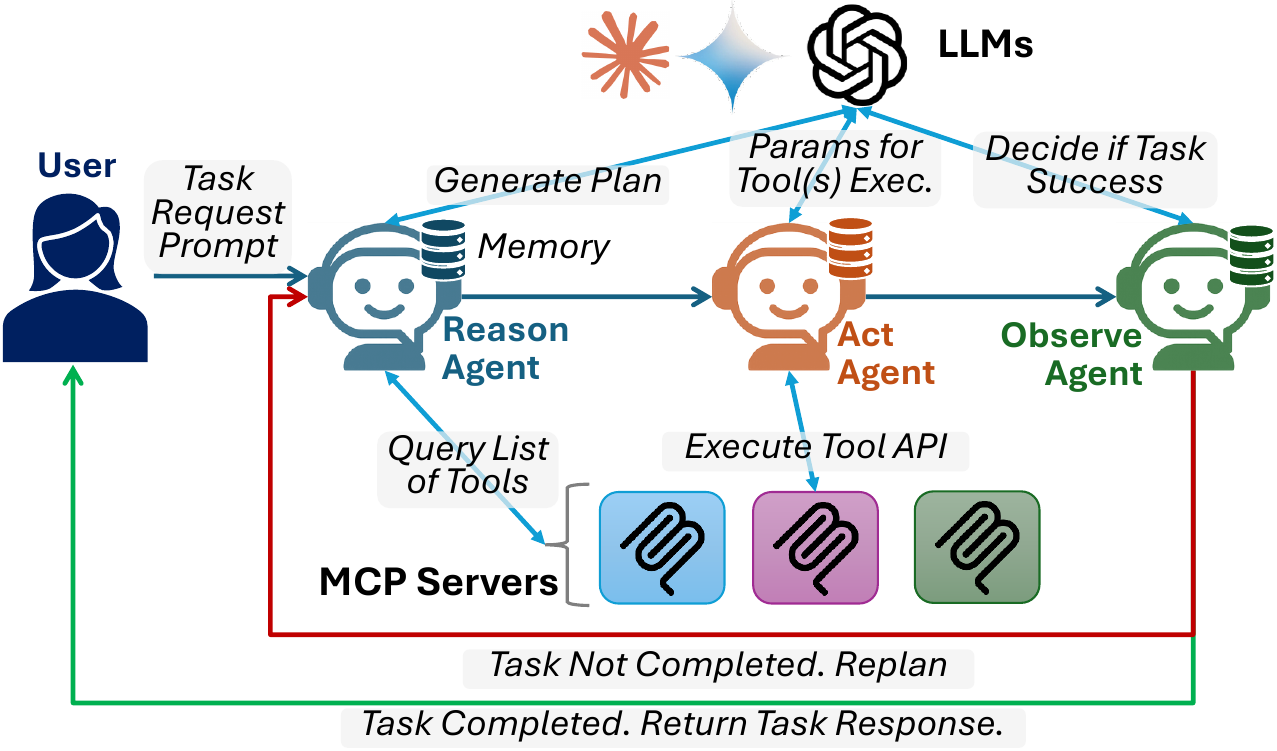}
\caption{Agentic workflow using the \react pattern. The agent uses memory and LLM reasoning to select MCP tools, observes tool outputs, and iterates until task completion.}
    \label{fig:react}
\end{figure}

\textit{Agentic workflows} built on Large Language Models (LLMs) combine reasoning, planning, memory and tool invocation to complete multi-step user tasks. These workflows are commonly expressed using recurring patterns such as \react~\cite{yao2023react}, \reflect~\cite{10.5555/3666122.3666499} and \llmc~\cite{llmcompiler}, and authored using graph-based frameworks such as LangGraph~\cite{langgraph_overview_docs} and AutoGen~\cite{wu2024autogen}. Fig.~\ref{fig:react} illustrates this execution model for \textit{\react}: the agent reasons over the user request and memory, invokes external tools, observes their outputs and iterates until the task completes. 
In practice, these tools are increasingly exposed through \textit{Model Context Protocol (MCP)} services, which act as endpoints for tool discovery and invocation~\cite{mcp_intro_docs,mo2025livemcpbench}. 
Thus, the services-computing problem is not only how to build an agent, but \textit{how to compose, manage and optimize stateful tool-using services under latency, cost and completion constraints}.

Always-on Virtual Machines (VMs), commonly used to host agentic frameworks as monolithic services~\cite{agenticapp2026}, and managed agent runtimes such as AWS Bedrock AgentCore~\cite{bedrock_AgentCore_Docs} simplify deployment, but are often inefficient for agentic workflows. \textit{Short-running workflows} are bursty across users and tenants, causing VM deployments to under-utilize provisioned capacity and managed runtimes to incur higher platform charges. \textit{Long-running workflows} expose a different failure mode: the agent and orchestrator may remain allocated while an external MCP tool executes for minutes to hours, even though they are only waiting for a result. \textit{Function-as-a-Service (FaaS)} offers elastic scaling and fine-grained billing~\cite{faas,faas-tpds-arxiv}, but its stateless and time-limited execution model conflicts with agent memory, large MCP payloads and long external tool calls.

These observations lead to three service-orchestration gaps. 
\textit{First}, agentic workflows require explicit state management (memory) across turns, across agent roles and within long-running iterations. Without this, later turns either fail due to missing context or repeat earlier tool calls, inflating latency and input tokens~\cite{sumers2024cognitive,langchain_memory_overview}. 
\textit{Second}, MCP service input/output must be treated as a first-class QoS concern. While MCP is the \textit{de facto} tool-interface for agentic systems, repeatedly moving large tool outputs through prompts increases LLM latency, context usage and cost, while na\"{i}ve FaaS wrapping of MCP servers can multiply cold starts and redundant tool invocations~\cite{mcp_intro_docs,mo2025livemcpbench}. 
\textit{Third}, long-running External MCP services require timeout-safe async invocation. FaaS platforms impose bounded function lifetimes, while durable workflows and managed runtimes greatly differ in how they bill wait-time, callbacks, checkpoints and replay~\cite{faas,aws_lambda_durable_docs,aws_durable_key_concepts,bedrock_AgentCore_Docs}.

We present \textbf{\fw}, a FaaS middleware that optimizes MCP-enabled agentic workflows for both short-running and long-running MCP services.
\fw decomposes agentic patterns into modular serverless roles, persists workflow state through agent memory injection, and reduces MCP input/output overhead using S3 handle passing and tool-output caching. For long-running EMCP tools, \fw checkpoints agent and orchestrator state, suspends execution while the tool runs, and resumes using callbacks. This makes the agent--tool and orchestrator--agent bridges timeout-safe, while avoiding billing of FaaS wait-times.

We makes these specific contributions in this article: 
\begin{enumerate}[wide] 
\item We identify service runtime trade-offs for MCP-enabled agentic workflows, contrasting always-on VM, AWS AgentCore managed-runtime, synchronous Step Functions, AWS Durable Functions (DF) and \fw deployments for bursty short-running and long-running service workloads (\S\ref{sec:bf}). 
\item We design \fw to map \react, \reflect, \llmc and \mlzero{} workflows to stateful composite services built from modular FaaS agent roles, cross-turn agent memory management and within-turn checkpoints (\S\ref{sec:arch}, \S\ref{sec:fpp}). 
\item We optimize MCP service invocation through automated FaaS wrapping, S3 handle passing, tool-output caching and callback-based async invocation for long-running EMCP services (\S\ref{sec:arch:mcp}, \S\ref{sec:fpp}). 
\item We evaluate \fw on 26 short-tool tasks over 10 MCP services and two long-running \mlzero{} workloads from MLE-Bench. \fw reduces short-tool invocation cost by $8$--$12\times$ relative to VMs and $43$--$106\times$ relative to managed runtimes, reduces input tokens by up to $88\%$ and is $3.64$--$4.24\times$ cheaper than DF at the agent level and $2.67$--$8.23\times$ cheaper at the orchestrator level for long-running workflows (\S\ref{sec:results}). 
\end{enumerate}

A preliminary version of this work appeared in IEEE ICWS 2026~\cite{kulkarni2026optimizingfaasplatformsmcpenabled}. It focused on short-running agentic applications, cross-turn agent memory injection, and FaaS MCP optimizations such as S3 handle passing and tool-output caching. This article adds a critical \textit{long-running service-orchestration layer} that is increasingly important. It identifies timeout and idle-billing failure modes (\S\ref{sec:motivation:failures}), introduces within-turn checkpointing and callback-based suspend/resume across both the agent--tool and orchestrator--agent bridges (\S\ref{sec:fpp}), contrasts \fw from DF and Step Functions through checkpoint/replay and timeout comparisons (\S\ref{sec:results:failures}), and evaluates two long-running \mlzero{} workloads from MLE-Bench (\S\ref{sec:wf}, \S\ref{sec:results:setup}).

\section{Background} \label{sec:bf}

\textit{Agentic workflows} decompose an LLM-driven task into repeated reasoning, tool invocation, observation and state update steps. In this paper, we evaluate four representative patterns: \textit{\react}, which iterates over plan--act--observe steps~\cite{yao2023react}; \textit{\reflect}, which adds self-reflection over failed trajectories~\cite{10.5555/3666122.3666499}; \textit{\llmc}, which compiles tool calls into a dependency graph~\cite{llmcompiler}; and \textit{\mlzero}, which uses iterative code generation and execution for machine-learning engineering tasks~\cite{fang2025mlzero}. These patterns differ in control flow, but all form composite services that depend on explicit workflow state and external MCP service invocation.

\textit{Agent memory} is the workflow state that allows an agent to use prior context rather than restarting from an empty prompt. We distinguish state across three scopes: (i) Within an invocation across reasoning and tool-use steps, (ii) Across turns within a user session, and (iii) Across sessions. Prior works identify memory as a core requirement for agentic systems~\cite{sumers2024cognitive,langchain_memory_overview}. In cloud deployments, this state must be externalized because FaaS functions are stateless. Otherwise, later turns can fail due to missing context or repeat earlier tool calls, increasing latency and LLM input tokens.

\textit{MCP services} provide the tooling plane for these workflows. It exposes tools for discovery and invocation, while the agent decides which tool to call and how to use the returned output~\cite{mcp_intro_docs}. This separation improves modularity, but also makes MCP input/output a systems bottleneck: large tool responses can inflate prompts and LLM context usage, repeated tool calls can increase latency, and long-running tools can block agent execution. Recent MCP benchmarks show that practical agentic workloads increasingly compose many MCP tools, making this plane important for runtime performance~\cite{mo2025livemcpbench,mcpmarket_directory}.

\textit{Function-as-a-Service (FaaS) platforms} such as AWS Lambda and Azure Functions provide event-driven execution, auto-scaling and fine-grained billing~\cite{faas,faas-tpds-arxiv}. These properties match bursty agentic traffic, but FaaS functions are stateless and time-limited. Thus, an agentic FaaS platform must externalize workflow state, avoid repeatedly moving large MCP payloads, and handle external tools whose execution exceeds the function invocation window.
\textit{AWS Lambda Durable Functions (DF)} address long logical executions by checkpointing progress and resuming from events or callbacks~\cite{aws_lambda_durable_docs,aws_durable_key_concepts}. However, their cost depends on the number of durable operations, callbacks, checkpoints and replay steps. This makes the execution runtime and billing model central to the design of FaaS-based agentic workflows.

\begin{figure}[t!]
\centering
  \subfloat[Research Paper Summarization with \react on VM, AgentCore, and \fw at different RPS.]{%
     \includegraphics[width=0.65\columnwidth]{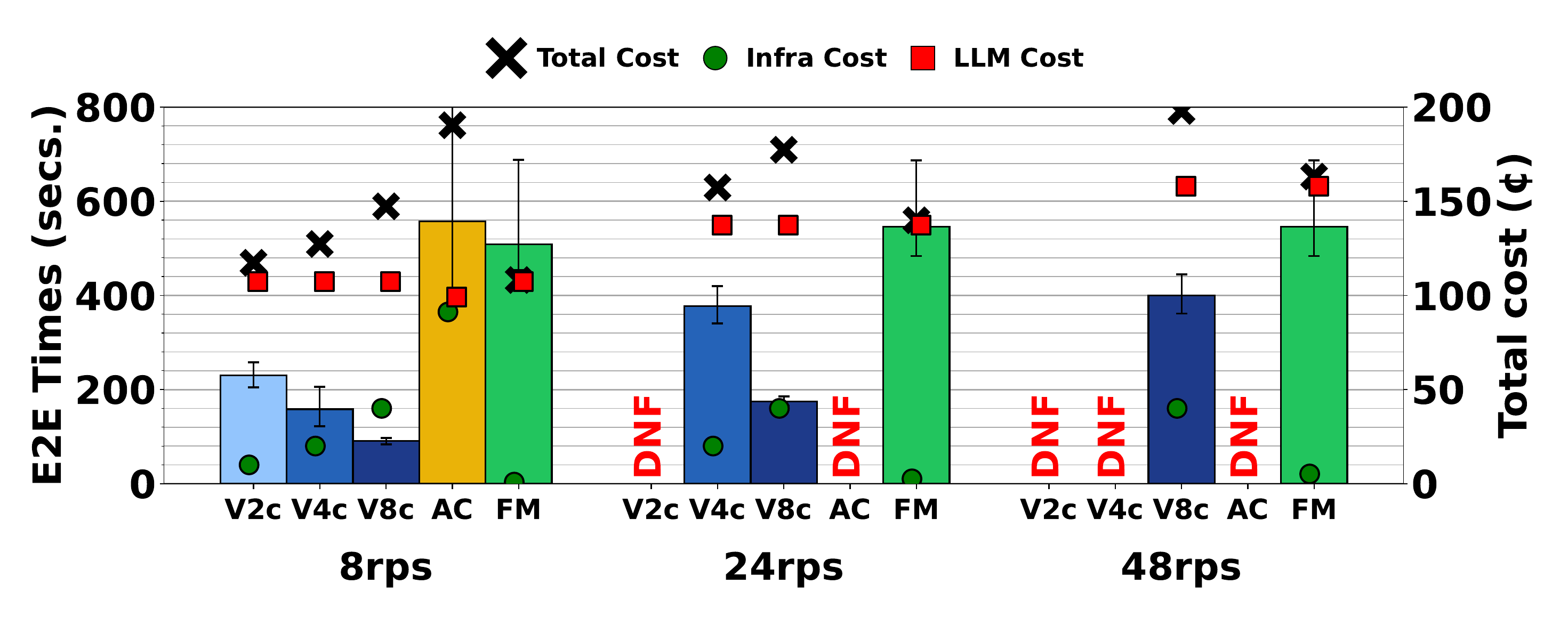}%
     \label{fig:motivation}%
  }\\
  \subfloat[Long-running \mlzero{} Tabular-2022 on Always-on (AC=AgentCore, VM) vs FaaS (SFN=Step Functions, DF=Durable Functions, \fw).]{%
     \includegraphics[width=0.65\columnwidth]{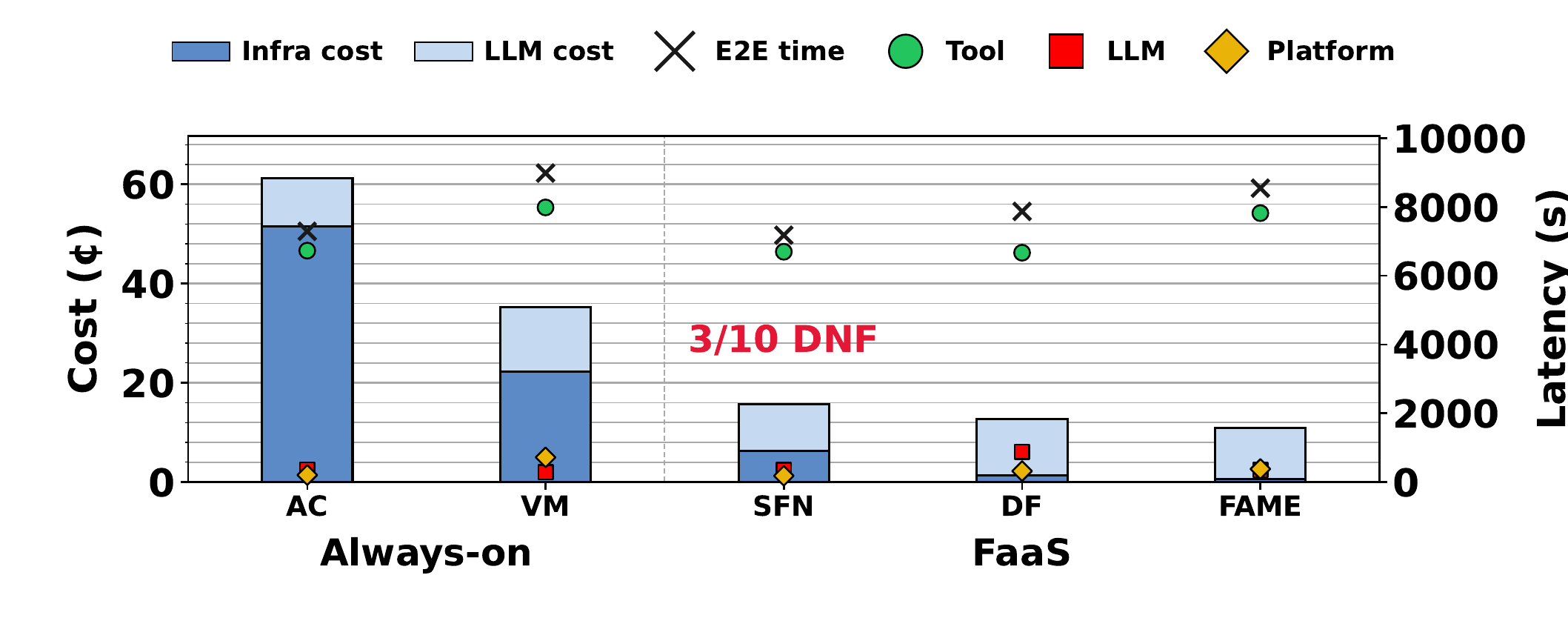}%
     \label{fig:claim1}%
  }
\caption{Cost--time trade-offs motivating FaaS for agentic workflows.
\fw and DF keep infra cost low during long tool waits; SFN times out on 3/10 coder iterations (DNF).}
\end{figure}

Next, we present studies that motivate the observed service runtime trade-offs. The complete workload definitions and platform configurations are described in \S\ref{sec:results:setup}.

\subsection{Why FaaS for Bursty Short-running Service Workflows?}
\label{sec:motivation:bursty} 
Agentic services often see bursty request arrivals across users and tenants. Fig.~\ref{fig:motivation} compares \textit{\fw} with \textit{EC2 VM-based} and \textit{AWS AgentCore} managed agentic runtimes, for a representative \textit{Research Paper Summarization} workflow using \textit{\react}. The workload issues bursts of $8$, $24$, and $48$ requests/s, followed by $180$\,s idle periods, repeated five times. 

\fw lowers platform cost by $8$--$12\times$ relative to different VM sizes and $43$--$106\times$ relative to AgentCore. This is because FaaS scales with the burst and stops billing during idle periods, while always-on VMs continue to pay for provisioned resources while managed runtimes charge a premium. Since total cost is still dominated by LLM token use, the platform-cost reduction translates into an overall $8$--$43\%$ cost reduction. \fw's latency remains comparable to the managed runtime, with modest overhead relative to the largest VM configuration.

\subsection{Why Async Invocation for Long-running External Services?} \label{sec:motivation:failures} 
\textit{Long-running agentic workflows} expose a different bottleneck. In \mlzero from MLE-Bench~\cite{fang2025mlzero}, the Coder agent invokes an external sandbox MCP service that trains or evaluates code for minutes to hours. During this period, a synchronous agent does no useful compute, but the platform may continue billing the agent and orchestrator runtime. 

Fig.~\ref{fig:claim1} shows this for the Tabular-2022 \mlzero{} workload. VM and AC runtimes complete the run, but retain billable resources while the sandbox executes, leading to $35$--$81\times$ higher cost than \fw.
A synchronous AWS Step Function deployment is cheaper than always-on runtimes, but it blocks inside the constituent Lambda function and times out on 3 of the 10 Coder iterations, at the 900\,s function limit. In contrast, our async FaaS \fw  runtime suspends during the external tool wait and resume from callbacks, avoiding both idle billing and function timeouts. 

These two observations motivate \fw: short-running service workflows need cross-turn state and MCP service I/O reduction to control prompt cost and latency, while long-running External MCP services need timeout-safe async invocation across both the agent-service and orchestrator-agent bridges.

\section{\fw Architecture\pgs{2.5}}
\label{sec:arch}

\textit{\underline{F}aaS \underline{A}gentic Workflow \underline{M}iddleware for MCP and Async \underline{E}xecution (\fw)} is our service-oriented middleware for deploying MCP-enabled stateful agentic FaaS workflows~\cite{mcp_intro_docs,mo2025livemcpbench}.
We view an agentic workflow as a composite service in which LLM agent roles provide decision services, MCP services provide tool access, and the orchestrator manages stateful service composition. 
This is analogous to QoS-aware service composition but with LLM-selected service invocations and MCP-specific state and payload concerns~\cite{zou2014qos,jatoth2017qos}.

Its design follows three principles.
\textit{First}, agentic patterns are decomposed into role-specific functions, so that each role can be configured, scaled and retried independently. 
\textit{Second}, workflow state is externalized from the FaaS runtime, allowing session to preserve context within and across turns without keeping functions alive~\cite{faas,faas-tpds-arxiv,sumers2024cognitive,langchain_memory_overview}. 
\textit{Third}, \fw optimizes the MCP service plane: short-running MCP services are wrapped as FaaS endpoints~\cite{mcp_intro_docs,faas,faas-tpds-arxiv}, while long-running or resource-heavy services remain external and are accessed through async MCP bridges with checkpointed suspend/resume~\cite{aws_lambda_durable_docs,aws_durable_key_concepts,chan2025mlebench}. 

Fig.~\ref{fig:arch} shows the long-running execution path used for External MCP (EMCP) tools, while Fig.~\ref{fig:fame_arch} in Appendix shows a \react workflow with short MCP tool-calls.
These principles support the short-tool optimizations evaluated in \S\ref{sec:results:e2e}--\ref{sec:results:benchmarks}; timeout-safe execution for long-running external tools is described separately in \S\ref{sec:fpp}.
\fw is open-sourced at \url{http://github.com/dream-lab/FAME}.

\subsection{Role-decomposed FaaS Orchestration} \label{sec:arch:faas}

\begin{figure}[t!]
\centering%
    \includegraphics[width=0.8\columnwidth, 
    clip]{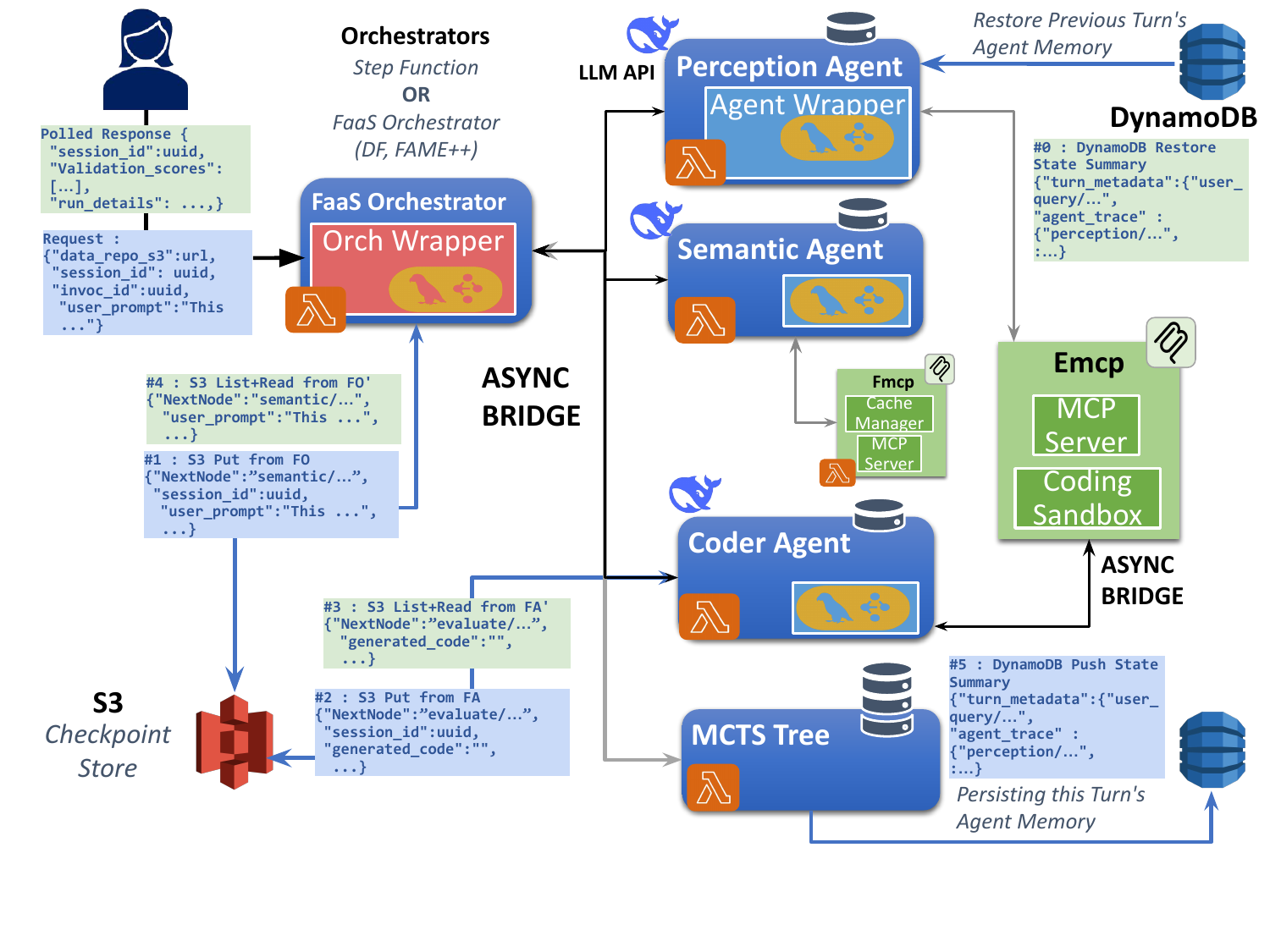}
\caption{Architecture of \fw for long-running External MCP (EMCP) workflows. The orchestrator and agent roles run as FaaS components, while long-running tools remain external. \fw checkpoints state and suspends execution across both the orchestrator-agent and agent-tool bridges, then resumes from callbacks when the downstream execution completes.}
    \label{fig:arch}
    
\end{figure}

\begin{table*}[t!]

\centering
\caption{Agentic patterns mapped to \fw. Implementation details are in Appendix~\S\ref{app:agent:implementation}.}
\label{tbl:patterns}
\footnotesize
\setlength{\tabcolsep}{2pt}
\renewcommand{\arraystretch}{0.9}
\begin{tabular}{L{2.15cm} L{3cm} L{5cm} L{5cm}}
\toprule
\textbf{Pattern} & \textbf{Roles (Lambda, MB)} & \textbf{Control Flow} & \textbf{Memory Model} \\
\midrule
\bf \react (RT)~\cite{yao2023react}
  & Planner, Actor, Evaluator (512)
  & Iterative plan$\to$act$\to$evaluate loop (Fig.~\ref{fig:react}); Evaluator routes failure feedback to Planner for replanning
  & Cumulative intra-session AM (default) \\
\midrule
\bf \reflect (RX)
~\cite{10.5555/3666122.3666499}
  & Actor, Evaluator, Self-Reflection (256)
  & Rather than replanning on failure, agent self-reflects on its trajectory; verbal feedback steers subsequent attempts
  & Two-tier: short-term trajectory + long-term reflections (App.~\S\ref{app:fame_memory:reflect}) \\
\midrule
\bf \llmc (LM)~\cite{llmcompiler}
  & Planner, Executor, Joiner (256); TFU~\texttt{(128)}
  & Compiles request into DAG of concurrent tool calls, exposing parallelism where dependencies permit
  & Default AM: Planner injects prior session context; Joiner saves summary, DAG structure, tool outputs \\
\midrule
\bf \mlzero ~\cite{fang2025mlzero}
  & Percep., Semantic, Coder (512); MCTS, SyncS3 (128)
  & MCTS optimizer~\cite{liang2025imcts} iteratively expands tree; Coder generates and executes code in sandbox
  & Default AM via Perception; MCTS tree grown per iteration (code, stdout/err, evals) \\
\bottomrule
\end{tabular}

\end{table*}

\fw maps the \textit{agents} of an agentic workflow to a set of \textit{stateless FaaS roles} choreographed by an \textit{orchestrator}. Each role contains a small \textit{LangGraph subgraph}~\cite{langgraph_overview_docs} with its own prompt, resource configuration and state interface. \textit{MCP tools} are invoked through a service plane that may be FaaS-wrapped for short-running tools or external for long-running ones~\cite{mcp_intro_docs,faas,faas-tpds-arxiv}. This separation avoids hosting the complete agent workflow as a monolithic service, while allowing the tool runtime to pick the layer best suited for its execution needs.

The \textit{orchestrator workflow} is itself a FaaS function in Fig.~\ref{fig:arch}: it invokes agent roles, routes workflow state and either waits synchronously for short-running roles or uses the async path in \S\ref{sec:fpp} for long-running roles.

Table~\ref{tbl:patterns} summarizes the four patterns we implement~\cite{yao2023react,10.5555/3666122.3666499,llmcompiler,fang2025mlzero}. They differ in control flow, from sequential plan--act--evaluate loops to DAG execution and MCTS search, but all form composite services that depend on workflow state and LLM-directed MCP service calls~\cite{zou2014qos,jatoth2017qos}.
\mlzero{} differs from the first three short-tool patterns because its Coder role invokes an EMCP sandbox whose execution can exceed the FaaS function lifetime. We therefore use \mlzero{} as the driving workload for the timeout-safe async execution design (\S\ref{sec:fpp}). The agent roles remain FaaS functions, while the long tool call remains outside FaaS and is accessed through checkpointed async bridges.

\subsection{Persistent Agent Memory across Turns}
\label{sec:arch:mem}

Agentic workflows require state even when each FaaS function is stateless. \fw separates three forms of state. \textit{Client Memory (CM)} is the user-visible request and response history~\cite{langchain_memory_overview}. \textit{Agent Memory (AM)} is the workflow execution state, including intermediate LLM messages, MCP inputs and outputs, tool handles and role-level decisions~\cite{sumers2024cognitive,langchain_memory_overview}. The same memory abstraction records tool outputs and handles from both FaaS-wrapped MCP tools and EMCP tools, so downstream roles can reuse prior results without depending on where the tool executed. Lastly, the \textit{actor memory prompt} is an instruction that tells the Actor when to reuse prior tool outputs rather than issuing redundant MCP calls.

Fig.~\ref{fig:mem} in Appendix illustrates these memory modes for \react: empty memory uses only the current request, client-native memory adds prior user-visible turns, and AM injects internal execution state to preserve context and avoid redundant tools.

\fw persists AM in DynamoDB using session and invocation identifiers. On each workflow invocation, \fw retrieves the relevant prior AM, compresses or replaces large tool payloads with S3 handles, and injects the resulting state into the role that needs it. On completion, the Evaluator or Joiner persists the updated state for later turns. This mechanism avoids the two failure modes seen with stateless execution: later turns can fail because earlier tool outputs are missing in the state, or they can repeat earlier tool calls and re-ingest large payloads into the prompt, stressing the LLM context size.

The memory model is specialized by pattern. \react and \llmc use cumulative intra-session AM. \reflect additionally stores short-term trajectories and long-term reflections to steer retries. \mlzero persists tree state across iterations, including generated code, execution output, evaluation results, and selected MCTS nodes. These variants use the same externalization mechanism, but differ in what state each pattern needs to resume or improve later steps.

\subsection{MCP Service Automation and I/O Optimizations}
\label{sec:arch:mcp}

\fw distinguishes between \textit{FaaS MCP services} and \textit{External MCP services}. 
FaaS MCP services are used for short-running service invocations that can execute within the function lifetime and benefit from elastic scaling, fine-grained billing, S3 handle passing, caching and consolidation.
EMCP services are used for long-running or resource-heavy service endpoints, such as code-execution sandboxes or scientific services, whose execution can exceed FaaS limits or require specialized resources~\cite{chan2025mlebench,gao2025democratizingaiscientistsusing}.

For \textit{EMCP tools}, \fw retains the service within their external hosting environment (e.g., VMs, baremetal). Instead, it optimizes the agent-side communication path using async dispatch, checkpointed suspension and callback-based resume.
For \textit{FaaS MCP tools}, \fw wraps MCP servers as FaaS endpoints by normalizing tool inputs and outputs for HTTP invocation, preserving MCP tool semantics and collecting telemetry such as execution duration and resource usage. This allows workflow developers to deploy short-running MCP tools as elastic cloud functions rather than maintaining long-running services/VMs for bursty interactions.

\fw applies two input/output \textit{optimizations} to FaaS MCP tools. First, \textit{S3 handle passing} stores large tool outputs in AWS S3 and passes object handles through prompts and downstream tools. Second, \textit{tool-output caching} stores deterministic MCP outputs under argument-derived keys in S3 
with developer-configured time-to-live  (TTL) values. These optimizations reduce repeated prompt payloads, redundant tool execution and LLM input tokens.

For FaaS MCP services, \fw supports both singleton and consolidated layouts~\cite{agentx}. \textit{Singleton deployments} assign one function to each MCP service or tool, improving isolation and resource sizing. \textit{Consolidated deployments} combine frequently co-invoked tools into one function service, amortizing cold starts and reducing steady-state invocation overhead.

\section{Timeout-safe Async Service Invocation for Long-running Workflows}
\label{sec:fpp}

Long-running agentic workflows expose a mismatch between FaaS execution limits and external service runtimes. 
In \mlzero{} workloads~\cite{fang2025mlzero,chan2025mlebench}, the Coder agent invokes an EMCP sandbox that may run training or evaluation jobs for minutes to hours. If the Coder Lambda function waits synchronously, the FaaS can hit the function lifetime limit. If the orchestrator waits synchronously for the Coder, the orchestrator also becomes a long-running FaaS execution, even though it is only waiting for the agent callback. Thus, timeout-safe service invocation must cover both the agent--tool-service bridge and the orchestrator--agent bridge.

Fig.~\ref{fig:arch} shows this async path in \fw. The agent--tool bridge dispatches the EMCP tool, checkpoints the agent FaaS's state and suspends the agent. The orchestrator--agent bridge dispatches the long-running agent role, checkpoints orchestration FaaS's state and suspends the orchestrator. Both bridges resume from callbacks, so neither the Coder agent nor the orchestrator are billed during the external tool wait.

\subsection{Agent--Tool Service Async Bridge} \label{sec:fpp:agent-tool}

\begin{figure}[t!]
\centering%
    \includegraphics[width=0.9\columnwidth, 
    clip]{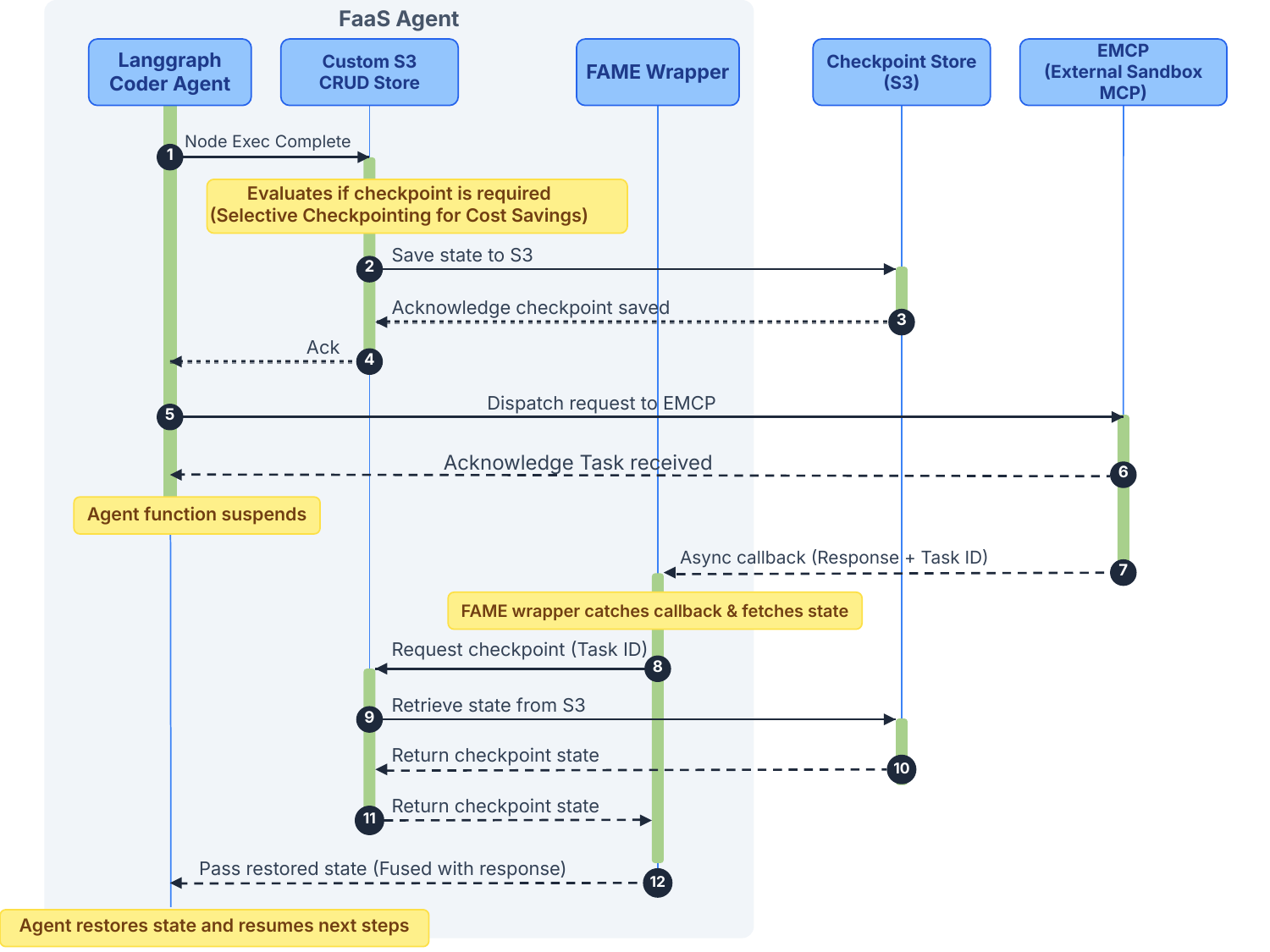}

\caption{\fw checkpoints state and suspends execution across agent-tool bridges, then resumes from callbacks when the downstream execution completes. The pattern is similar for the orch-agent bridge}
    \label{fig:arch_concise}
    
\end{figure}

The lower path in Fig.~\ref{fig:arch} and Fig.~\ref{fig:arch_concise} detail how an agent invokes a long-running \textit{External MCP (EMCP)} service without remaining active during the wait. When the agent dispatches an EMCP call, the service returns an ACK and a MCP \texttt{TaskId}. \fw uses this \texttt{TaskId} as the LangGraph \texttt{ThreadId} for checkpoint correlation. The agent's LangGraph state, agent memory, pending tool-call arguments, service metadata, callback target, and execution metadata are serialized through the LangGraph checkpointer and stored using a custom S3 CRUD backend. The execution metadata also records the \texttt{NextNode}, which identifies where the agent graph should continue after the tool result arrives. Once the checkpoint is committed, the Lambda call returns, suspending the agent FaaS execution.

The EMCP service returns results through a callback to the agent's Lambda \texttt{FunctionUrl}. This avoids polling, whose cost grows with the number of status checks, and instead makes response delivery independent of the tool duration. The callback enters a \fw resume-mode path in the Lambda handler, uses the returned \texttt{TaskId}/\texttt{ThreadId} to locate the latest S3 checkpoint, deserializes the LangGraph state, attaches the tool result or S3 result handle, and resumes execution from the stored \texttt{NextNode}. Thus, the agent consumes FaaS time only for dispatch, checkpoint, callback, restore and resume.

\subsection{Orchestrator--Agent Async Bridge} \label{sec:fpp:orchestrator-agent} 

The upper path in Fig.~\ref{fig:arch} applies the same suspend/resume pattern between the orchestrator and a long-running agent role such as the Coder, analogous to Fig.~\ref{fig:arch_concise} but with orchestrator being the FaaS and Coder being the long duration invocation. The orchestrator internally uses a LangGraph wrapper for the long-running agent call. Before dispatching the Coder, \fw checkpoints the orchestration state, pending agent invocation metadata, callback target, and continuation state through the same S3-backed checkpointing interface. The orchestrator then invokes the Coder asynchronously and returns after receiving an ACK, rather than waiting while the Coder blocks on the EMCP sandbox.

When the Coder completes, it invokes the orchestrator callback. 
The callback restores the orchestration checkpoint from S3, attaches the Coder result to the workflow state, and resumes the orchestrator from the resume point. This separates the billable lifetime of the orchestrator from the billable time of the Coder and the external sandbox, while preserving the same logical workflow order as synchronous execution.

\subsection{Selective Checkpointing}

\fw builds on the LangGraph checkpointer but replaces the default checkpoint store with a custom S3 CRUD backend~\cite{langgraph_checkpoint_api}. A na\"{i}ve implementation checkpoints every graph node, which increases S3 Put/List/Get operations and callback-restore overhead. \fw instead adds a \texttt{checkpoint\_required} state flag. Nodes that can dispatch a long-running service set this flag before execution; other nodes remain on the synchronous inline path.

The long/short decision is made from the MCP service profile by the workflow designer rather than by waiting at runtime; short-running FaaS MCP services execute synchronously.
EMCP services are marked async when they use external resources, have tool-side timeouts that can approach/exceed the Lambda limit, or are known from the workflow design to run for minutes to hours, e.g., the \mlzero{} sandbox. 

Selective checkpointing is the key distinction between \fw and durable execution platforms such as AWS DF, which records durable operations and replays execution on resume~\cite{aws_lambda_durable_docs,aws_durable_key_concepts}. \fw stores explicit LangGraph state only at service suspension boundaries, i.e., the two bridges. 
Implementation details are in Appendix~\S\ref{app:agent:implementation}.

\subsection{Cost Implications} \label{sec:fpp:cost}

The async design changes the cost driver from wait duration to resume operations. In synchronous VM, AgentCore or Step Functions, platform cost scales with the EMCP wait because the agent or orchestrator remains allocated. Polling also scales with wait duration through repeated billed checks. \fw instead uses callbacks: EMCP invokes the agent \texttt{FunctionUrl}, and the agent invokes the orchestrator callback, so cost is independent of tool runtime after dispatch.

A \fw checkpoint set uses (cheaper) S3 \texttt{Put}, \texttt{List} and \texttt{Get}: storing the serialized LangGraph checkpoint, finding the latest checkpoint for the \texttt{ThreadId}, and restoring it on callback. DF instead charges durable \texttt{invoke}, \texttt{callback}, \texttt{step}, data-written and replay operations. Thus, \fw may store larger explicit checkpoints but for fewer durable operations, and avoids replaying intermediate graph steps.

This explains the long-running results in \S\ref{sec:results:failures}: Step Functions times out on synchronous waits, DF and \fw avoid the timeout, and \fw is cheaper because it checkpoints only at service suspension boundaries using the LangGraph state and S3-backed checkpoint store.

\section{Service Workloads and Experimental Setup}
\label{sec:wf}
\label{sec:results:setup}

We evaluate \fw along two service-execution dimensions. 
First, we study \textit{short-running MCP service invocations} using \react, \reflect and \llmc across four applications and five memory/cache configurations. These workloads stress cross-turn service context, MCP service input/output reduction, FaaS MCP wrapping, caching and S3 handle passing. 
Second, we study \textit{long-running EMCP services} using \mlzero{} on two MLE-Bench tasks. 
These stress timeout-safe async service invocation, within-turn checkpointing, callback-based suspend/resume and idle-billing across service runtimes.
Accordingly, short-running workflows are orchestrated with AWS Step Functions, whereas long-running EMCP workflows use \fw's callback-driven async orchestrator.

All experiments run on AWS \texttt{ap-south-1}. Short-tool workflows use AWS Lambda and Step Functions, with all agent roles calling the OpenAI \textit{GPT-4o-mini} endpoint. Long-running \mlzero{} workflows use \textit{Deepseek-V4-pro} through OpenRouter across all platform configurations; both use temperature$=0$. For each comparison, prompts, services, model settings and task inputs are fixed. So measured differences reflect service runtime, memory, MCP service I/O and checkpointing behavior rather than prompt or service variation.

\subsection{Short-running MCP Service Workloads} \label{sec:wf:short}
The short-running workloads span multi-turn applications, benchmark tasks, payload-heavy service interactions and payload-light database lookups. The MCP tools are exposed through FaaS MCP service deployments, allowing \fw to apply S3 handle passing, service-output caching and singleton or consolidated MCP layouts. In total, these exercise 26 service tasks over 10 MCP services and 54 exposed tools, of which 24 tools are invoked by the evaluated tasks.

\subsubsection{Research Paper Summarization (RS)}\label{sec:wf:app:rs}

This is a three-turn session that summarizes a paper from its title~\cite{pan2025experiencesmodelcontextprotocol}. The Q1--Q3 prompts ask for the introduction and contributions, methods and analysis, and conclusions and implications. The expected Q1 execution downloads the paper PDF using an \textit{ArXiv MCP service}~\cite{arxiv_mcp} and summarizes sections using a \textit{RAG MCP service}~\cite{rag_mcp}. The user does not repeat the paper title in Q2 or Q3, so later turns must recall the Q1 paper context through memory injection rather than re-discovering/processing the PDF. We evaluate three papers (P1--P3) of sizes 5.6\,MB, 2.1\,MB, and 3.7\,MB~\cite{kulkarni2026optimizingfaasplatformsmcpenabled}. This workload is payload-heavy because tools fetch and process large PDF artifacts.

\subsubsection{Log Analytics (LA)}\label{sec:wf:app:log}

This is a three-turn session over system logs~\cite{kulkarni2026optimizingfaasplatformsmcpenabled}. The prompts count error states, summarize timestamp statistics for the most frequent error, and generate comparative statistics with a visualization. The workflow uses 3 MCP servers: \textit{Log Analyzer} for fetching and filtering logs~\cite{logscribe_mcp}, \textit{Calculator} for aggregate statistics~\cite{calculator_mcp_server}, and \textit{Visualization} for generating and executing plotting code~\cite{visualization_mcp_server}. We evaluate Apache, Hadoop and OpenSSH three log files (F1--F3) of sizes 170\,KB, 380\,KB and 220\,KB from LogHub~\cite{loghub}. \fw's pass by reference optimization using S3 handles is to avoid repeatedly embedding log contents in prompts.

\subsubsection{TauBench (TB)}
\label{sec:wf:app:tb}
This evaluates multi-turn agentic workflows in task-oriented domains~\cite{taubench}. We use 10 read-only tasks from the \textit{airline domain}, covering reservation retrieval, membership checks and policy enforcement. The database is hosted on an EC2 VM, 
and \textit{database operations} are exposed as MCP tools. These tasks are payload-light, so caching has limited benefit; memory primarily avoids redundant tool calls and preserves prior task context.

\subsubsection{LiveMCPBench (LM)}
\label{sec:wf:app:livemcp}
This evaluates agentic workflows that coordinate across multiple MCP servers for real-world tasks~\cite{mo2025livemcpbench}. We select 10 tasks covering \textit{report generation}, \textit{trend analysis} over GitHub and arXiv, \textit{web reconnaissance} and \textit{news/financial aggregation}. These are single-turn tasks but often payload-heavy, producing PDF/Markdown/JSON artifacts. They stress S3 handle passing and MCP input/output reduction, while cache hits are less common because task-local tool calls usually have unique signatures.

\subsection{Long-running External MCP Service Workloads} \label{sec:wf:long}

We use \mlzero{} for long-running EMCP service evaluation~\cite{fang2025mlzero}. It decomposes \textit{Machine-Learning Engineering (MLE)} into Perception, Semantic, Coder and MCTS roles. Coder invokes an EMCP sandbox service that executes generated code, trains models and returns evaluation results. This tool execution can exceed a single FaaS invocation window, so the sandbox remains external while \fw leverages async 
operations across the bridges.

We evaluate two MLE-Bench tasks~\cite{chan2025mlebench}. \textit{Tabular-2022} trains a binary classifier over a 0.57\,MB dataset and runs for 10 MCTS iterations. \textit{Dog-Breed} trains a multi-class image classifier over a 0.75\,MB dataset and runs for 3 iterations. The EMCP sandbox runs in a Docker container on an EC2 \texttt{m6i.xlarge} instance and enforces a 1\,hour tool-side timeout. The service has four tools: \texttt{exec\_sandbox}, \texttt{write\_file}, \texttt{read\_file} and \texttt{kill\_processes}. Each configuration has its own sandbox to avoid interference.

\subsection{Service Runtime Baselines} \label{sec:wf:baselines}
For the bursty short-tool study, we compare \fw against two service runtime classes. The \textbf{VM} baseline hosts the same agent graphs and MCP servers on EC2 \texttt{c7i.*} instances with 2/4/8 vCPUs. The managed-runtime uses AWS Bedrock AgentCore (\textbf{AC}) with distributed agent runtimes. All baselines use identical tools, prompts, model settings and task inputs.

For long-running EMCP workflows, we compare five service runtime configurations. \textbf{VM} deploys the same agent graphs on a single EC2 \texttt{c7i.large}; only the agent-side EC2 wall-clock time is counted as VM cost. \textbf{AC} deploys the distributed agent graphs on AgentCore. \textbf{Step Functions} uses AWS Lambda functions, which lack checkpointing, so the Coder Lambda waits synchronously for EMCP completion. \textbf{DF} deploys the orchestrator and Coder using AWS Durable Functions with native checkpointing~\cite{aws_lambda_durable_docs,aws_durable_key_concepts}. 
\textbf{\fw} enables selective checkpointing and suspend/resume for both async bridges.
Appendix Table~\ref{tab:runtime-comparison} contrasts these runtimes.

\subsection{Configurations} \label{sec:wf:configs}
For \textit{short-running service workloads}, we evaluate five memory/cache configurations. \textbf{E} uses empty memory and provides no prior context across turns. \textbf{N} uses client-native memory, consisting of prior user prompts and final responses. \textbf{C} enables MCP service-output caching without agent memory. \textbf{M} enables agent memory injection, where prior execution state and S3 handles are retrieved and injected into the workflow. \textbf{MC} combines agent memory and MCP caching. These separate the effects of user-visible history, internal service execution state and deterministic service reuse (Details in Appendix~\S~C-A).

For \textbf{C} and \textbf{MC}, \fw wrappers enable S3-based MCP caching with developer-set TTL; we default to a 1-day TTL, and the S3 MCP cache is flushed between independent runs. The \textit{actor memory prompt} is enabled for \textbf{M} and \textbf{MC}. It instructs the Actor to inspect prior service messages and reuse matching service outputs before issuing new MCP calls.

For \textit{long-running EMCP service workloads}, we exercise the runtime platforms rather than memory/cache variants: \textbf{VM}, \textbf{AC}, synchronous \textbf{Step} Functions, \textbf{DF} and \textbf{\fw}; we use the \textbf{M} variant for \fw.

\subsection{Metrics and Cost Accounting} \label{sec:wf:metrics}

For \textit{short-running service workloads}, we report end-to-end (E2E) latency per session, per-turn latency, LLM input and output tokens, monetary cost, Actor-level latency splits, MCP service invocations and cache hits. Monetary cost includes LLM tokens and FaaS components such as Lambda, Step Functions, DynamoDB and S3. Deterministic tasks such as LA and TB are validated against expected outputs, while non-deterministic ones such as RS are checked using a predefined LLM rubric. We report averages over three independent runs.

For \textit{long-running EMCP service workloads}, we isolate agent and orchestrator behavior, and 
report infrastructure cost, checkpoint operation cost, storage cost, billed duration, infrastructure latency and Did Not Finish (DNF) outcomes; 
LLM, external service execution and application-function costs are common across runtimes.
Stacked bars decompose infrastructure cost into cold-start overhead, communication/compute and checkpoint operations/storage, while markers report latency and billed duration. All long-running values use the median run, selected by E2E latency across three trials.

While AC does not bill idle CPU during pure I/O waits, long-running sessions are billed for retaining memory footprint and active background work such as waiting for long tool-calls \cite{bedrock_AgentCore_Docs}.
In contrast, \fw fully suspends the FaaS agent and orchestrator and resumes only on callback, mitigating billed time.

Costs are computed from metered durations, storage operations and token counts using provider pricing for the same region and time window. Absolute costs are as listed by provider, but the relative trends are driven by service runtime semantics: always-on runtimes bill idle waits, synchronous FaaS can hit function lifetime limits, and async FaaS shifts cost to dispatch, checkpoint, callback, restore and resume.

\section{Experimental Results \pgs{3.5}}\label{sec:results}

We first evaluate how service runtime choice affects cost and latency for bursty short-running MCP service workflows (\S\ref{sec:results:baselines}), then quantify how agent memory and MCP service I/O optimizations affect completion, latency, tokens and cost (\S\ref{sec:results:e2e}--\ref{sec:results:tokens}). We next examine generalization across agentic benchmarks (\S\ref{sec:results:benchmarks}), and finally evaluate timeout-safe async service invocation for long-running EMCP workflows (\S\ref{sec:results:failures}). We close with a discussion of limitations (\S\ref{sec:results:limitations}).

\subsection{Service Runtime Choice Affects Cost and Latency} \label{sec:results:baselines}

Fig.~\ref{fig:motivation} summarizes the bursty RPS sweep described in \S\ref{sec:motivation:bursty}. \fw achieves $8$--$12\times$ lower platform cost than VM deployments and $43$--$106\times$ lower platform cost than AC. This is because \fw scales FaaS capacity with the burst and stops billing during idle intervals, while VM deployments continue paying for provisioned instances and AC charges managed runtimes during the workflow.

The platform-cost reduction translates into an $8$--$43\%$ reduction in total cost, since LLM token charges still dominate the end-to-end bill. Latency remains comparable to AC and has a modest $1.4$--$2.2\times$ overhead relative to the largest VM configuration. This is the expected trade-off: the largest VM keeps capacity warm and can reduce latency, but it pays higher for that capacity even during idle periods.

These results support using FaaS as the default service runtime for bursty short-running agentic service workflows. The benefit is strongest when request arrivals are sparse or bursty and when the VM baseline is sized for peak demand. The remaining cost is dominated by LLM input tokens, which motivates the memory and MCP input/output optimizations evaluated next.

\begin{figure}[t!]

\centering%
  \subfloat[Research Summary \textit{(RS/P1)}]{
     \includegraphics[width=0.45\columnwidth]{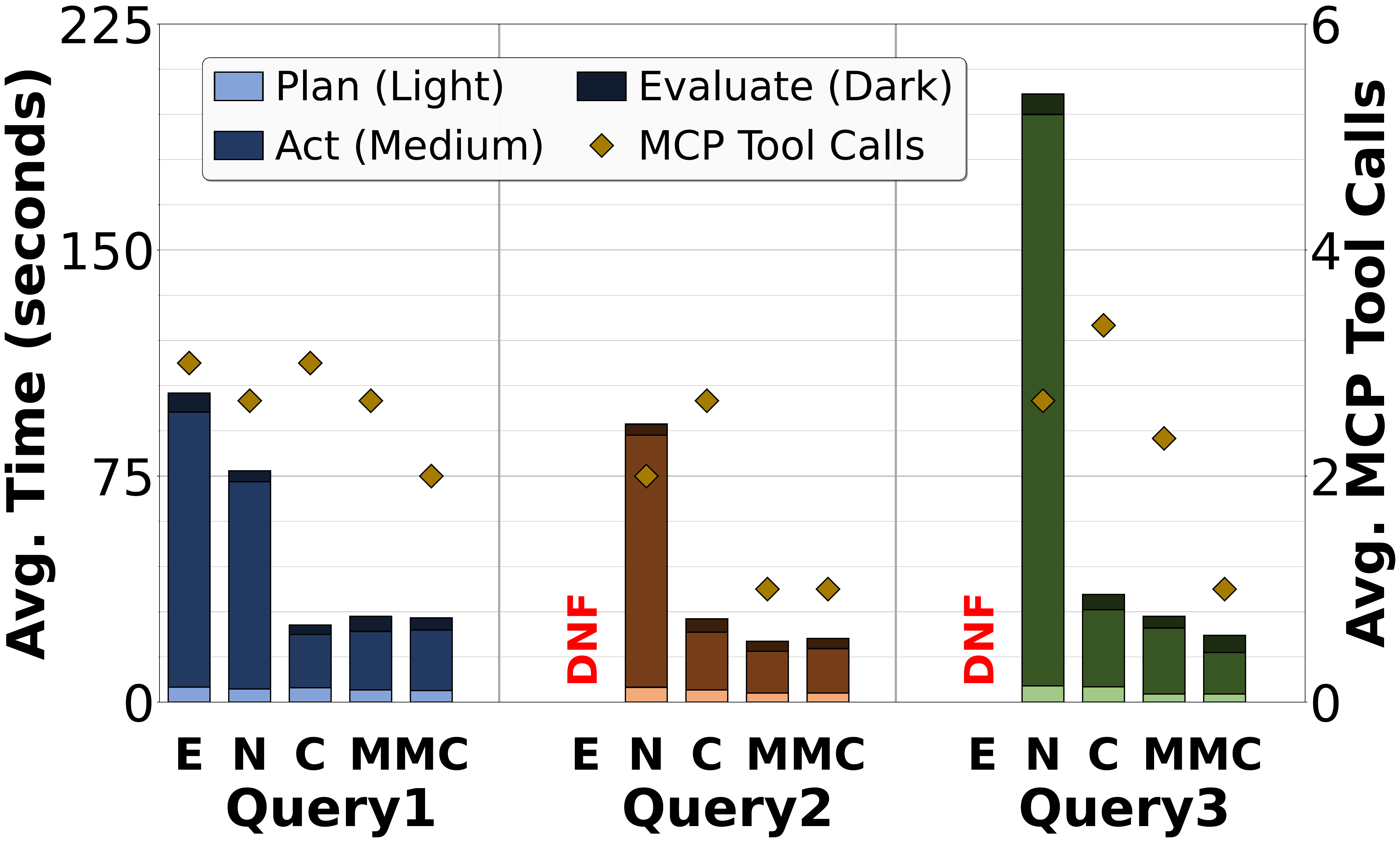}
     \label{fig:arxiv-d-s1-e2e}
  }~~
  \subfloat[Log Analytics \textit{(LA/F1)}]{
     \includegraphics[width=0.45\columnwidth]{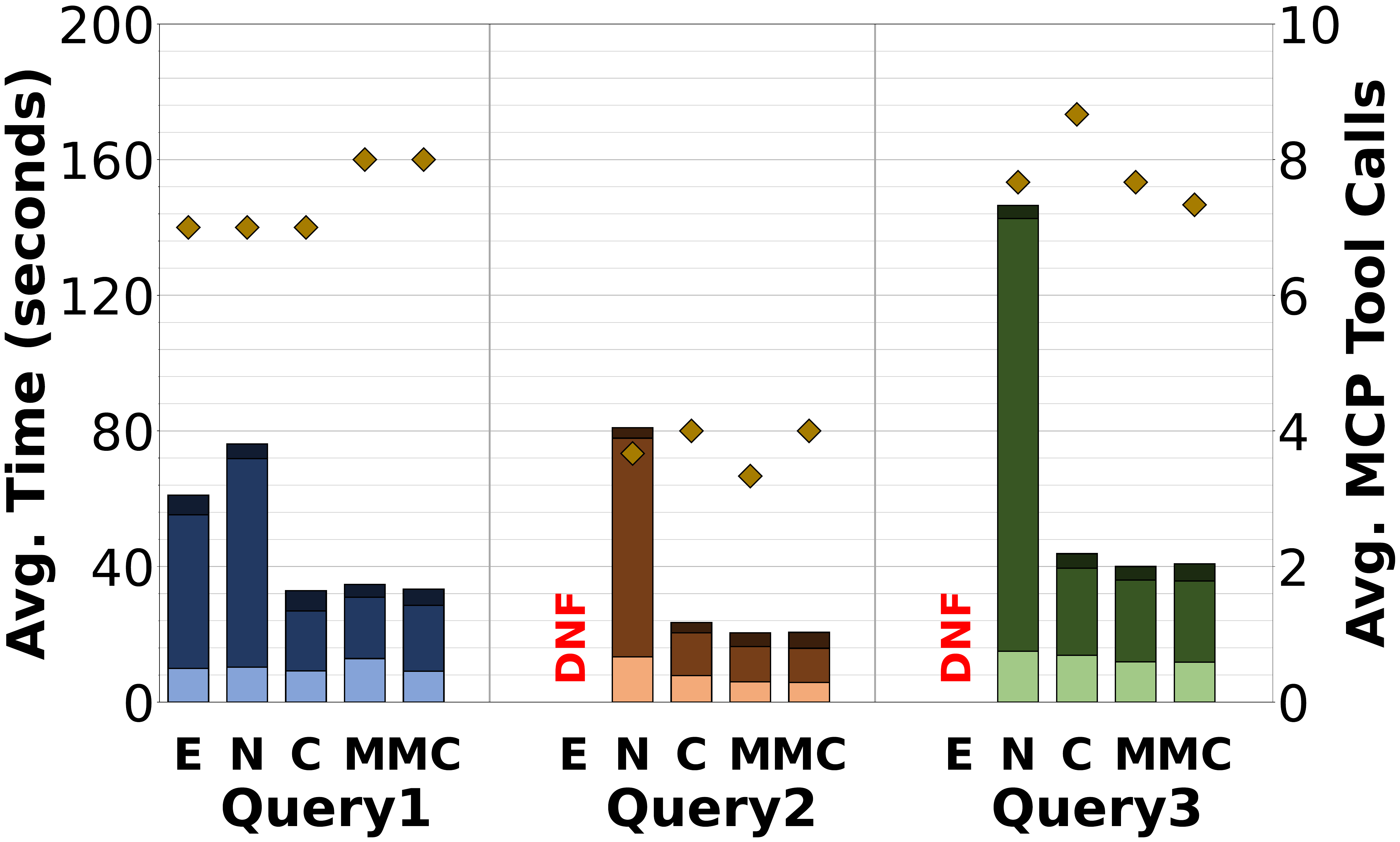}
     \label{fig:logger-d-s1-e2e}
  }\\  
  \subfloat[Research Summary \textit{(RS/P2)}]{
    \includegraphics[width=0.45\columnwidth]{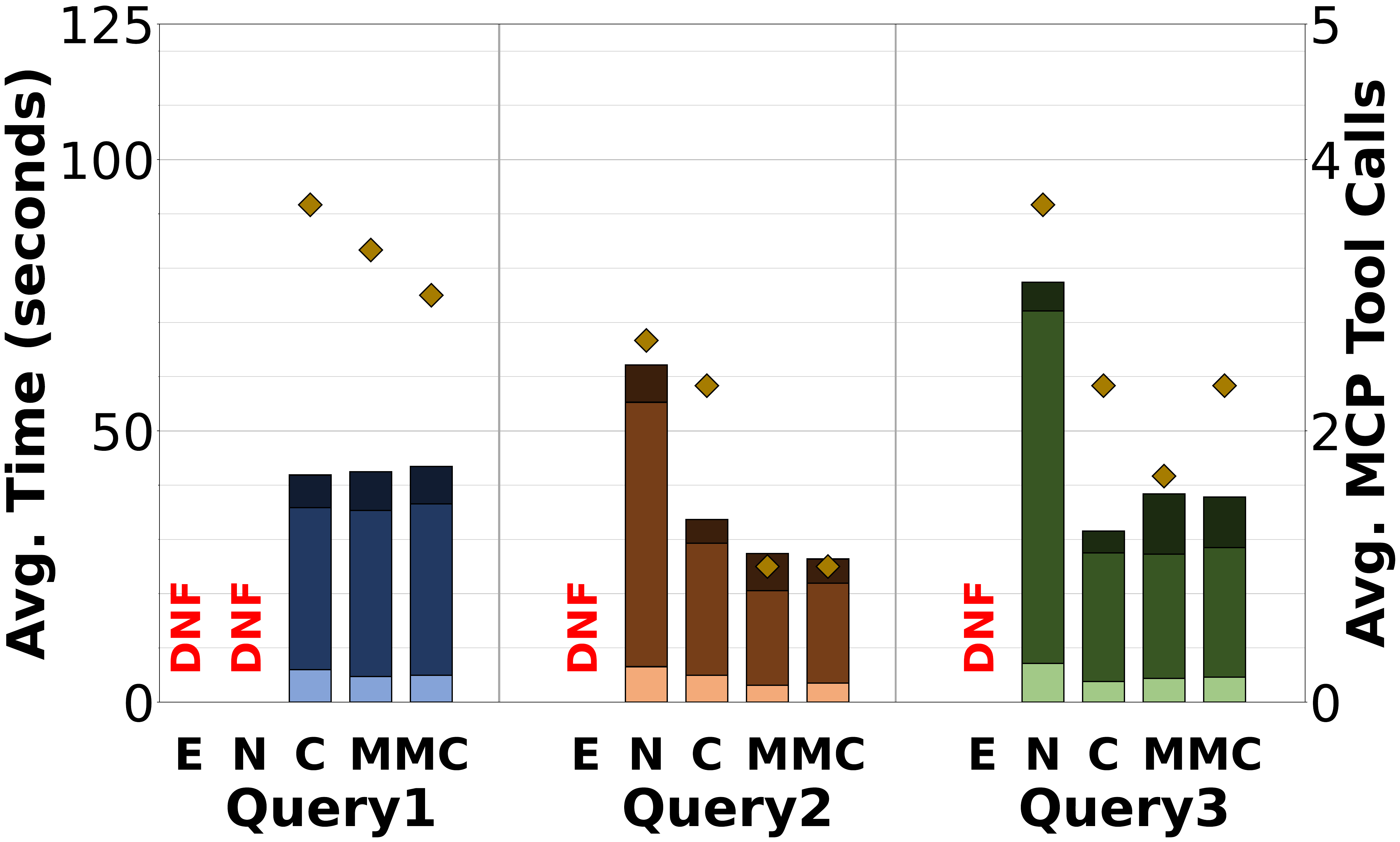}
     \label{fig:arxiv-d-s2-e2e}
  }~~
  \subfloat[Log Analytics \textit{(LA/F2)}]{
    \includegraphics[width=0.45\columnwidth]{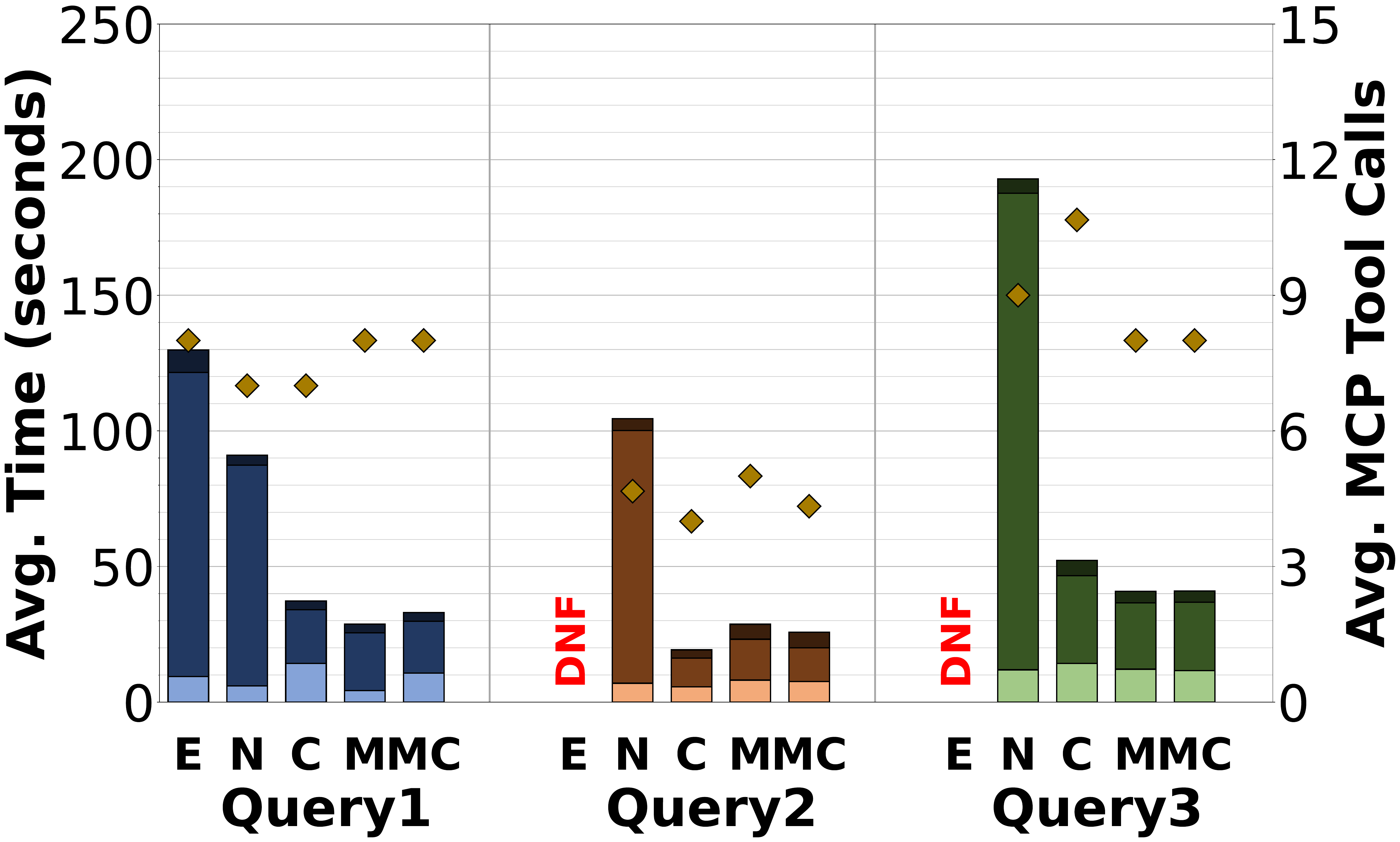}
     \label{fig:logger-d-s2-e2e}
  }\\
\caption{Average per-turn latency for \react on Research Summarization and Log Analytics. Memory and MCP optimizations reduce later-turn latency by avoiding repeated tool calls and large prompt payloads.}
\label{fig:plot-e2e-latencies}

\end{figure}

\subsection{Stateful Service Context Improves Completion and Latency} \label{sec:results:e2e}

\takeaway{Stateful service context is needed for multi-turn sessions.} Fig.~\ref{fig:plot-e2e-latencies} shows per-turn latency for \react on RS and LA. Without persisted context, \textbf{E} DNFs on later turns for all four input files, i.e., queries Q2/Q3 for RS/P1, RS/P2, LA/F1 and LA/F2. \textbf{N} completes more often because it includes prior user-visible turns, but it still repeats work because it lacks the internal service trace and S3 handles stored in AM. In contrast, \textbf{M} and \textbf{MC} inject prior execution state and reuse tool outputs, improving the success rate and avoiding redundant MCP calls.
Plots and analysis for P3/F3 are in Appendix~\ref{app:results:react-detail}. 

\takeaway{Memory and MCP optimizations reduce latency on payload-heavy workflows.} On RS/P1, Q1 latency drops from about $100$\,s under \textbf{E} and $75$\,s under \textbf{N} to about $25$\,s under \textbf{C/M/MC} (Fig.~\ref{fig:arxiv-d-s1-e2e}). The later-turn effect is larger: RS/P1 Q3 is DNF under \textbf{E}, about $200$\,s under \textbf{N}, and about $20$--$35$\,s under \textbf{C/M/MC}. LA shows the same pattern: LA/F2 Q3 falls from about $190$\,s under \textbf{N} to about $40$--$55$\,s under \textbf{C/M/MC} (Fig.~\ref{fig:logger-d-s2-e2e}). This is because caching and S3 handle passing avoid re-ingesting large artifacts, while AM avoids repeated tool calls.

\takeaway{The benefit depends on the agentic pattern.} \reflect follows the same trend as \react, with up to $17\times$ lower E2E latency by using persisted trajectories and reflections to steer retries. \llmc shows smaller gains on RS and weaker results on LA because single-shot DAG planning can mis-reference intermediate outputs, trigger executor failures and cause expensive replans; detailed \reflect and \llmc plots are in Appendix~\S\ref{app:results:reflect-llmc}. This indicates that \fw's optimizations behave as intended, while the realized gain depends on the control-flow robustness of the agentic pattern.

\para{Summary} For multi-turn short-tool workflows, agent memory is both a correctness mechanism and a performance optimization. It preserves the internal execution state needed by later turns, while S3 handles and caching reduce repeated MCP input/output. The gains are largest when workflows move large files and modest when tool outputs are already compact.

\subsection{Token and Cost Savings} \label{sec:results:tokens}
\begin{figure}[t!]
\centering%
  \subfloat[Research Summary \textit{(RS/P1)}]{
     \includegraphics[width=0.45\columnwidth]{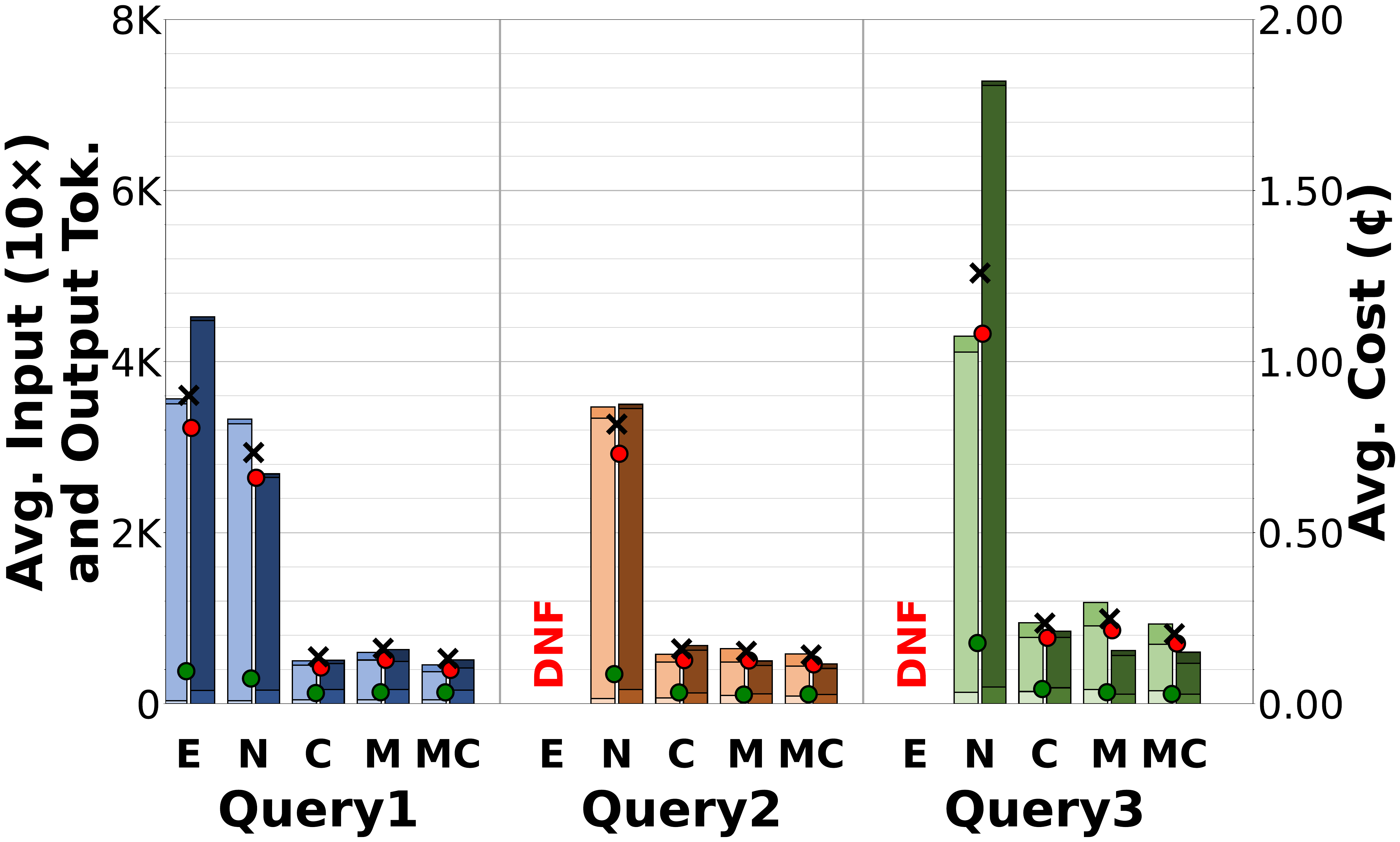}
     \label{fig:arxiv-d-s1-token}
  }~~
  \subfloat[Log Analytics \textit{(LA/F1)}]{
     \includegraphics[width=0.44\columnwidth]{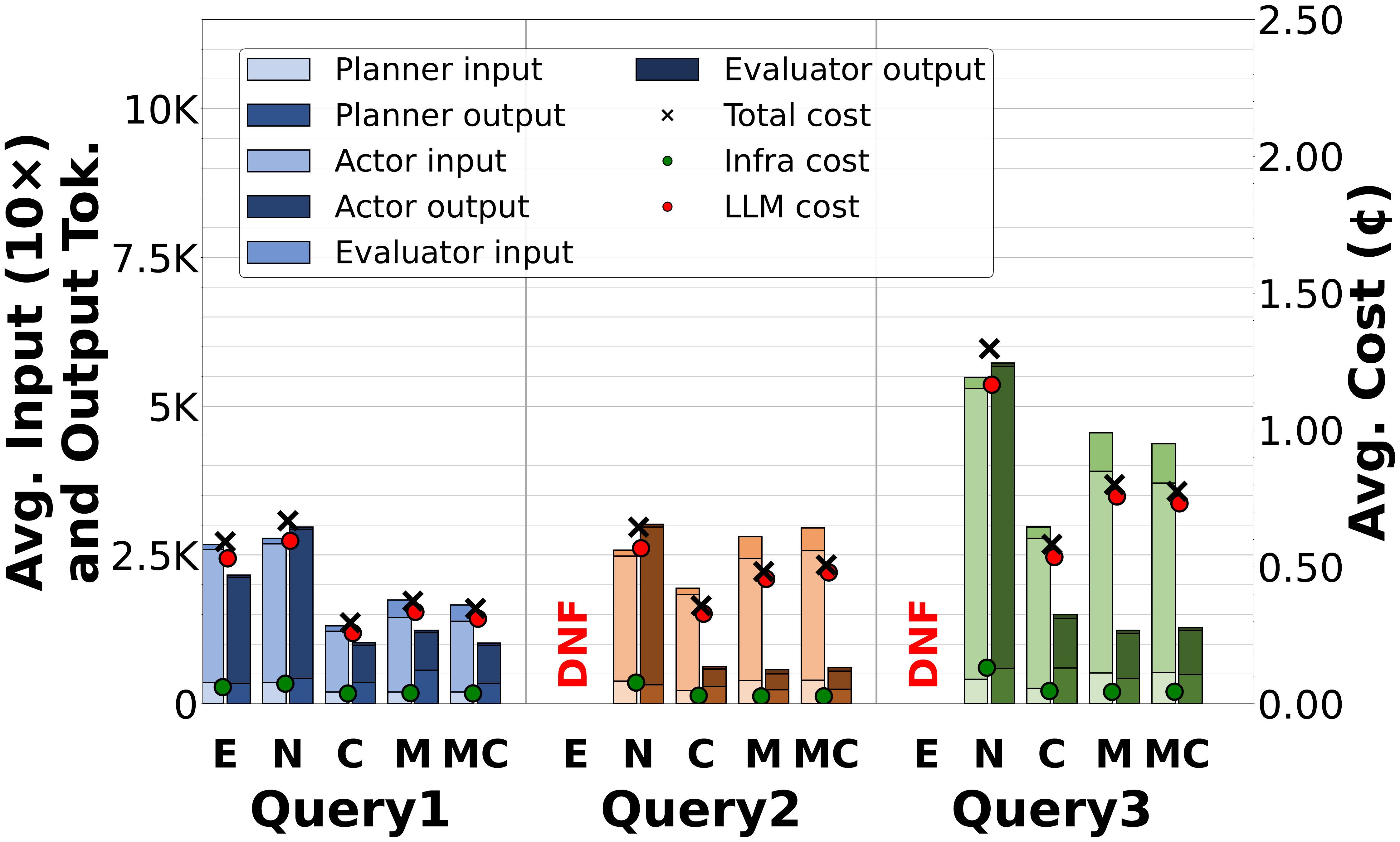}
     \label{fig:logger-d-s1-token}
  }\\  
  \subfloat[Research Summary \textit{(RS/P2)}]{
    \includegraphics[width=0.45\columnwidth]{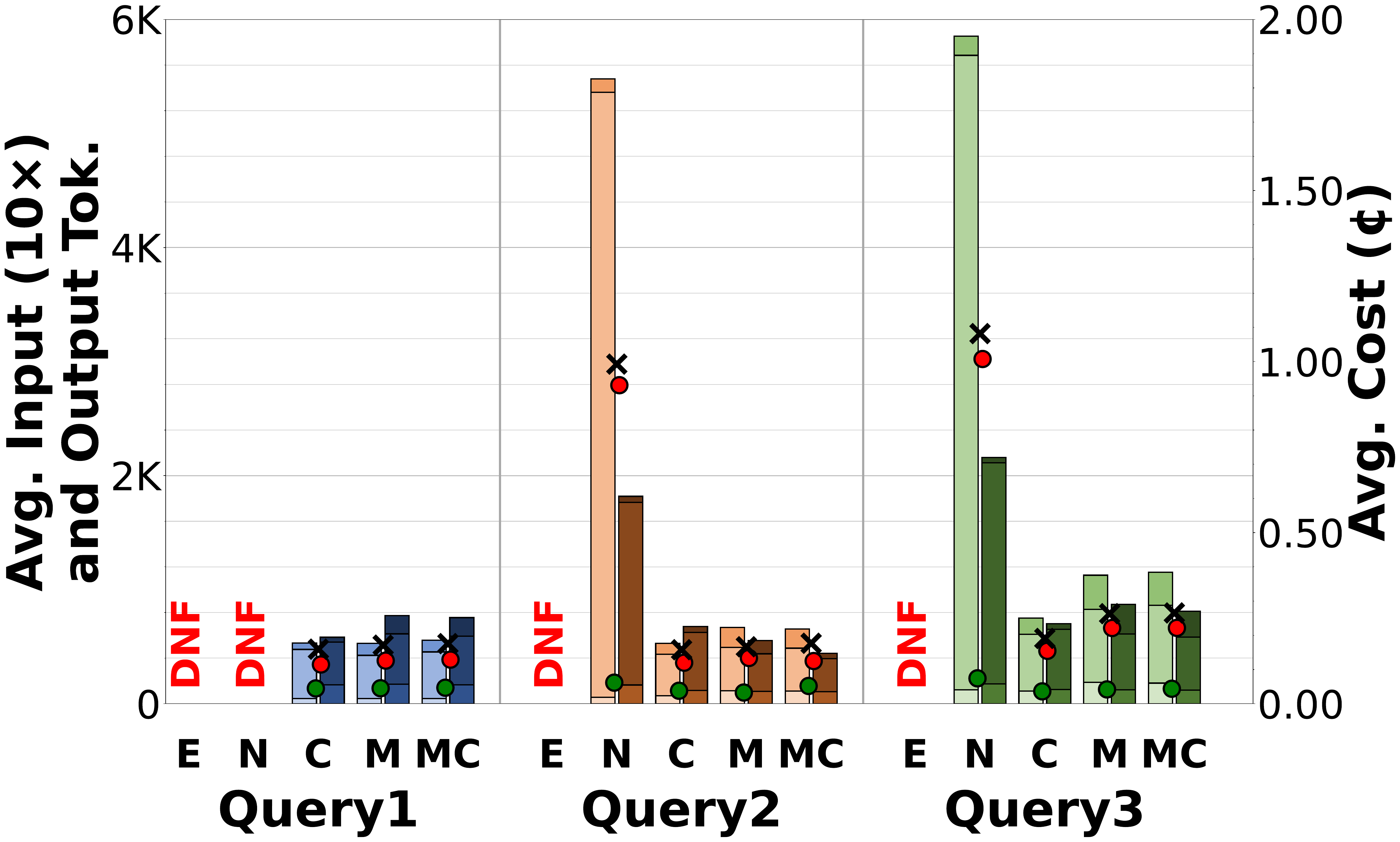}
     \label{fig:arxiv-d-s2-token}
  }~~
  \subfloat[Log Analytics \textit{(LA/F2)}]{
    \includegraphics[width=0.44\columnwidth]{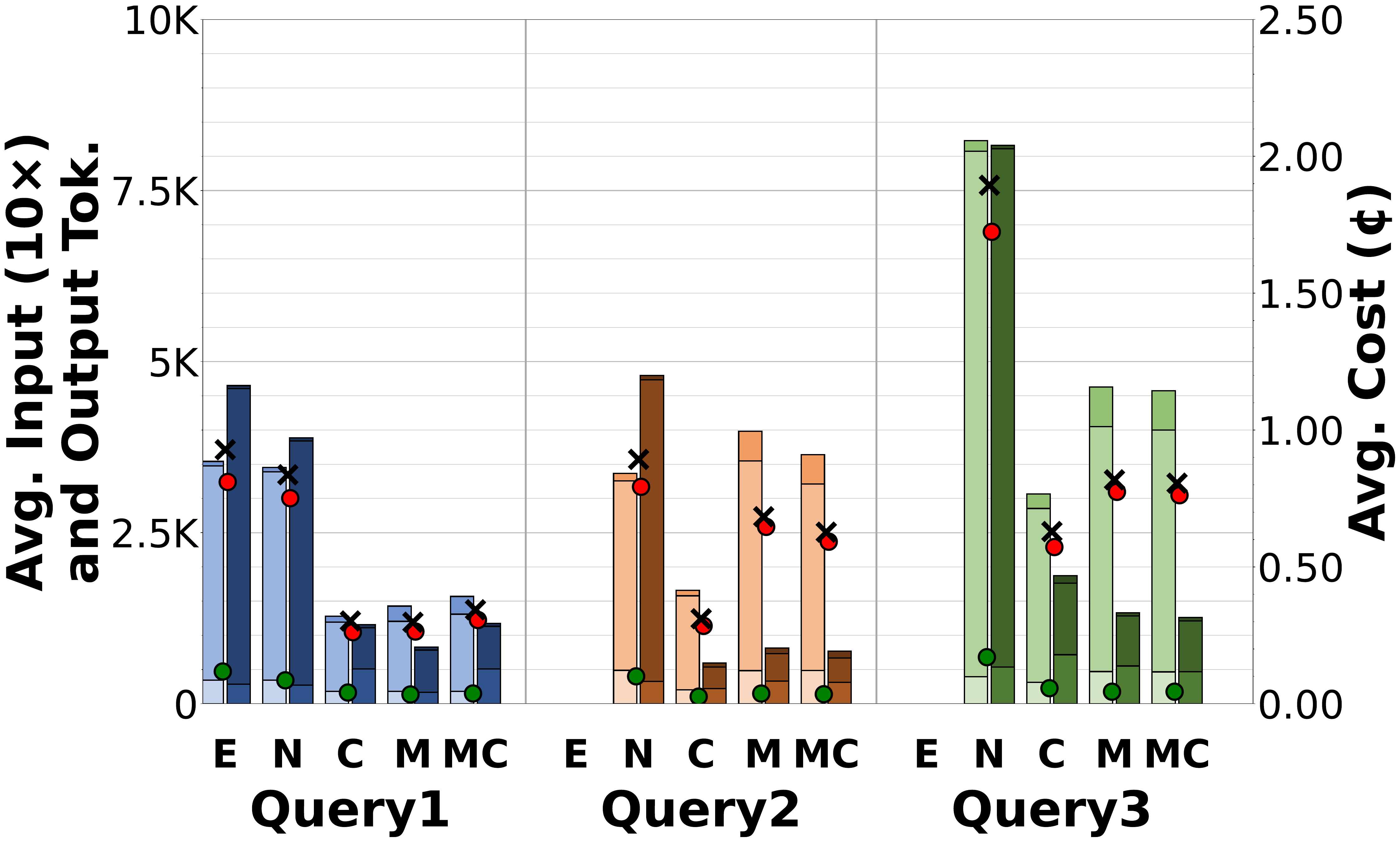}
     \label{fig:logger-d-s2-token}
  }

\caption{Input/output tokens and cost for \react. Each pair of bars reports input tokens, scaled by $10\times$, and output tokens; markers report LLM, FaaS agent and FaaS MCP cost.}
\label{fig:plot-LLM-IO-token-count}
\end{figure}

\takeaway{Reducing MCP service payloads is the dominant cost lever.} Fig.~\ref{fig:plot-LLM-IO-token-count} shows that input tokens dominate output tokens for the evaluated \react workloads. On RS/P1, the Q3 \textbf{N} input bar is the largest in the panel, while \textbf{C/M/MC} reduce it by roughly an order of magnitude (Fig.~\ref{fig:arxiv-d-s1-token}). Aggregated over the session, \textbf{E} and \textbf{N} consume $35{,}646$ and $33{,}277$ input tokens, while \textbf{M/C/MC} reduce this to $4{,}536$--$6{,}013$ input tokens. This is an $\approx85\%$ reduction from reusing AM and S3 handles instead of repeatedly inserting large tool outputs into prompts.
Detailed latency and cost/token results for \reflect and \llmc are in Appendix~\S\ref{app:results:reflect-llmc}.

\takeaway{Token savings translate into end-to-end cost savings.} For RS, the token reduction lowers total cost by $\approx5.2\times$ on P1 and $\approx9.3\times$ on P3 relative to \textbf{N}. The same qualitative trend appears in LA: for LA/F1 Q3, \textbf{N} has the largest input-token bar in the panel, while \textbf{C/M/MC} are visibly lower and also reduce the LLM-cost marker (Fig.~\ref{fig:logger-d-s1-token}). \reflect follows the same trend, with up to $\approx84\%$ fewer input tokens and order-of-magnitude cost reduction on RS. \llmc reduces tokens by $66$--$74\%$ on RS, but its gains do not always transfer to LA because plan-quality errors trigger replans and inflate token use.

\takeaway{LLM cost dominates FaaS platform cost for short-tool workflows.} Across short-tool applications, LLM cost accounts for $\approx61$--$94\%$ of total cost. Thus, reducing input tokens through AM, S3 handle passing and caching has a larger effect on total cost than reducing Lambda or Step Functions execution time alone. E.g., $50,000$ input tokens with the GPT model cost \textcent$0.75$, while $\approx50$\,s of Step Functions execution costs \textcent$0.085$, about $9\times$ lower.

\para{Summary} For short-tool workflows, \fw reduces cost primarily by avoiding repeated prompt payloads rather than by reducing FaaS execution cost alone. Agent memory removes repeated tool/action traces across turns, while S3 handles and caching prevent large artifacts from re-entering prompts. This is why \textbf{MC} gives the strongest savings on payload-heavy workloads: up to $88\%$ input-token reduction and up to $66\%$ total-cost reduction for the evaluated short-tool workloads.

\subsection{Applicability Across Agentic Service Benchmarks} \label{sec:results:benchmarks}

\begin{figure}[t]
\centering%
  \subfloat[LiveMCP Bench \react]{%
     \includegraphics[width=0.45\columnwidth]{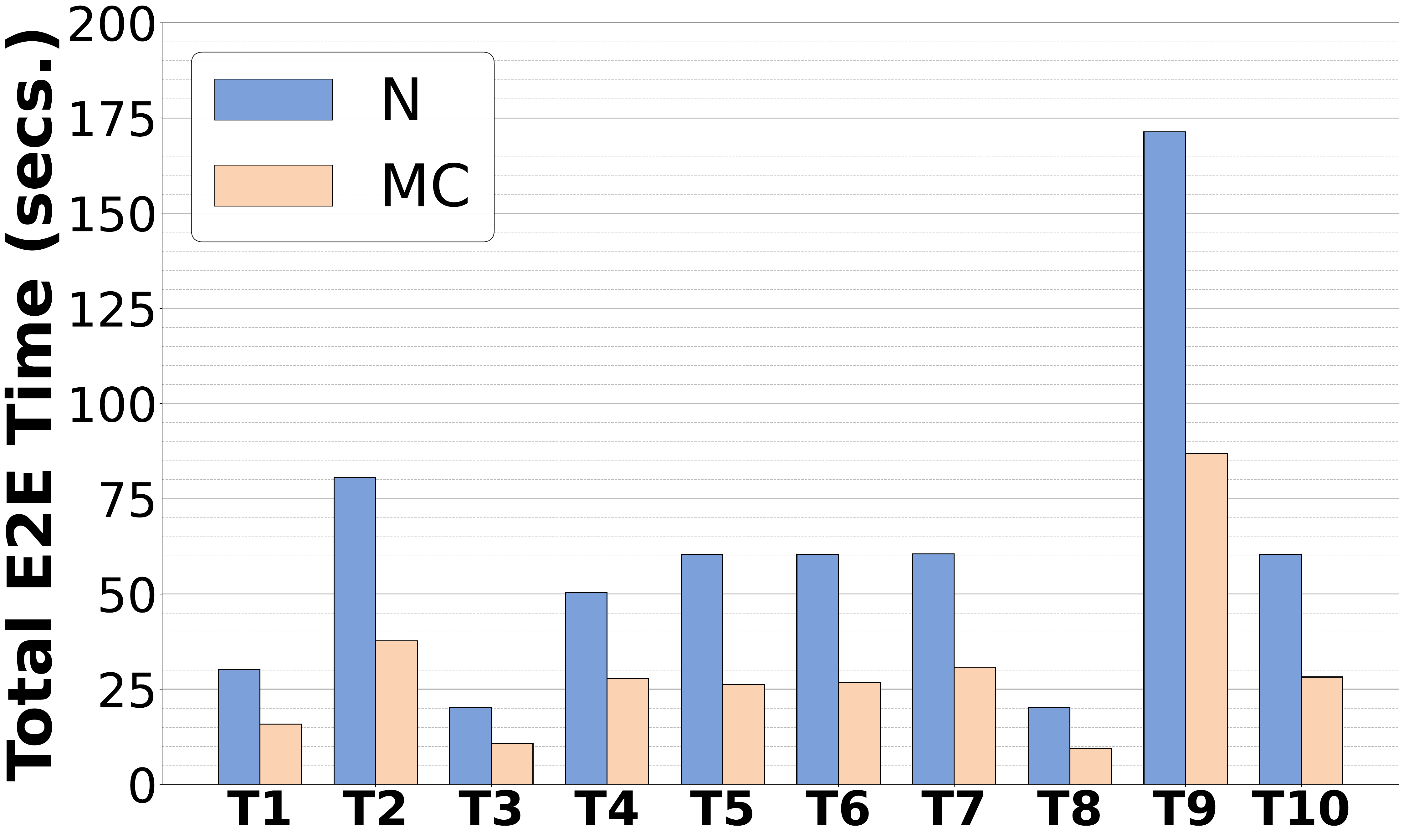}%
     \label{fig:livemcp-react}%
  }~~
  \subfloat[Tau Bench \react]{%
    \includegraphics[width=0.45\columnwidth]{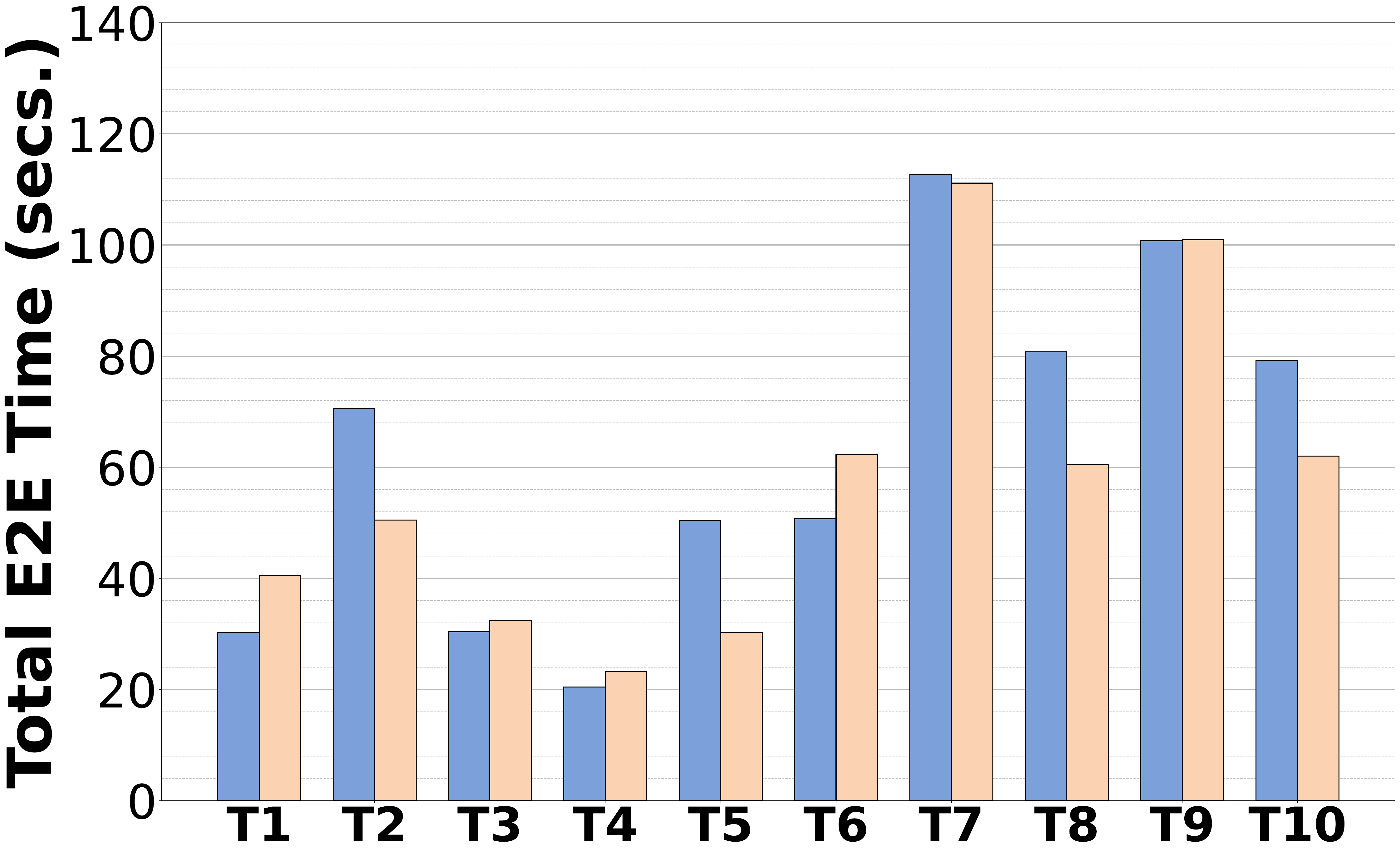}%
     \label{fig:tau-react}%
  }
\caption{Per-task latency on LiveMCPBench and TauBench under \react, comparing client-native memory \textbf{N} against agent memory with caching \textbf{MC}.}
\label{fig:other-benchmarks}
\end{figure}

\takeaway{Payload-heavy benchmark tasks benefit most from \fw.} Fig.~\ref{fig:livemcp-react} shows that \textbf{MC} reduces latency across most LiveMCPBench tasks. For example, task T9 falls from $\approx 170$\,s under \textbf{N} to $\approx 85$\,s under \textbf{MC}, and T2 drops from $\approx 80$\,s to $\approx 40$\,s. These tasks often coordinate multiple MCP tools and produce large PDF/Markdown/JSON artifacts. S3 handle passing and AM reduce the amount of artifact data reintroduced into prompts, so the gains are largest for document generation, report generation and trend-aggregation tasks.

\takeaway{Payload-light benchmark tasks show smaller and mixed latency effects.} TauBench tasks are primarily database-style lookups and policy checks with compact tool outputs. Fig.~\ref{fig:tau-react} shows that \textbf{MC} improves some tasks, such as T8, which falls from $\approx 80$\,s to $\approx 60$\,s, and T10, which falls from  $\approx 80$\,s to $\approx 60$\,s. However, some tasks are similar or slightly slower under \textbf{MC}, such as T7, where both configurations are  $\approx 110$\,s, and T6, where \textbf{MC} is modestly higher. Thus, TauBench confirms functional applicability and bounded overhead on payload-light tasks, but does not show the large latency gains seen on payload-heavy workloads.

\begin{table}[t!]
\centering
\caption{\react task success and cache behavior under memory/cache configurations. Hits are cache hits over MCP lookups under \textbf{C} and \textbf{MC}.}
\label{tbl:correctness-mem-cache}
\footnotesize
\setlength{\tabcolsep}{4pt}
\renewcommand{\arraystretch}{0.95}
\begin{tabular}{lccc}
\hline
\textbf{App} & \textbf{Success} & \textbf{Cache hits (C)} & \textbf{Cache hits (MC)} \\
\hline\hline
RS (3 runs) & 3/3 & 6/22 (27\%) & 0/16 (0\%) \\
LA (3 runs) & 3/3 & 11/54 (20\%) & 10/55 (18\%) \\
TauBench (10 tasks) & 10/10 & 0/46 (0\%) & 0/9 (0\%) \\
LiveMCPBench (10 tasks) & 10/10 & - & 0/24 (0\%) \\
\hline
\end{tabular}
\end{table}

\takeaway{Caching helps only when deterministic reuse exists.} Table~\ref{tbl:correctness-mem-cache} shows that cache hits occur for RS and LA, where repeated deterministic tool calls arise across turns. On RS, \textbf{C} records $27\%$ cache hits, while \textbf{MC} records no cache hits because AM prevents the repeated tool calls from being issued. On LA, both \textbf{C} and \textbf{MC} observe cache hits at about $18$--$20\%$ because later turns reuse intermediate log-processing outputs. In contrast, TauBench and LiveMCPBench show $0$ cache hits, consistent with unique lookup keys or single-turn task-local tool calls.

\para{Summary} The benchmark results show that \fw is not tied to one application. It provides the largest gains for payload-heavy workflows that exchange large artifacts through MCP tools, while maintaining successful execution on payload-light workflows where the latency effect is smaller or neutral. The relative contribution of each optimization depends on workload structure: AM helps multi-turn context reuse, S3 handles help large artifacts, and caching helps repeated calls.

\subsection{\fw Cost Analysis for Long-running Workflows} \label{sec:results:failures}

\begin{figure}[t!]
\centering%
  \subfloat[\textit{Tabular-2022}]{%
     \includegraphics[width=0.45\columnwidth]{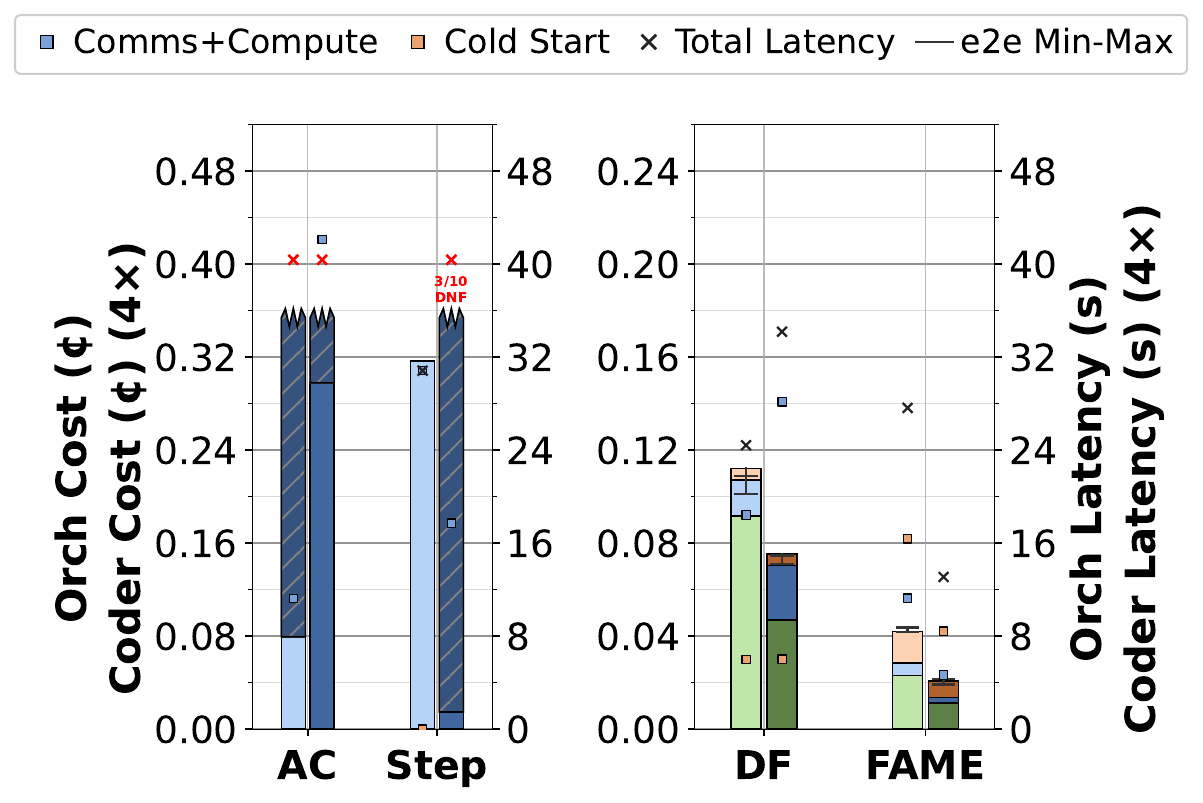}%
    \label{fig:claim23a}%
  }~~
  \subfloat[\textit{Dog-Breed}]{%
     \includegraphics[width=0.45\columnwidth]{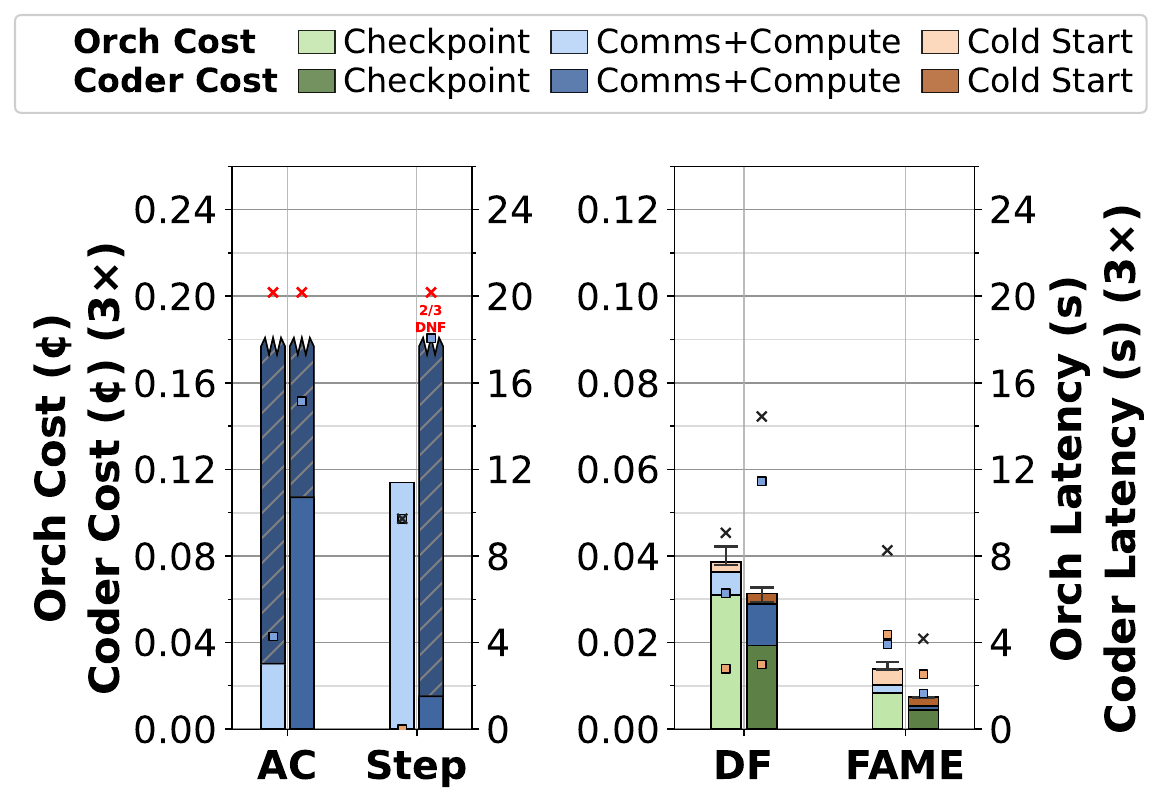}%
    \label{fig:claim23b}%
  }%
\caption{\mlzero{} orchestrator and Coder-agent cost-time trade-offs across AgentCore (AC), synchronous Step Functions, Durable Functions (DF), and \fw. Synchronous service runtimes either retain billable resources during EMCP waits or time out, while async service runtimes suspend during EMCP execution.}
\label{fig:claim23}
\end{figure}

\begin{figure}[t!]
\centering%
  \subfloat[\textit{Tabular-2022}]{%
     \includegraphics[width=0.45\columnwidth]{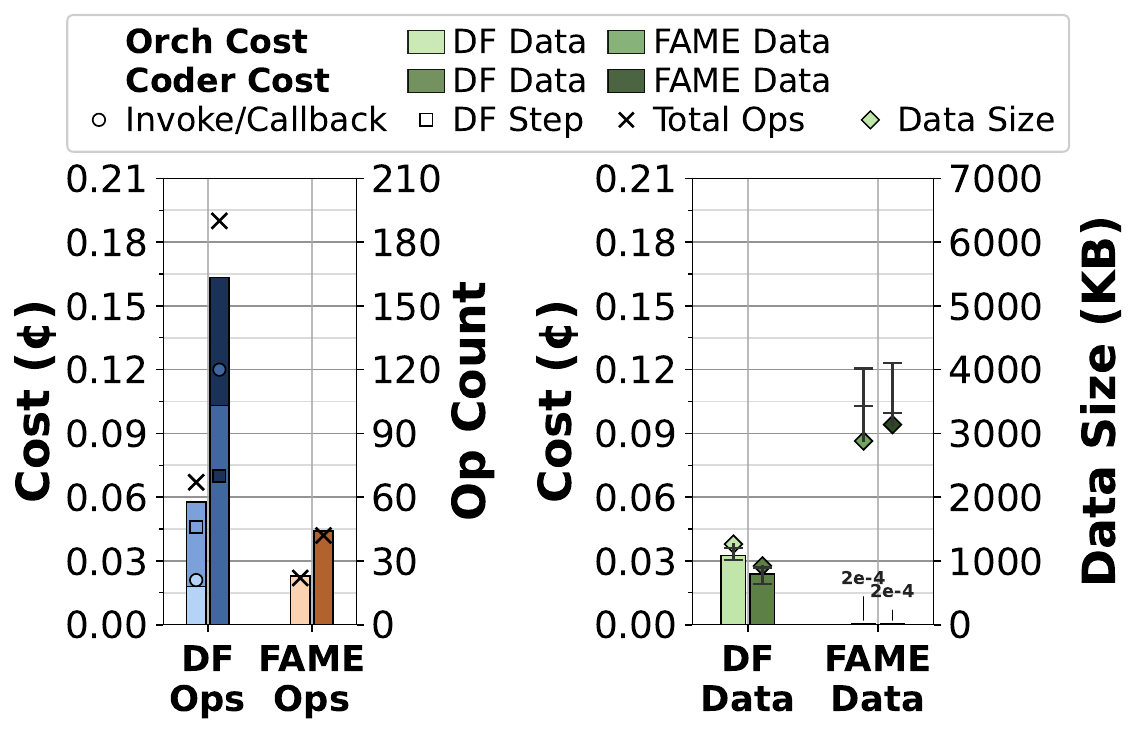}%
    \label{fig:claim23a:checkpoint}%
  }~~
  \subfloat[\textit{Dog-Breed}]{%
     \includegraphics[width=0.45\columnwidth]{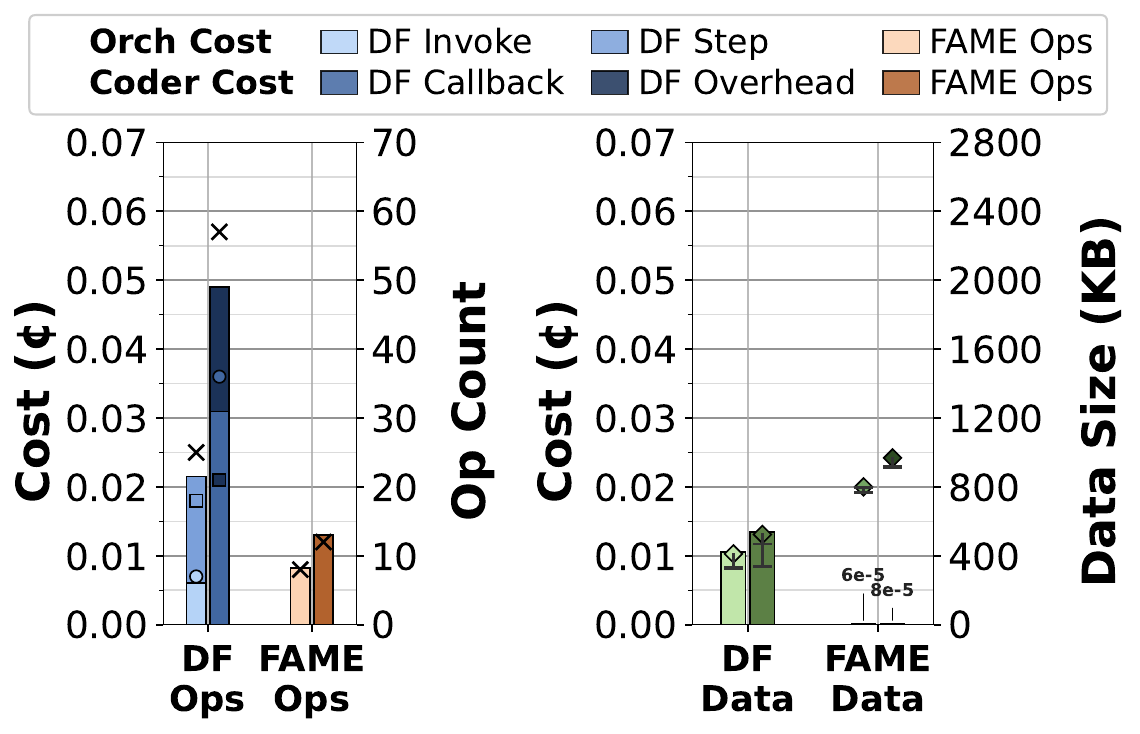}%
    \label{fig:claim23b:checkpoint}%
  }%
\caption{\mlzero{} checkpoint operation and storage breakdown for DF and \fw. \fw reduces operation cost through selective checkpointing and reduces storage cost despite storing larger checkpoint objects.}
\label{fig:claim23:checkpoint}
\end{figure}

\takeaway{Synchronous service invocation either times out or bills idle waits.}
Fig.~\ref{fig:claim23} shows the cost-time trade-off for the Coder agent and orchestrator on the two \mlzero{} workloads. Synchronous Step Functions times out when the Coder Lambda waits beyond the FaaS lifetime, failing on 3 of 10 Coder iterations for \textit{Tabular-2022} (Fig.~\ref{fig:claim23a}) and 2 of 3 iterations for \textit{Dog-Breed} (Fig.~\ref{fig:claim23b}).
Appendix Fig.~\ref{fig:claim1-per-iter} shows the per-iteration Tabular-2022 behavior.
AC completes the runs, but its billed cost is dominated by the long active-task window: for Tabular-2022, the Coder performs 168\,s of useful compute over 6,810\,s of billed session activity, and the orchestrator performs 11.2\,s of useful compute over 7,279\,s of billed session activity.

\takeaway{Async service invocation reduces Coder-agent cost.}
Among async deployments, \fw reduces Coder-agent infrastructure cost relative to DF. On \textit{Tabular-2022}, \fw costs 0.083\textcent{} compared to DF's 0.30\textcent{}, making \fw $3.64\times$ cheaper; its runtime latency is also lower at 52.3\,s compared to 136.7\,s for DF (Fig.~\ref{fig:claim23a}). On \textit{Dog-Breed}, \fw costs 0.022\textcent{} compared to DF's 0.094\textcent{}, making \fw $4.24\times$ cheaper, with 12.5\,s latency compared to 43.3\,s for DF (Fig.~\ref{fig:claim23b}).
This gain comes from checkpointing only at suspension points rather than using log-based replay for the full durable execution. E.g., in Tabular-2022, FAME checkpoints only at the 40 async call nodes while DF checkpoints after every node/step resulting in an additional 60 steps.

\takeaway{Async orchestration reduces orchestrator cost.} 
The orchestrator shows the same trend. On \textit{Tabular-2022}, \fw costs 0.042\textcent{} at the orchestrator level, compared to 0.11\textcent{} for DF and 0.32\textcent{} for Step Functions, giving $2.67\times$ and $7.25\times$ lower cost respectively (Fig.~\ref{fig:claim23a}). On \textit{Dog-Breed}, \fw costs 0.014\textcent{}, compared to 0.039\textcent{} for DF and 0.11\textcent{} for Step Functions, giving $2.79\times$ and $8.23\times$ lower cost respectively (Fig.~\ref{fig:claim23b}).
The corresponding runtime latencies are comparable across async orchestrators, which indicates that the main benefit is on cost rather than latency alone.

\takeaway{Checkpoint operation cost explains the difference from DF.} Fig.~\ref{fig:claim23:checkpoint} decomposes the checkpoint cost. 
For the Agent--Tool bridge on \textit{Tabular-2022}, \fw's S3 Put/List/Get operation set costs 0.044\textcent{}, compared to 0.103\textcent{} for DF callbacks alone; after adding DF step overhead, \fw's operation cost is about $3.7\times$ lower (Fig.~\ref{fig:claim23a:checkpoint}).
For \textit{Dog-Breed}, the same trend holds, with \fw operation cost about $3.8\times$ lower than DF after step overhead (Fig.~\ref{fig:claim23b:checkpoint}).
Although \fw stores larger checkpoint objects, S3 storage is cheaper in this regime, yielding $95.2\times$ lower storage cost for Tabular-2022 and $175\times$ lower storage cost for Dog-Breed at the Coder level.
Appendix~\S\ref{app:results:long-cost} gives the Tabular-2022 operation and storage breakdown.
The agent--tool checkpoint storage costs are $95.2\times$ cheaper for \fw than DF despite being larger in size by $3.4\times$. The orchestrator checkpoint is similarly $142.1\times$ cheaper on storage costs than DF despite $2.28\times$ larger size. These gains are a direct consequence of FAME leveraging S3, whose storage costs are $324\times$ cheaper than DF.

\para{Summary} For long-running EMCP workflows, timeout-safe execution requires async handling at both bridges. Suspending only the Coder avoids the agent-side wait, but the orchestrator still remains billable unless the orchestrator--agent bridge also suspends. \fw applies selective checkpointing at both levels, so its cost scales with checkpoint, callback and resume operations rather than with the external tool wait-time.

\subsection{Discussion} \label{sec:results:limitations} 

Our evaluation focuses on service execution behavior rather than agent-quality benchmarking. We fix prompts, tools, model settings, decoding parameters and task inputs across configurations, and report systems metrics such as latency, tokens, infrastructure cost, checkpoint cost, billed duration and DNF outcomes. This isolates service runtime, memory, MCP service I/O and checkpointing effects, but does not claim that \fw improves the semantic quality of the underlying LLM agent.

The results are scoped to AWS \texttt{ap-south-1}, 26 short-running service tasks over 10 MCP servers, and two long-running \mlzero{} workloads from MLE-Bench. Absolute costs depend on provider pricing and implementation choices, but the main trends follow from service runtime semantics: VM and managed runtimes bill idle and background task waits, synchronous FaaS can hit function lifetime limits, and async FaaS shifts cost to checkpoint, callback and resume operations. \fw provides timeout-safe async service invocation, not exhaustive fault tolerance: callbacks are keyed to suspended executions for idempotent resume, but recovery from lost callbacks, storage outages, checkpoint corruption and adversarial MCP outputs remains future work.

\section{Related Work\pgs{0.5}}
\label{sec:related}

\subsubsection{Agentic workflows on FaaS} 
Prior works examine QoS-aware service composition, service placement, and FaaS function delivery for conventional service workloads~\cite{zou2014qos,jatoth2017qos,yu2023faasdeliver}.
But few explore agentic workflow service orchestration on FaaS runtimes. 
AgentX~\cite{agentx} proposes using Step Functions to orchestrate agents, but its design focuses on basic serverless control flow and does not evaluate realistic MCP-enabled workflows, cross-turn memory, FaaS MCP optimizations, or long-running external tools. In contrast, \fw implements and evaluates \react, \reflect, \llmc and \mlzero{} on FaaS, separates agent roles from the MCP tool plane, and quantifies latency, token, cost, timeout and checkpoint behavior across short-tool and long-tool workloads.

Serverless workflow systems and FaaS orchestration have been studied extensively for event-driven cloud applications~\cite{faas,faas-tpds-arxiv}. These systems provide elasticity and fine-grained billing, but the agentic setting introduces two additional requirements: LLM-driven workflows must preserve conversational and execution state across turns, and tool calls can produce large prompt payloads or block for long durations. \fw addresses these requirements by combining cross-turn agent memory, S3 handle passing, MCP tool-output caching, and callback-based suspend/resume. Thus, \fw is not only a mapping of agents to functions, but a service execution runtime specialized for MCP-enabled agentic workflows.

\subsubsection{Agentic frameworks and memory} 

Agentic frameworks such as LangGraph~\cite{langgraph_overview_docs}, AutoGen~\cite{wu2024autogen} and related graph-based approaches~\cite{GoT,autogen_graph_flow} provide abstractions for composing LLM calls, tools and control-flow edges. These simplify application authoring, but they do not by themselves determine an efficient cloud runtime. A LangGraph workflow can be hosted monolithically on a VM, deployed through managed runtimes or decomposed into FaaS roles. \fw takes the latter path and adds the missing deployment mechanisms: per-role FaaS orchestration, durable cross-turn state, MCP I/O reduction and async bridges for long-running tools.

Memory is a core requirement for agentic systems~\cite{sumers2024cognitive,langchain_memory_overview}. Prior framework support typically focuses on what memory is exposed to the LLM, such as conversational history, summaries or tool messages. \fw instead treats memory as a systems primitive. It distinguishes client-visible history from internal agent memory, persists execution state outside the FaaS runtime, injects only the state needed by subsequent roles, and replaces large tool outputs with S3 handles. This makes memory both a correctness mechanism for later turns and a cost mechanism for reducing prompt and tool call bloat.

\subsubsection{MCP protocols and tool-plane optimization} 
MCP standardizes how agents discover and invoke tools~\cite{mcp_intro_docs}. Recent MCP benchmarks and directories focus on tool-use capability, task coverage, and benchmark construction~\cite{mo2025livemcpbench,luo2025mcpuniverse,wang2025mcpbench}. These efforts are complementary to \fw: they help define realistic tool-using workloads, while \fw studies how such workloads should be deployed and optimized on cloud execution substrates. In particular, \fw measures the systems cost of MCP invocation, large tool outputs, repeated calls, FaaS cold starts and long external waits.

Other work targets API-to-MCP conversion or protocol-level interoperability~\cite{mastouri2025restapismcp,liao2025agentmaster}. \fw addresses a different layer. It assumes MCP as the tool service interface and optimizes the execution path beneath it. For short-running tools, \fw wraps MCP servers as FaaS endpoints and applies S3 handle passing, caching and singleton/consolidated deployments. For long-running or resource-heavy tools, \fw does not force tool execution into FaaS; instead, it treats the MCP service as external and optimizes the agent-side and orchestrator-side async bridges.

\subsubsection{Long-running agentic and scientific workflows} 
Long-running agentic workflows are increasingly appearing in scientific, machine-learning and compute-bound settings. Tool collections and MCP wrappers expose external services such as sandboxes, HPC systems, simulations and scientific data tools to LLM agents~\cite{chan2025mlebench,gao2025democratizingaiscientistsusing,pan2025experiencesmodelcontextprotocol,zhao2026polyjarvisllmorchestratedagentautomated,Shiraishi_2026,zhang2026el}. These works primarily focus on tool availability and domain capability. They do not evaluate how the agent runtime and orchestrator behave while tools execute for minutes to hours.

\fw focuses on this execution-fabric gap. The long-running tool itself remains external, but the agent and orchestrator are FaaS components that must avoid timeouts and idle billing. Durable Functions provide one possible async execution model through durable operations and replay~\cite{aws_lambda_durable_docs,aws_durable_key_concepts}, while Step Functions supports callback-based orchestration at the workflow level~\cite{beswick2021integrating}. \fw differs by applying selective LangGraph checkpointing at both the agent-tool and orchestrator-agent bridges. This reduces durable operation and replay overhead while preserving timeout-safe suspend/resume for External MCP workflows.

\section{Conclusions and Future Work}
\label{sec:conclusion}

This article presents \fw, a FaaS middleware for MCP-enabled agentic workflows. \fw decomposes agentic patterns into modular serverless roles, externalizes workflow state through agent memory, and reduces MCP input/output overhead using S3 handle passing and tool-output caching. Across 26 short-tool tasks over 10 MCP servers, \fw reduces latency by up to $17\times$, input tokens by up to $88\%$, and total cost by up to $66\%$.

For long-running External MCP workflows, \fw provides timeout-safe async execution across both the agent-tool and orchestrator-agent bridges. It checkpoints selected LangGraph state, suspends FaaS execution during external tool waits, and resumes from callbacks. On two MLE-Bench \mlzero{} workloads, \fw avoids synchronous Step Functions timeouts and is $3.64$--$4.24\times$ cheaper than Durable Functions at the agent level and $2.67$--$8.23\times$ cheaper at the orchestrator level.

Future work will study adaptive placement and consolidation for FaaS MCP tools, memory summarization with provenance and TTL policies, and broader evaluation on write-heavy MCP tools, long-running scientific services, and multi-cloud FaaS deployments.

\section*{Acknowledgment}

This work was supported by the IBM-IISc Hybrid Cloud lab.

The authors used AI-assisted tools only for language refinement and clarity. All ideas, methods, experiments, results and conclusions are the authors' own, and the final manuscript was reviewed and approved by the authors.

\bibliographystyle{plain}
\bibliography{paper}

\clearpage
\appendix

\noindent {\huge \textsf{Appendix}}

\section{Cross-Turn Agent Memory}
\label{app:fame_memory}

\fw persists agent memory (AM) outside the FaaS runtime so that later turns can reuse internal execution state rather than relying only on user-visible conversation history. AM is stored in DynamoDB using \texttt{session\_id} for the user session and \texttt{invocation\_id} for a workflow execution. Each record stores an \texttt{expireAt} field and an optimized \texttt{messages} object. On the first invocation of a session, \fw commits the Evaluator's accumulated state to DynamoDB. On later invocations, \fw retrieves the prior AM, compresses or replaces large tool outputs with S3 handles, and injects the resulting state into the Planner or Actor before normal execution. After each run, only the new delta produced by the Evaluator is appended as the next memory state.

\begin{figure}[t]
\centering%
    \includegraphics[width=.8\columnwidth]{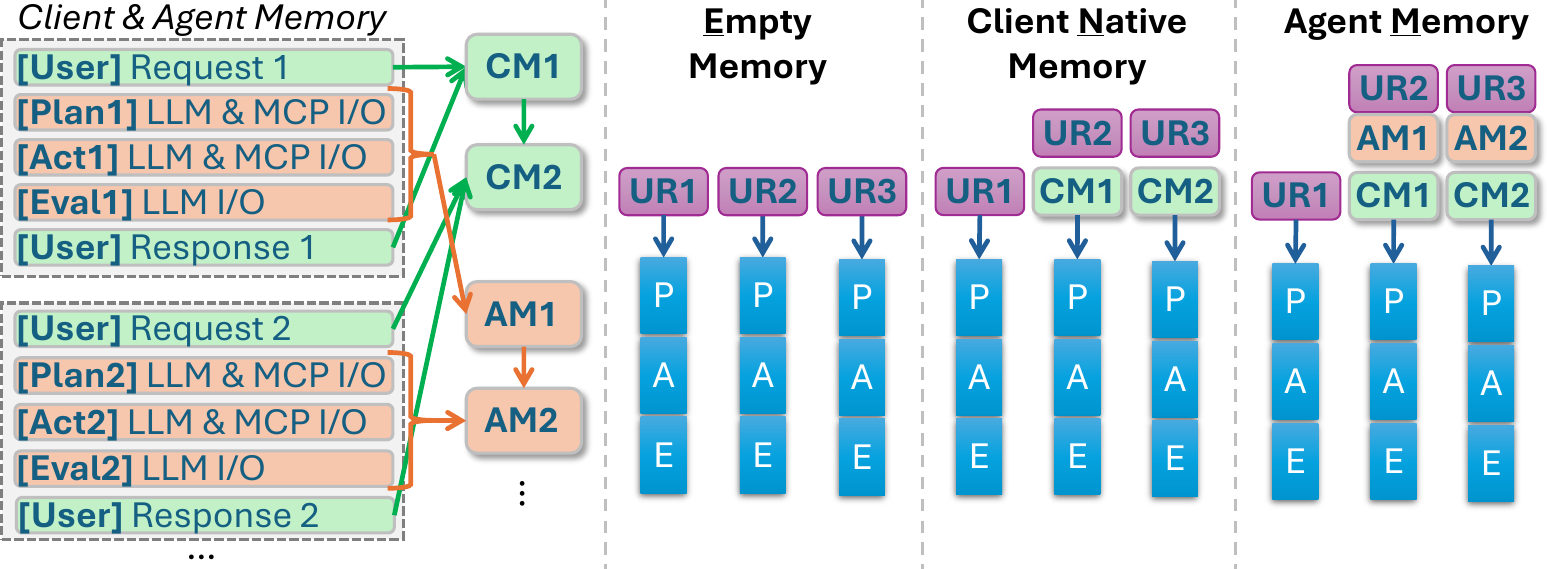}
\caption{Client Memory (CM) and Agent Memory (AM) injection into a \react workflow, together with the current User Request (UR).}
    \label{fig:mem}
\end{figure}

Fig.~\ref{fig:mem} illustrates the three memory modes used by \react. Empty memory provides only the current User Request (UR). Client memory (CM) adds prior user-visible requests and responses. Agent Memory (AM) additionally injects internal execution state, including prior tool calls, tool outputs, and S3 handles. This AM injection, combined with the actor-memory prompt, is the mechanism used by \fw to avoid missing context and redundant MCP calls in later turns.

\begin{figure}[b!]
\centering%
    \includegraphics[width=.85\columnwidth, 
    clip]{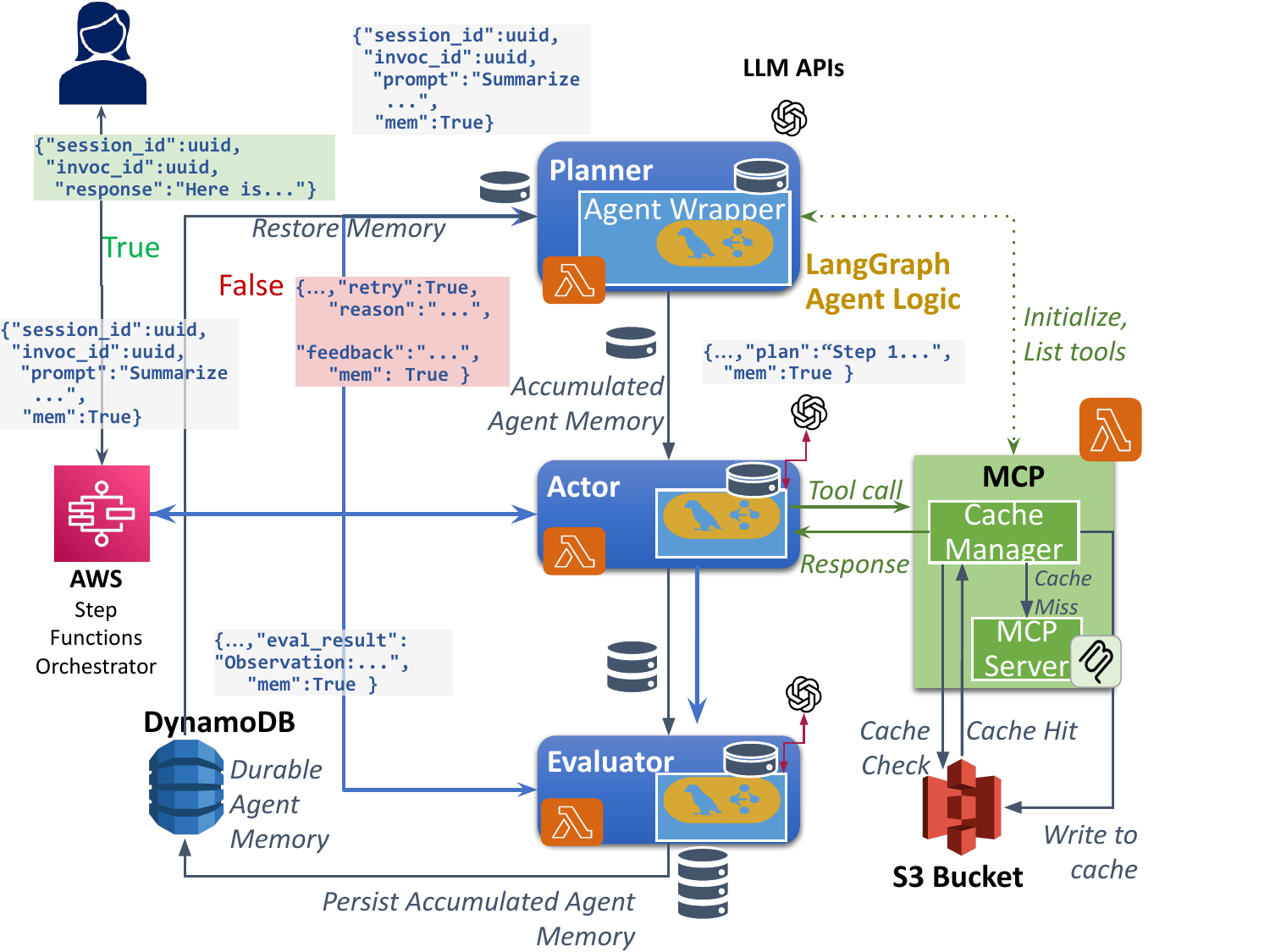}
\caption{Architecture of \fw for \react Workflow with short tool-call MCP services orchestated by AWS Step Functions.}
    \label{fig:fame_arch}
\end{figure}

\fw runs under at-least-once Lambda execution semantics. Step Functions provide workflow-level execution management, while AM persistence occurs inside role Lambdas using idempotent upserts keyed by \texttt{session\_id:invocation\_id}. Transient FaaS faults are retried with exponential backoff. Agentic failures are handled by bounded replanning, with \texttt{max\_iterations=3} in our implementation.

\subsection{Memory in \reflect} 
\label{app:fame_memory:reflect}
\reflect extends the cumulative intra-session AM used by \react with a two-tier memory model. 

First, \textit{Short-Term Memory (STM)} stores the Actor's per-attempt trajectory, including LLM reasoning, tool calls, and tool outputs, as \texttt{current\_trajectory}. The Evaluator uses this trajectory to score the Actor's output, and the Self-Reflection role uses it to diagnose failures. STM is cleared between retries after the Self-Reflection role consumes it.

Second, \textit{Long-Term Memory (LTM)} stores a bounded sliding window of natural-language reflections, e.g., \texttt{MAX\_REFLECTIONS=3}. These reflections summarize prior failure modes and are injected into the Actor's prompt during subsequent attempts.

\begin{lstlisting}[language={},breaklines=true,breakindent=0pt,mathescape=true]
$\textbf{[REFLECTIONS FROM PREVIOUS ATTEMPTS]:}$ 1. The agent hallucinated a value... 2. Premature termination...
\end{lstlisting}

The default \fw AM remains responsible for session-level context reuse. It retains the optimized intra-session history, drops unnecessary system messages, replaces large tool payloads with S3 URIs, and enforces a size cap of 350\,KB before injection. Thus, \react, \reflect, and \llmc share the same external memory substrate, while \reflect adds trajectory and reflection state to improve retry behavior.

\section{Agentic Pattern Implementation details}
\label{app:agent:implementation}
\subsection{\react Workflow (RT)}
\fw composes \react as a Step Functions workflow whose agent roles are AWS Lambda functions. The workflow state is passed between roles as messages, while persistent agent memory is handled separately as described in Appendix~\S\ref{app:fame_memory}. Each role is implemented as a small LangGraph subgraph with its own prompt, resource configuration, and state interface.

The Planner and Evaluator each contain an LLM-call node. The Planner receives the user request, injected memory, and MCP tool descriptions, and produces a plan for the Actor. The Actor graph contains an LLM-call node and a tool-call node connected by conditional edges. The Actor iterates over LLM reasoning and MCP tool invocation until the plan is complete or a loop limit is reached; our implementation uses a loop limit of 25. The Evaluator validates the Actor output and either returns the final response or routes feedback to the Planner for replanning.

\subsection{\reflect Workflow (RX)}

\reflect extends the \react execution model with an explicit Self-Reflection role. The Actor first executes a full \react-style inner loop. The Evaluator then checks whether the Actor's output satisfies the user request. If the output is valid, the workflow returns the response to the user. If the output is invalid, the Self-Reflection role analyzes the execution trajectory and the Evaluator feedback, such as wrong tool use, hallucinated output, or premature termination.

The Self-Reflection role produces verbal feedback that is stored as long-term reflection memory and injected into subsequent Actor attempts. This retry loop continues until the task succeeds or the bounded trial count is exhausted, at which point a \texttt{Fail} state terminates the workflow. The memory structure used by \reflect is described in Appendix~\S\ref{app:fame_memory:reflect}.

\subsection{\llmc Workflow (LM)} \label{app:agent:implementation:llmc}
\fw realizes \llmc with Planner, Executor, and Joiner agent roles, plus a Task Fetching Unit (TFU) support function. The Planner produces a numbered task list with explicit data dependencies, such as \texttt{Task 3 -> Task 5}. The TFU parses this plan, maintains a dependency tracker, and emits the next ready batch once all predecessor tasks have completed. 

Ready task batches are executed using a Step Functions \texttt{Map} state with up to 10 parallel Executors. Each Executor first attempts the planned MCP call directly. If the call fails or the arguments do not match the MCP service signature, the Executor falls back to a minimal LangGraph repair agent that revises the parameters and retries the call. 

The workflow has two nested loops. The inner \texttt{TFU -> ExecuteBatch -> TFU} loop drains the task DAG across dependency levels. The outer \texttt{Joiner -> Planner} loop allows limited global replanning when the Joiner determines that the compiled task graph did not produce a valid final answer; our implementation allows up to 3 replans. Per-batch concurrency limits, task timeouts, and idempotent task identifiers are used to keep execution compatible with serverless workflow constraints.

\subsection{\mlzero Workflow} \label{app:agent:implementation:mlzero}

Each \mlzero{} session uses a single coding sandbox. At the beginning of the run, the orchestrator uploads the problem artifacts, including the problem description and train/test datasets, to the sandbox and invokes the Perception agent. The Perception agent scans the problem description and dataset files to extract task structure, dataset schema, ML problem type, and suitable implementation libraries. This information initializes the root node of the Monte Carlo Tree Search (MCTS).

During each MCTS iteration, the orchestrator selects and expands a node in the search tree, invokes the Semantic agent to retrieve relevant documentation, and passes the resulting context to the Coder agent. The Coder generates a training script and invokes the EMCP sandbox to execute the code. The sandbox returns execution output and evaluation results, which are used to update the expanded node and backpropagate the reward through the MCTS tree. This process repeats until the search budget is exhausted.

Unlike the short-running patterns, \mlzero{} invokes an external sandbox whose execution can exceed a single FaaS invocation window. This is why \mlzero{} is used as the driving workload for the timeout-safe async service invocation design in \S\ref{sec:fpp}.

\begin{figure}[t]
\centering%
    \includegraphics[width=0.65\columnwidth]{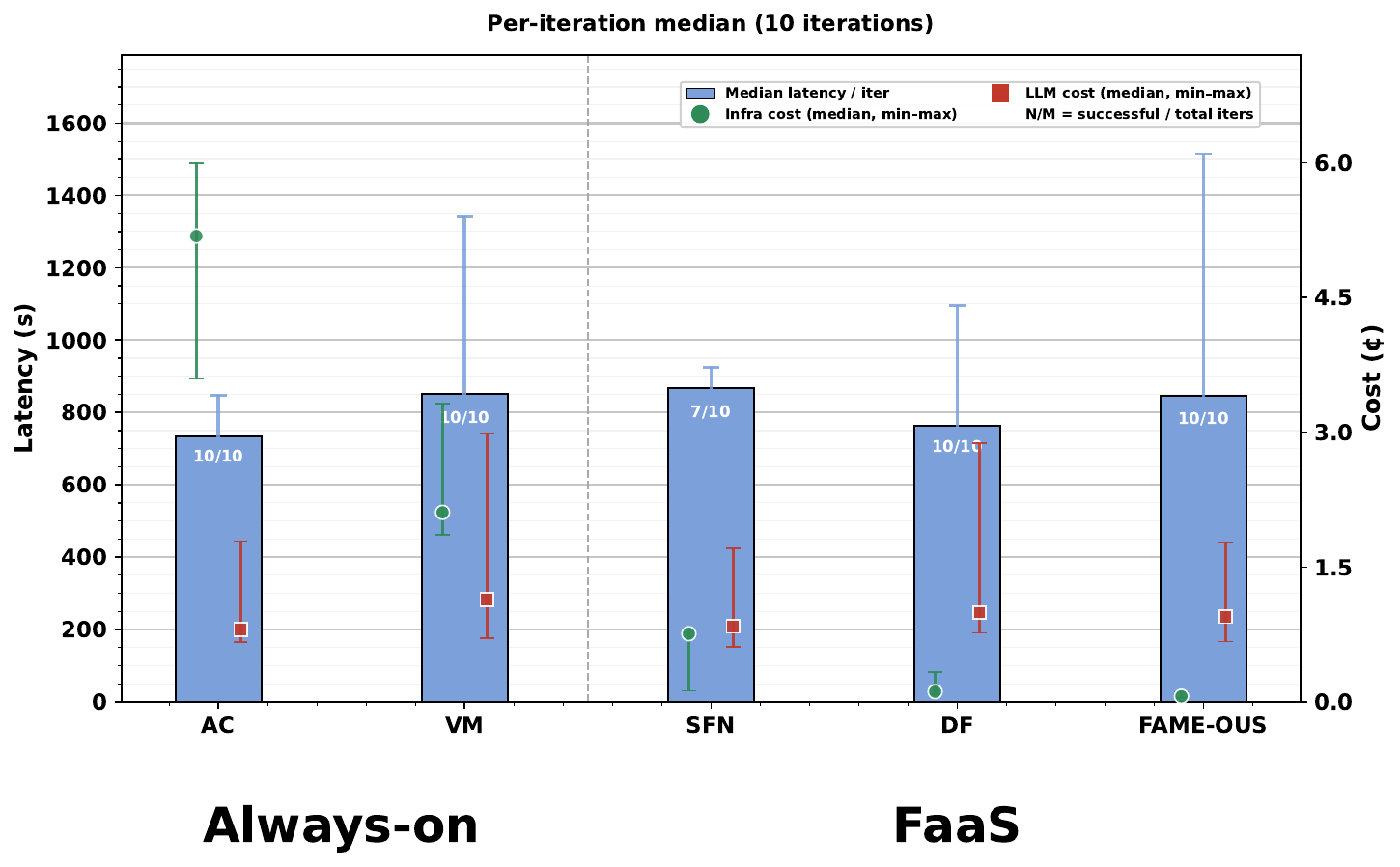}
\caption{Per-iteration median latency and cost for the Tabular-2022 \mlzero{} run. Synchronous Step Functions does not finish 3 of 10 Coder iterations; the other platforms finish all 10 iterations.}
    \label{fig:claim1-per-iter}
\end{figure}

Fig.~\ref{fig:claim1-per-iter} shows the per-iteration behavior behind the aggregate Tabular-2022 result in Fig.~\ref{fig:claim1}. Synchronous Step Functions fails on three Coder iterations because the Lambda invocation waits beyond the platform lifetime limit, while VM, AC, DF, and \fw complete all iterations. The completed iterations also show that the long-running cost gap is not caused by one outlier iteration; rather, the same pattern repeats across the MCTS search, with always-active or synchronous runtimes retaining resources across the sandbox wait while async runtimes resume only after the sandbox returns.

\section{Workflow Configurations and MCP Deployment} \label{app:wf-config}

\subsection{Memory and Cache Configurations} \label{app:mem-config}

Table~\ref{tbl:mem-config} summarizes the short-running workflow configurations used in \S\ref{sec:wf:configs}. \textbf{E} uses no prior memory, \textbf{N} uses client-visible history, \textbf{C} enables MCP cache reuse, \textbf{M} enables agent memory injection, and \textbf{MC} combines agent memory with MCP caching.

\begin{table}[t!]
\centering
\caption{Short-running Workflow Experiment Configurations}
\label{tbl:mem-config}
\footnotesize
\begin{minipage}[b]{0.62\columnwidth}%
\begin{tabular}{p{2.7cm}|p{5cm}}%
\hline
\bf Param. & \bf Values\\
\hline\hline
Agentic Patterns & \react (\textbf{RT}), \reflect (\textbf{RX}), \llmc (\textbf{LC})\\\hline
Agentic Apps., \#~Tasks & Research Paper Summarization, $3\times$ (\textbf{RS}); Log Analytics $3\times$ (\textbf{LA}); TauBench (\textbf{TB}), $10\times$; LiveMCPBench (\textbf{LM}), $10\times$\\\hline
Memory Configs. & \textbf{E, N, C, M, MC} \\
\hline
\end{tabular}%
\end{minipage}%
\begin{minipage}[b]{0.385\columnwidth}%
\centering%
\begin{tabular}{C{1cm}|C{1cm}C{1.5cm}C{1cm}}%
\hline
\textbf{Cfg.} & \textbf{Client Mem.} & \textbf{Agentic Mem.} & \textbf{MCP Cache} \\
\hline\hline
\bf E  & \xmark & \xmark & \xmark \\
\bf N  & \cmark & \xmark & \xmark \\
\bf C  & \cmark & \xmark & \cmark \\
\bf M  & \cmark & \cmark & \xmark \\
\bf MC & \cmark & \cmark & \cmark \\
\hline
\end{tabular}
\end{minipage}

\end{table}

\subsection{AgentCore Implementation} \label{app:agentcoreimplementation}
Our AgentCore baseline deploys the orchestrator and agent roles as distributed AgentCore runtimes. The orchestrator invokes the Coder agent through \texttt{invoke\_agent\_runtime}; the Coder then invokes the sandbox that executes the generated training script. Since the sandbox execution can last for minutes to hours, the AgentCore implementation must keep the agent session alive while the external work progresses.

To avoid AgentCore idle-timeout behavior, the Coder places the long-running sandbox call on a background task. The runtime registers the task using \texttt{app.add\_async\_task()}, causing \texttt{/ping} to report \texttt{HealthyBusy} while the task exists. The main request path returns an \texttt{ACCEPTED} response with a job identifier, and the caller polls a \texttt{check\_status} action every 30\,s. Intermediate results are stored in an in-memory dictionary protected by a \texttt{threading.Lock}. When the sandbox finishes, \texttt{app.complete\_async\_task()} removes the task and \texttt{/ping} returns to \texttt{Healthy}.

This implementation keeps the AgentCore session and polling path active across the sandbox wait, while the external sandbox is billed separately. In contrast, \fw persists state through checkpointing and resumes only on callback, so the FaaS agent and orchestrator do not remain active during the EMCP wait.

\subsection{Agentic Application Prompts} \label{app:agenticappprompts}

The short-running workloads use fixed prompts across configurations. For Log Analytics, the three turns request error counting, timestamp statistics, and comparative visualization.

\begin{lstlisting}[language={},breaklines=true,breakindent=0pt,mathescape=true]
$\textbf{LOG ANALYTICS PROMPT:}$
$\textbf{Q1:}$ Count the occurrences of error states $\langle$STATE$\rangle$ in the log file $\langle$FILE$\rangle$; $\textbf{Q2:}$ Find the mean and standard deviation of timestamps for the most frequent error; $\textbf{Q3:}$ Find the min/max/mean/median timestamps with visualization and comparison between error states.
\end{lstlisting}

The actor-memory prompt is enabled for \textbf{M} and \textbf{MC}. It instructs the Actor to reuse prior tool outputs when the prior call matches the current need.

\begin{lstlisting}[language={},breaklines=true,breakindent=0pt,mathescape=true]
$\textbf{ACTOR MEMORY PROMPT:}$ Check previous ToolMessage responses in conversation history before making new tool calls. Extract data from previous tool outputs instead of calling tools again with the same parameters. Only make new calls if data is unavailable or parameters differ.
\end{lstlisting}

For MLE-Bench, we add task-specific constraints to bound the search budget and obtain valid submissions within the evaluation window. The original MLZero setup allows much larger search budgets and GPU-enabled sandboxes; our prompts limit training time and library choices so that the systems comparison focuses on runtime behavior rather than further agent tuning.

\begin{lstlisting}[language={},breaklines=true,breakindent=0pt,mathescape=true]
$\textbf{TABULAR-2022 USER PROMPT:}$ Solve this ML problem with an appropriate choice of Deep Learning Model. This is a binary classification problem on tabular data.  Follow these strict requirements:
1. Use the autogluon.tabular library. 
2. Use preset "medium_quality"  with a "time_limit" = 600. These are all parameters that you can pass inside the .fit() method.
\end{lstlisting}

\begin{lstlisting}[language={},breaklines=true,breakindent=0pt,mathescape=true]
$\textbf{DOB-BREED USER PROMPT:}$ Solve this ML problem with an appropriate choice of Deep Learning Model. This is a Multi-class image classification problem. Follow these strict requirements: 1. Use the machine_learning tool. 2. Use smaller models for transfer learning(e.g MobileNetV3-Small). 3. Filter the training data first (using `os.path.exists`) as some image IDs in `labels.csv` are missing. 4. Limit training time to under 1800 secs and predict class probabilities.
\end{lstlisting}

\subsection{Automated FaaS Wrappers for MCP Services} \label{sec:app:arch:mcp}

The elasticity and cost benefits of exposing agentic patterns as FaaS workflows are fully realized only if the MCP service plane can scale in the same manner. \fw therefore provides wrappers that expose MCP tools as Lambda endpoints while preserving MCP tool semantics.

\begin{lstlisting}[language=Python]
@mcp.tool()      # Decorator for Anthropic MCP Tool
@fame.wrapper()  # Decorator for FAME MCP Wrapper
def fetch(url: str, max_length: int = 5000):
    # Original tool call logic of MCP
    content = fetch_url_content(url)
    result = process_content(content, max_length)
    return result
\end{lstlisting}

The wrapper normalizes HTTP invocation, records execution telemetry, and enables the MCP optimizations described below. This allows short-running MCP services to run as elastic cloud functions instead of long-running servers.

\subsection{Optimizations for FaaS MCP}\label{sec:arch:mcp:opt}
\fw provides optimizations to reduce latency/costs for FaaS MCP deployments. 

\subsubsection{Invocation Caching}
\label{sec:mcp-cache}

For deterministic tools, repeated MCP calls with identical arguments can inflate latency and LLM input tokens, e.g., repeatedly downloading the same arXiv paper. \fw introduces a cache manager that derives a cache key from the tool name and arguments, stores the tool response in S3, and returns cache hits directly. Cache entries use developer-defined TTL values to avoid cache bloat; for example, volatile tools can use short TTLs, while stable document identifiers can use long TTLs. Cacheability is opt-in so that non-deterministic tools are not reused incorrectly.

\subsubsection{S3-based File Handling} \label{sec:mcp-s3}
Tools that return large inline responses can overwhelm agent memory and the LLM context window. \fw supports storing large tool outputs in S3 and returning the S3 URI instead of the full payload. Downstream MCP tools can receive the URI as an argument and download the object locally. For example, a downloader tool can store a paper PDF in S3 and pass the URI to a downstream RAG tool, avoiding repeated insertion of the file content into prompts.

\subsubsection{Singleton vs. Consolidated MCP Servers} \label{sec:mcp-cons-methods}
\fw supports both per-tool or per-server \textit{singleton} Lambdas and \textit{consolidated} Lambdas that expose multiple MCP tools, as suggested by prior work~\cite{agentx}. Singleton deployments improve isolation and allow tool-specific resource sizing. Consolidated deployments combine frequently co-invoked tools into one function, amortizing cold starts and reducing steady-state invocation overhead. Both modes are supported by the wrapper, allowing workflow developers to choose per application.

\subsubsection{Discussion}
The MCP wrapper closes the deployment gap between FaaS-hosted agents and MCP-hosted tools. Combined with agent memory injection, S3 handle passing, and tool-output caching, it reduces repeated payload movement, redundant tool invocations, and LLM prompt growth for short-running service workflows.

\section{Additional Short-running Results} \label{app:results:short}

This appendix provides supporting results for the short-running service workloads discussed in \S\ref{sec:results:e2e}--\S\ref{sec:results:benchmarks}. It includes additional \react inputs, detailed \reflect and \llmc plots, benchmark extensions, the AgentCore comparison, and MCP deployment-strategy measurements.

\subsection{\react Detailed Per-Turn Analysis: RS/P3 and LA/F3} \label{app:results:react-detail}

\begin{figure}[t]
\centering
  \subfloat[Research Summary \textit{(RS/P3)}]{
    \label{fig:arxiv-d-s3-e2e}
    \includegraphics[width=0.45\columnwidth]{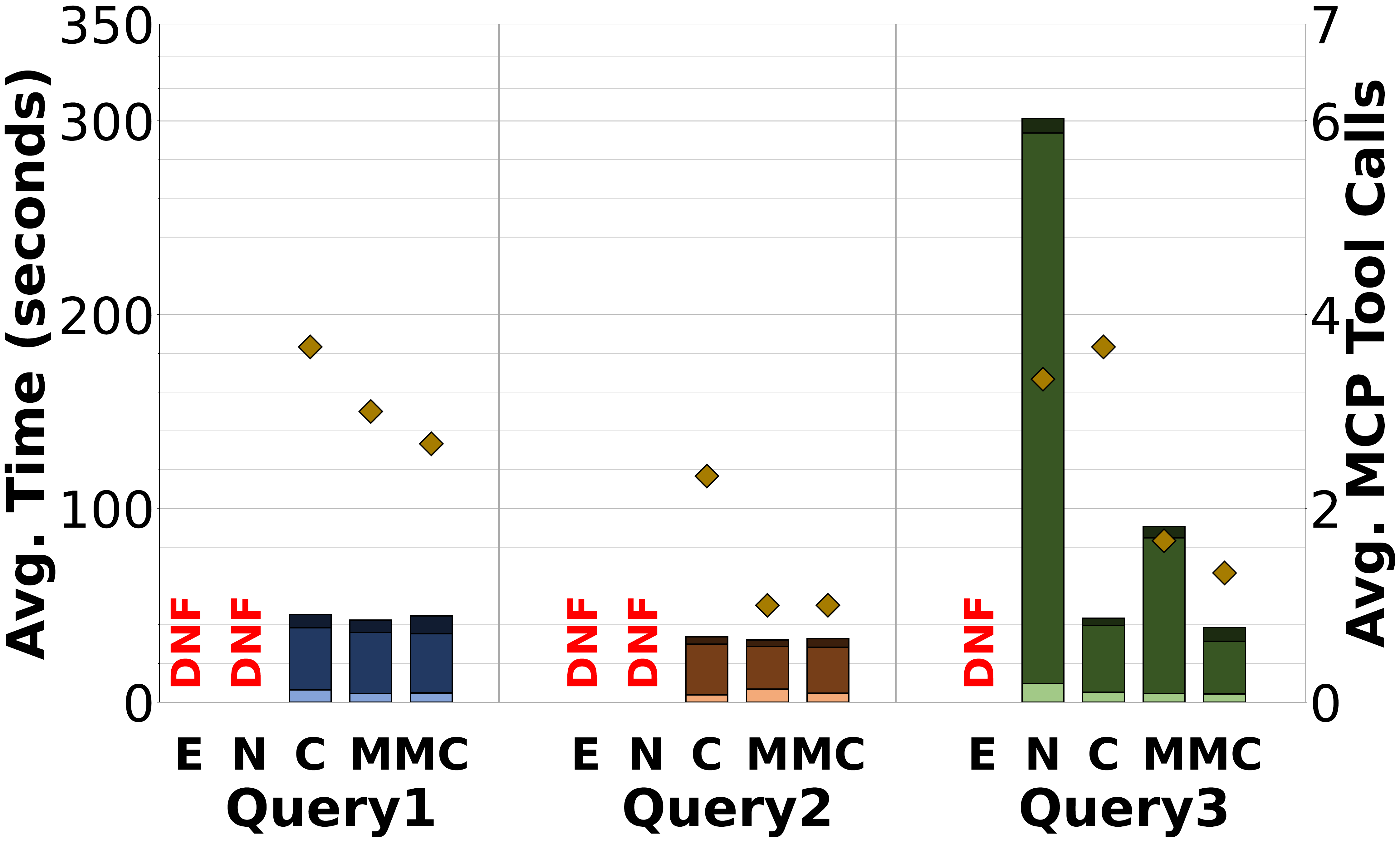}
  }~~
  \subfloat[Log Analytics \textit{(LA/F3)}]{
    \includegraphics[width=0.45\columnwidth]{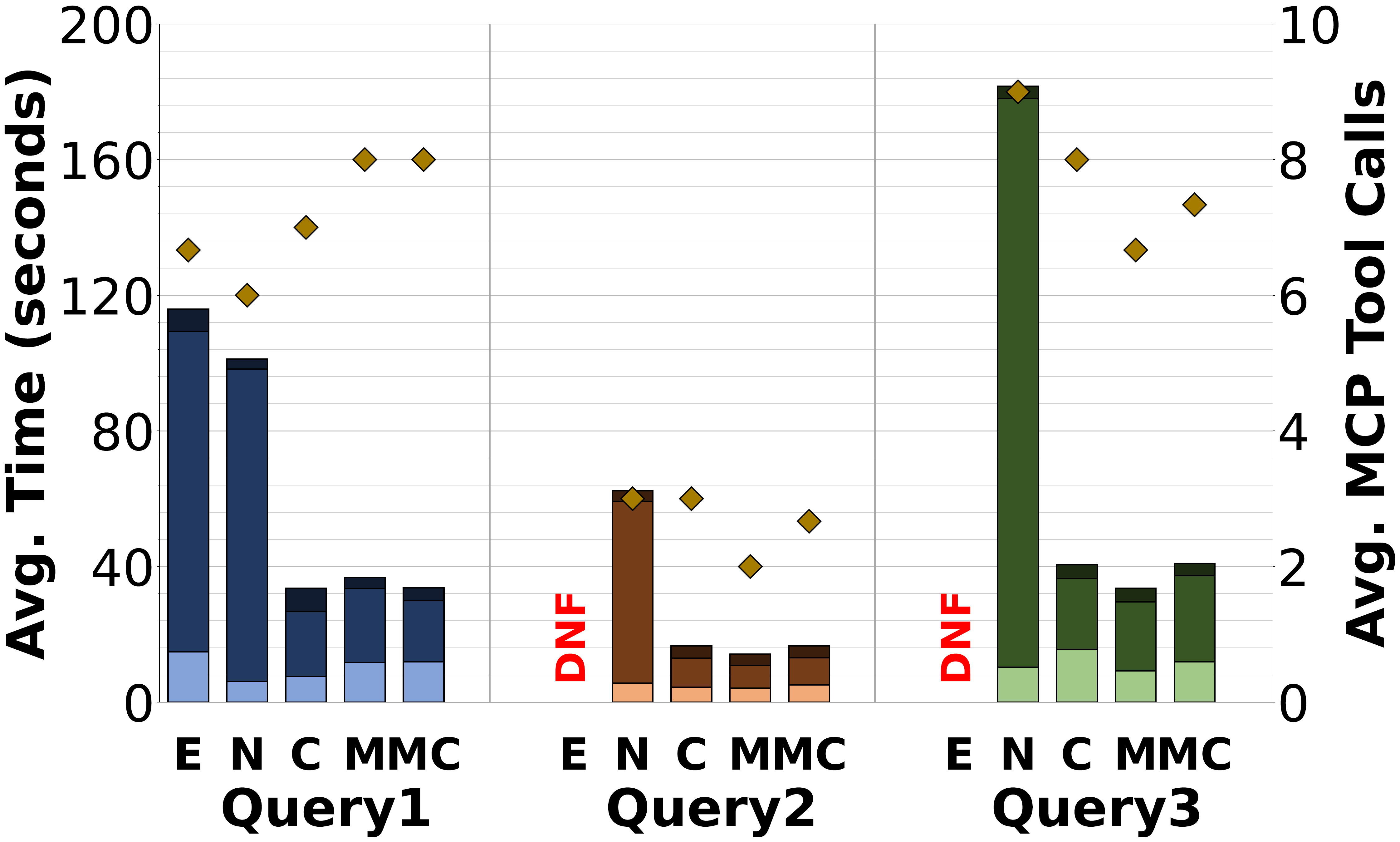}
     \label{fig:logger-d-s3-e2e}
  }
\caption{Average per-turn latency for \react on the third inputs, RS/P3 and LA/F3. The trends match the main-text \react results in Fig.~\ref{fig:plot-e2e-latencies}.}
\label{fig:plot-e2e-latencies-p3}
\end{figure}

Fig.~\ref{fig:plot-e2e-latencies-p3} complements the main-text \react latency results by showing the third RS and LA inputs. As in Fig.~\ref{fig:plot-e2e-latencies}, configurations with agent memory and MCP I/O optimizations reduce later-turn latency by avoiding repeated tool calls and repeated prompt insertion of large artifacts.

\subsection{\reflect and \llmc Results} \label{app:results:reflect-llmc}

\begin{figure}[t]
\centering%
  \subfloat[\reflect \textit{(RS/P1)}]{
     \includegraphics[width=0.45\columnwidth]{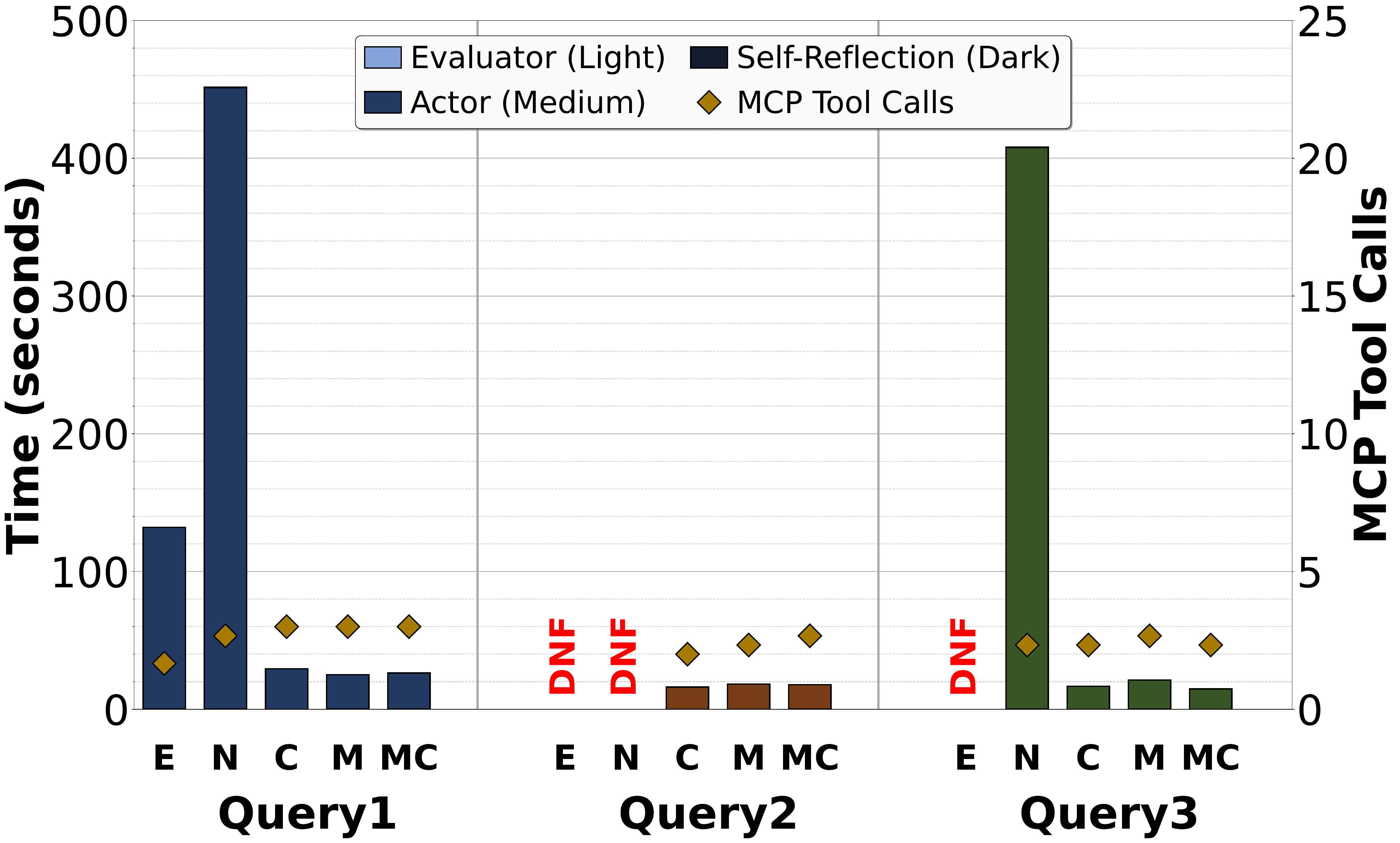}
     \label{fig:arxiv-d-s1-e2e-refl}
  }~~
  \subfloat[\reflect \textit{(LA/F1)}]{
    \includegraphics[width=0.45\columnwidth]{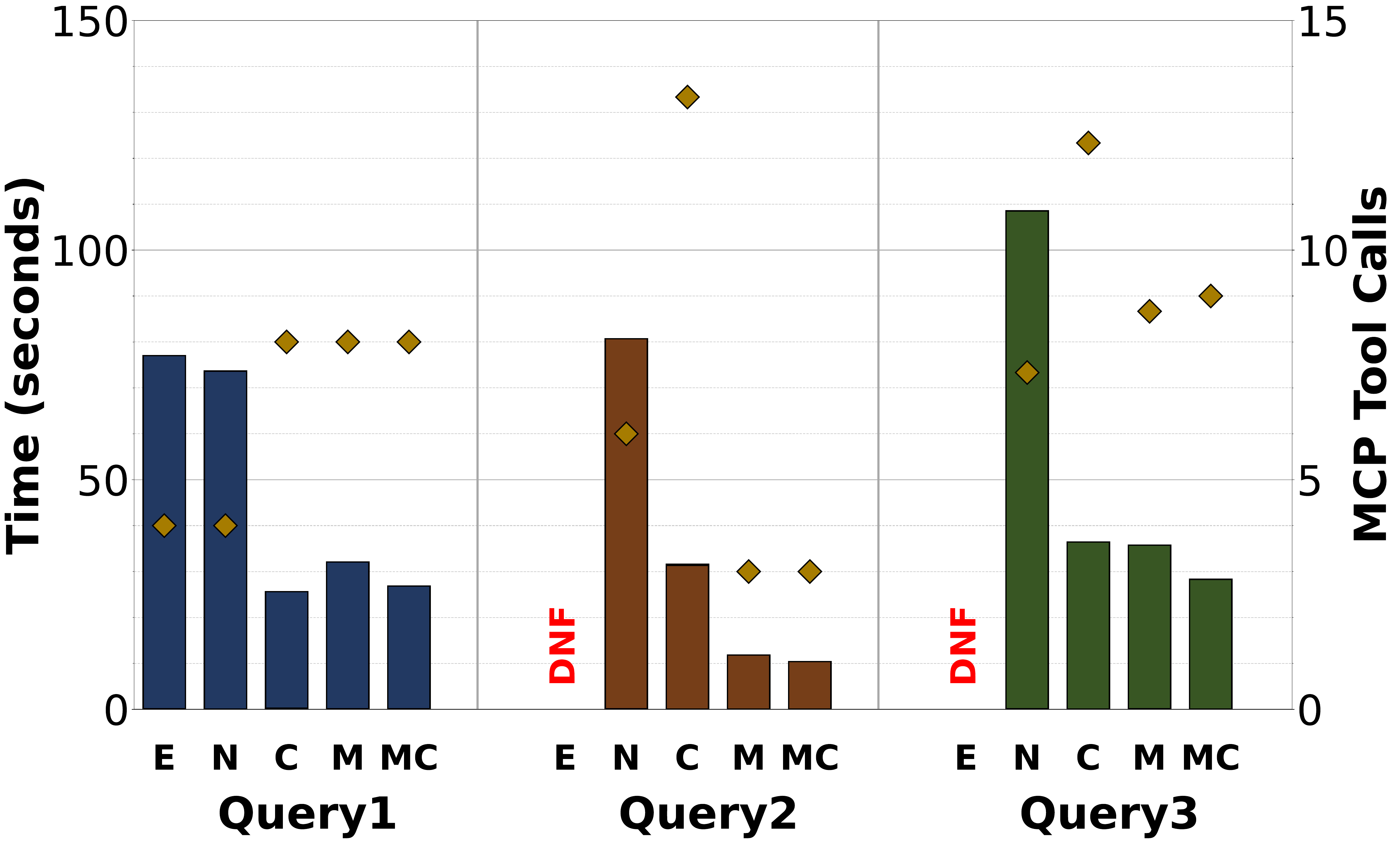}
     \label{fig:logger-d-s3-e2e-refl}
  }\\
  \subfloat[\llmc \textit{(RS/P1)}]{
     \includegraphics[width=0.45\columnwidth]{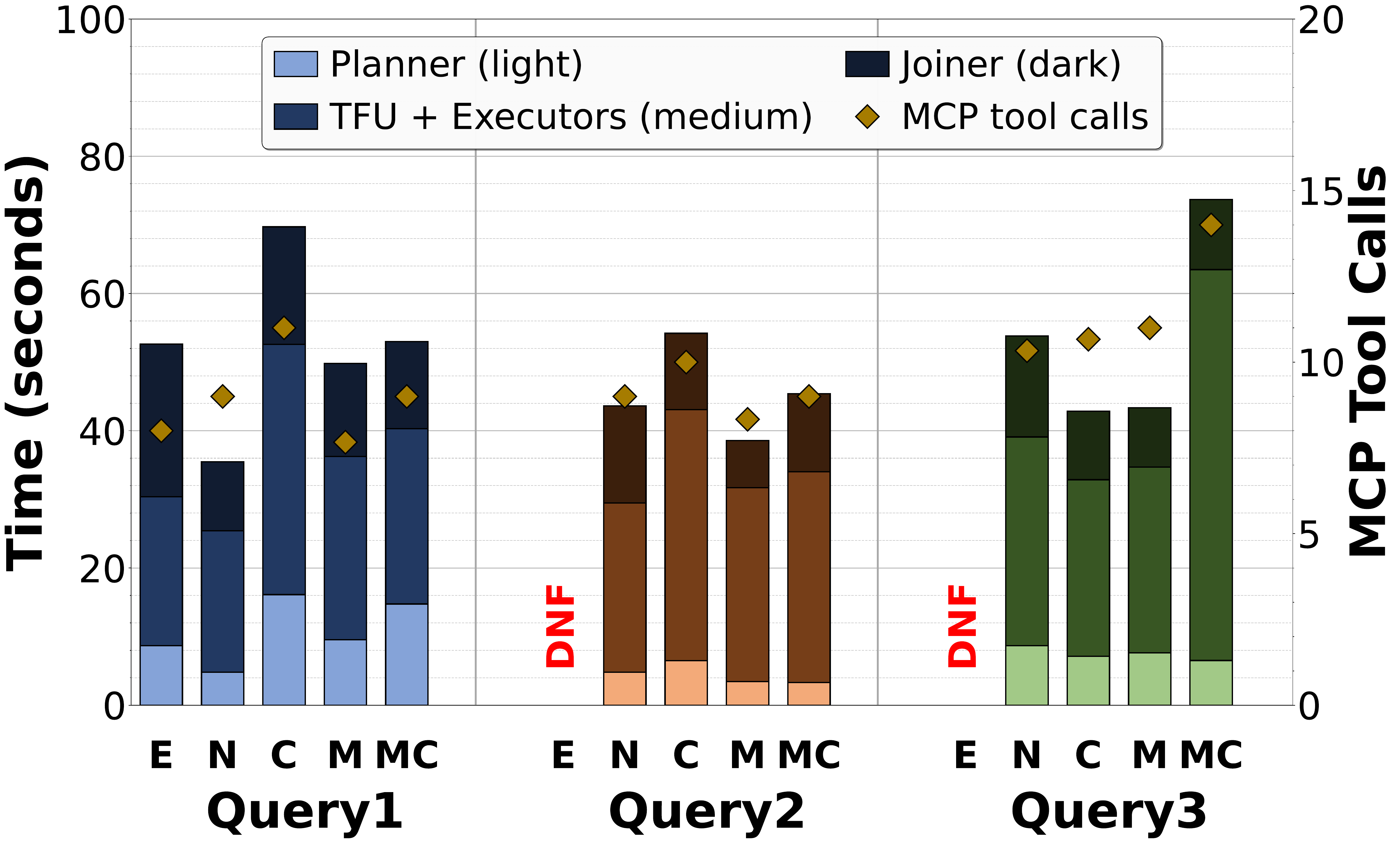}
     \label{fig:arxiv-d-s1-e2e-llmc}
  }~~
  \subfloat[\llmc \textit{(LA/F1)}]{
    \includegraphics[width=0.45\columnwidth]{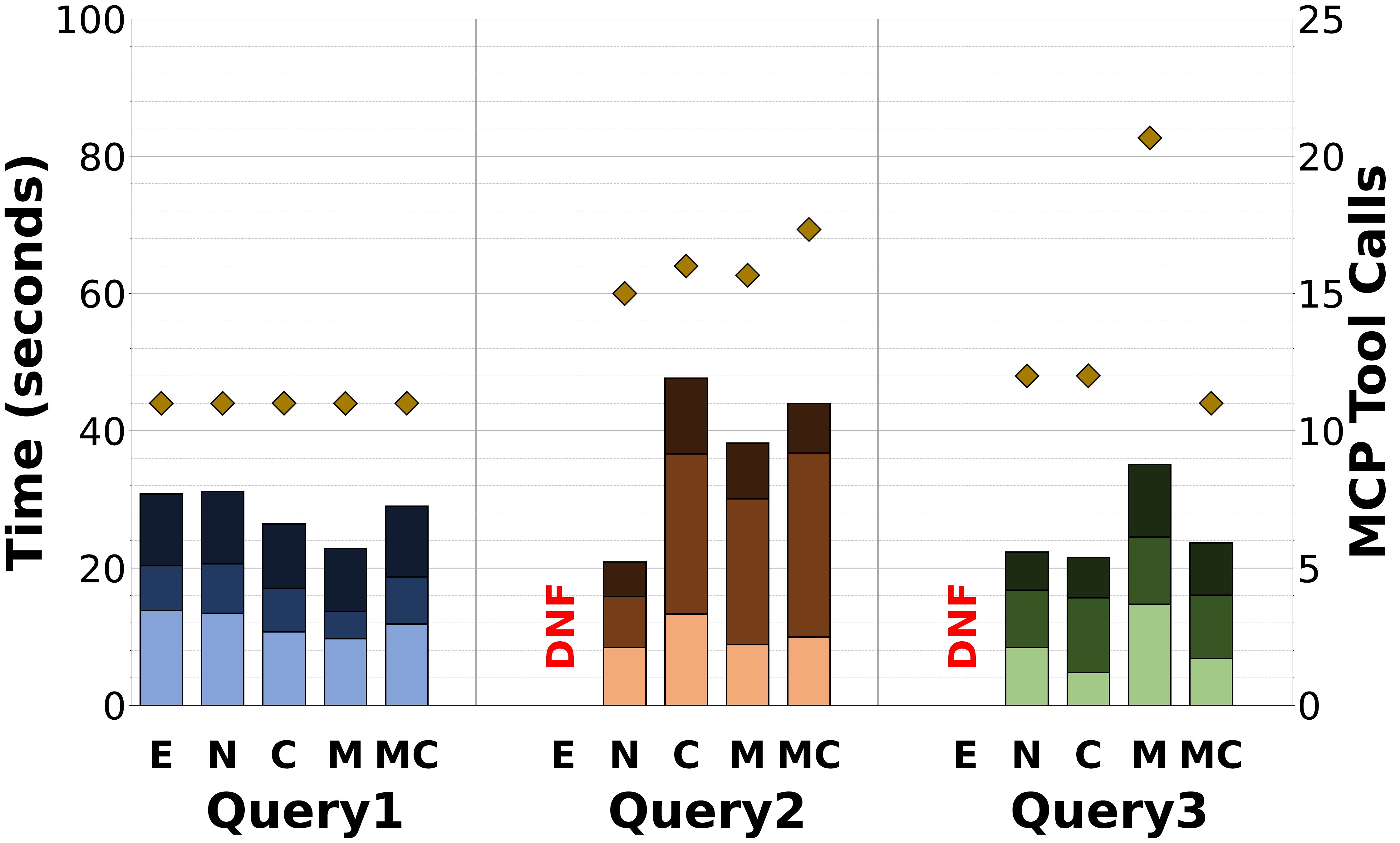}
     \label{fig:logger-d-s1-e2e-llmc}
  }
\caption{Per-turn latency for \reflect and \llmc on RS/P1 and LA/F1.}
\label{fig:plot-latency-reflexion-llmc}
\end{figure}

\paragraph{\reflect.}
With \reflect, failed turns become targeted retries guided by persisted trajectories and reflections in DynamoDB. On RS/P1, optimized configurations achieve up to $17\times$ lower latency, while LA/F1 shows smaller but consistent reductions (Fig.~\ref{fig:arxiv-d-s1-e2e-refl}, Fig.~\ref{fig:logger-d-s3-e2e-refl}). The reasons mirror \react: memory reduces repeated work across turns, and S3 handles and caching reduce large prompt payloads.

\paragraph{\llmc.}

\llmc exposes workflow-level parallelism, but the benefit depends on the structure and quality of the generated task DAG. On RS/P1, latency gains are modest because the workflow is mostly linear across turns (Fig.~\ref{fig:arxiv-d-s1-e2e-llmc}). On LA/F1, plan-quality issues can cause executor failures and replans, increasing latency for optimized configurations relative to \textbf{N} (Fig.~\ref{fig:logger-d-s1-e2e-llmc}). These failures arise from the \llmc planning pattern rather than from the \fw deployment mechanism.

\subsection{Token and Cost (RS, LA)}
\begin{figure}[t]
\centering%
  \subfloat[\reflect \textit{(RS/P1)}]{
    \includegraphics[width=0.45\columnwidth]{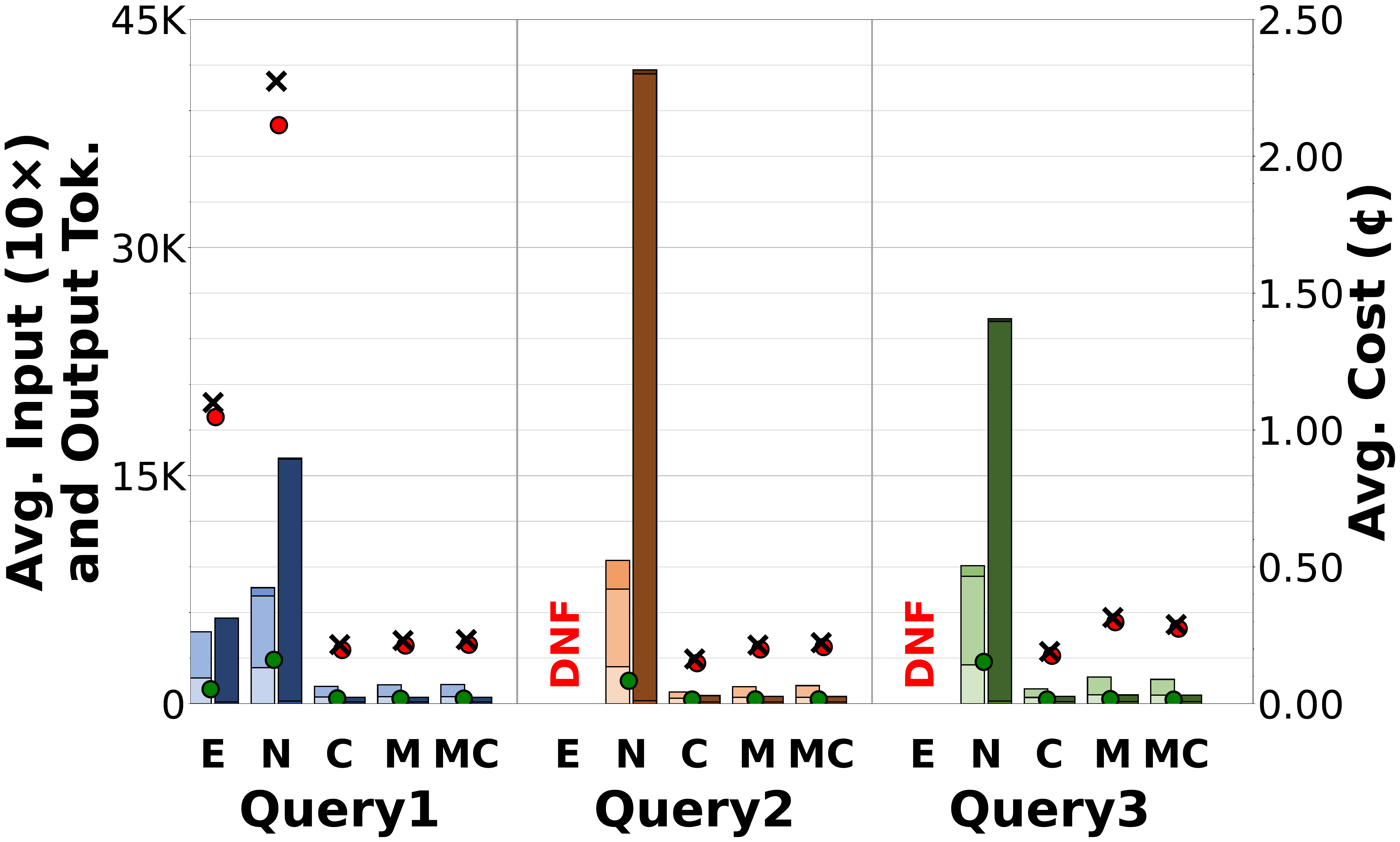}
     \label{fig:arxiv-d-s2-io-refl}
  }~~
  \subfloat[\reflect \textit{(LA/F1)}]{
    \includegraphics[width=0.45\columnwidth]{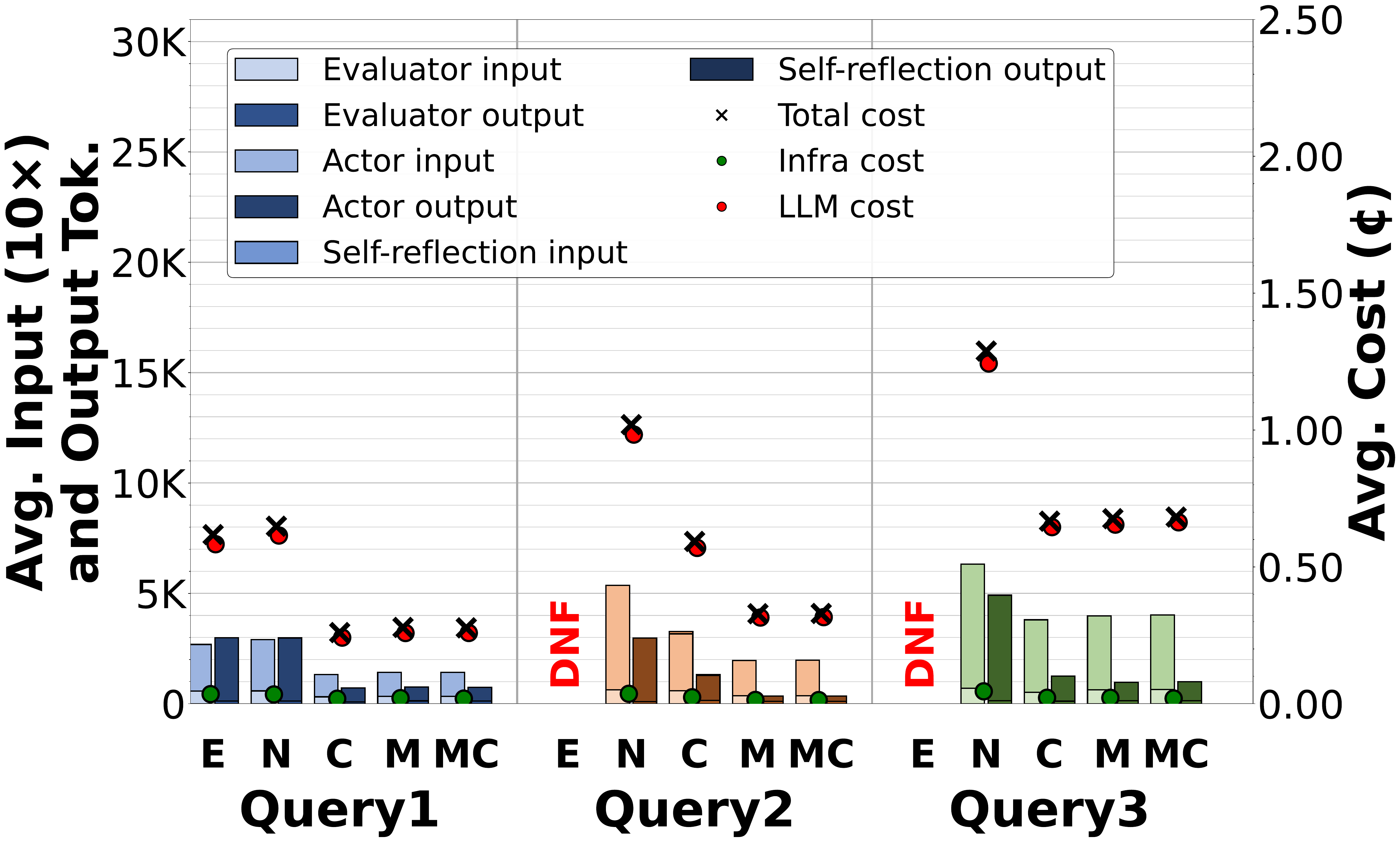}
     \label{fig:logger-d-s3-costs-refl}
  }\\
    \subfloat[\llmc\textit{(RS/P1)}]{
    \includegraphics[width=0.45\columnwidth]{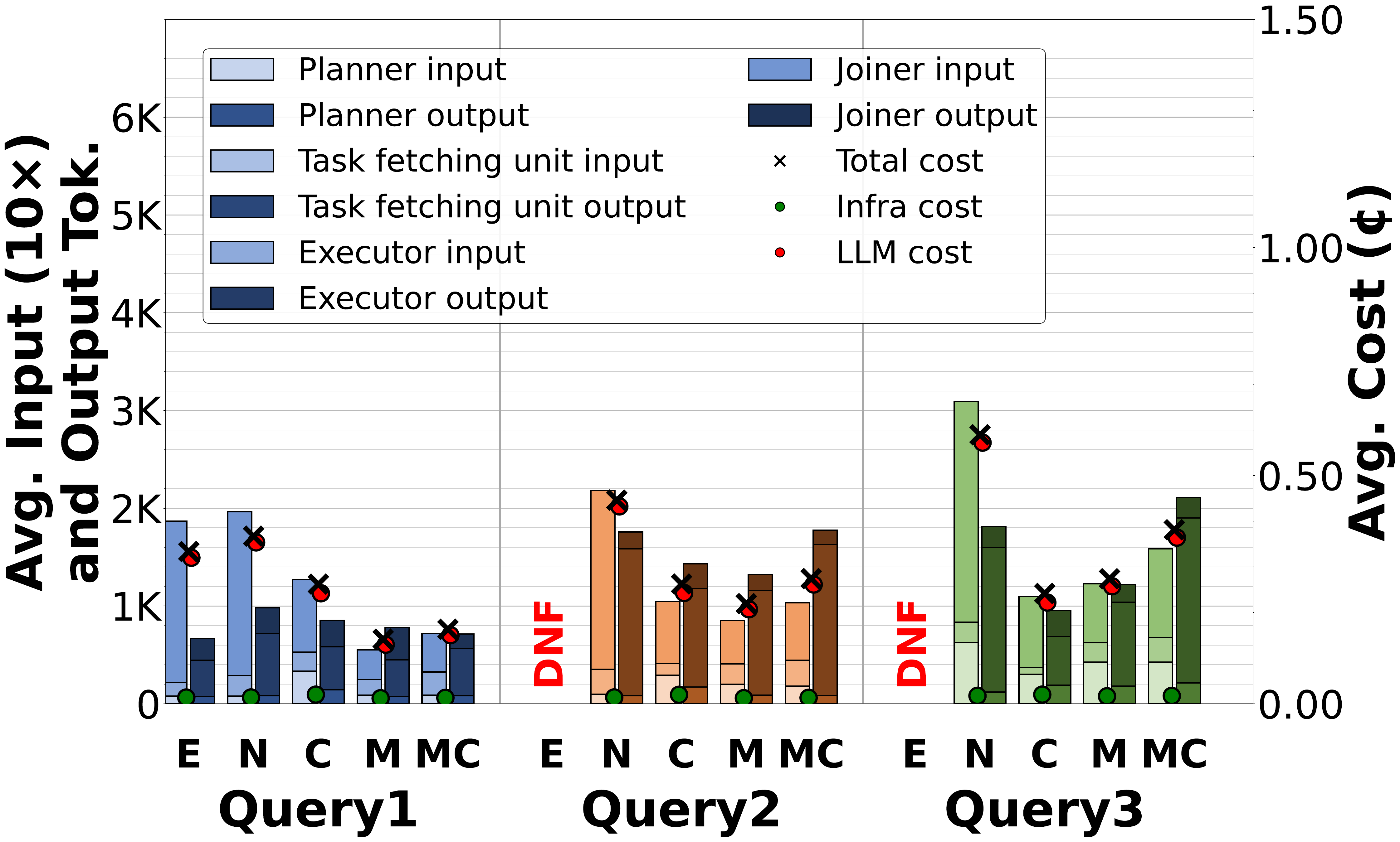}
     \label{fig:arxiv-d-s2-io-llmc}
  }~~
  \subfloat[\llmc \textit{(LA/F1)}]{
    \includegraphics[width=0.45\columnwidth]{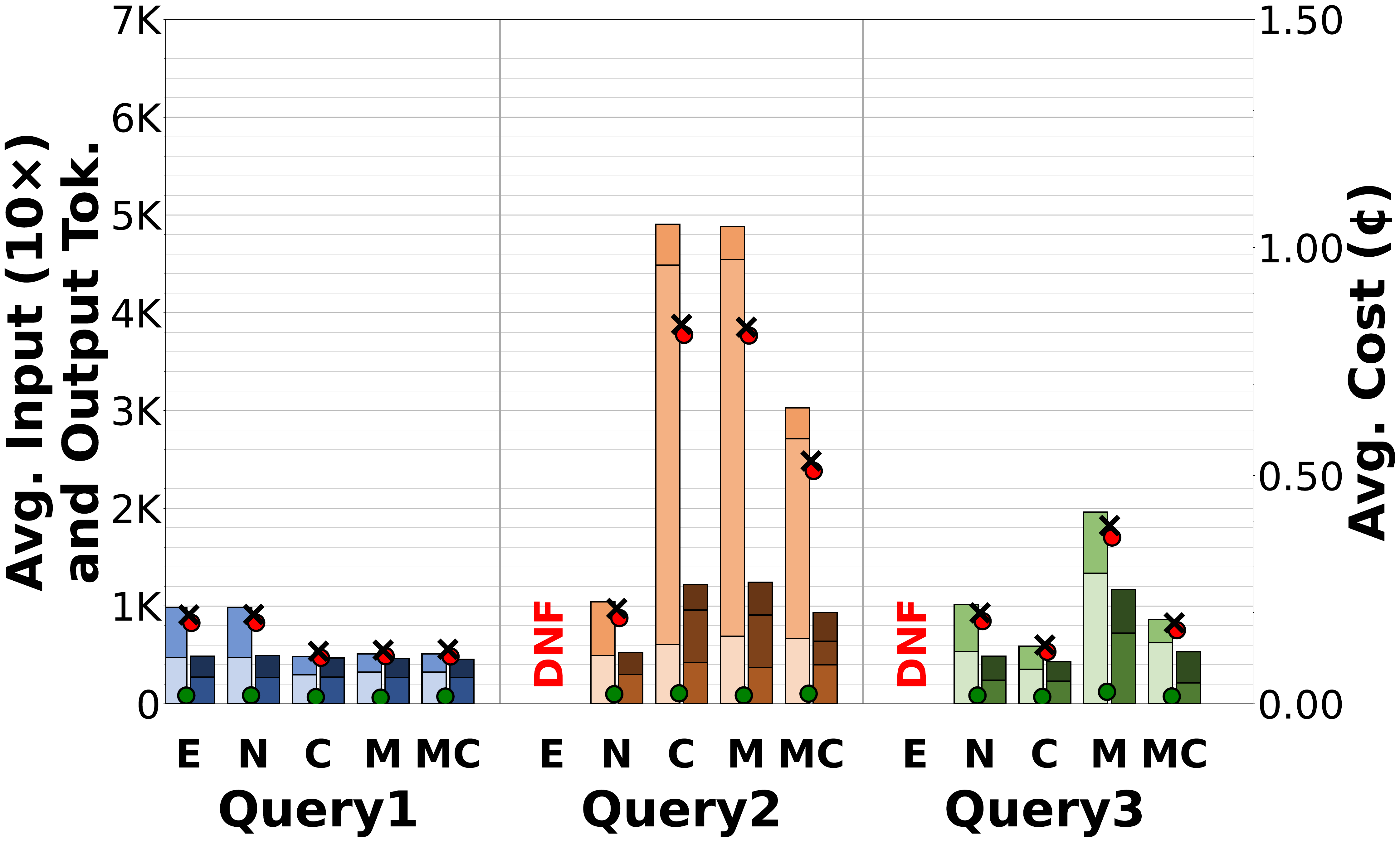}
     \label{fig:logger-d-s1-io-llmc}
  }
\caption{Token use and cost for \reflect and \llmc. Bars show input/output tokens and markers show monetary cost.}
\label{fig:plot-cost-reflexion-llmc}
\end{figure}

\paragraph{\reflect.} \reflect follows the same token and cost trends as \react. On RS/P1, optimized configurations reduce input tokens by up to $\approx84\%$ relative to \textbf{N}, with corresponding cost reductions (Fig.~\ref{fig:arxiv-d-s2-io-refl}). LA/F1 shows the same qualitative trend, with lower token use and lower cost under memory and MCP I/O optimizations (Fig.~\ref{fig:logger-d-s3-costs-refl}).

\paragraph{\llmc.} For \llmc on RS/P1, \textbf{M} reduces token use by $\approx74\%$ and cost by $3.16\times$, while \textbf{C/MC} reduce token use by $\approx66$--$67\%$ with $2.4$--$2.7\times$ cost reductions (Fig.~\ref{fig:arxiv-d-s2-io-llmc}). On LA/F1, the plan-quality failures discussed above increase token use and weaken cost gains (Fig.~\ref{fig:logger-d-s1-io-llmc}).

\subsection{Benchmarks: \reflect on TauBench and LiveMCPBench} \label{app:results:reflect-benchmarks}

\begin{figure}[t]
\centering
  \subfloat[LiveMCP Bench Reflection]{%
     \includegraphics[width=0.45\columnwidth]{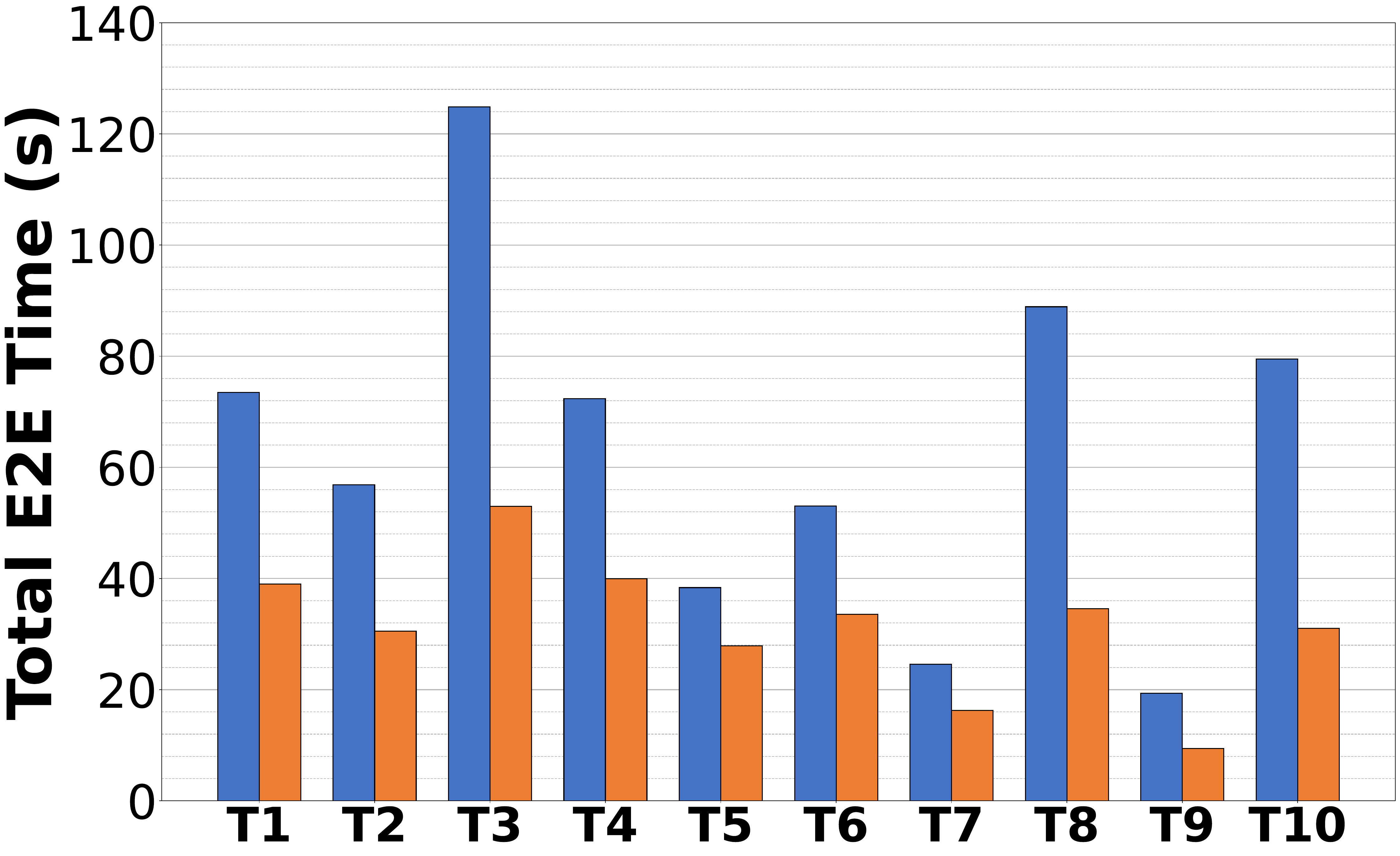}%
     \label{fig:livemcp-reflx}%
  }~~
  \subfloat[Tau Bench \reflect]{%
    \includegraphics[width=0.45\columnwidth]{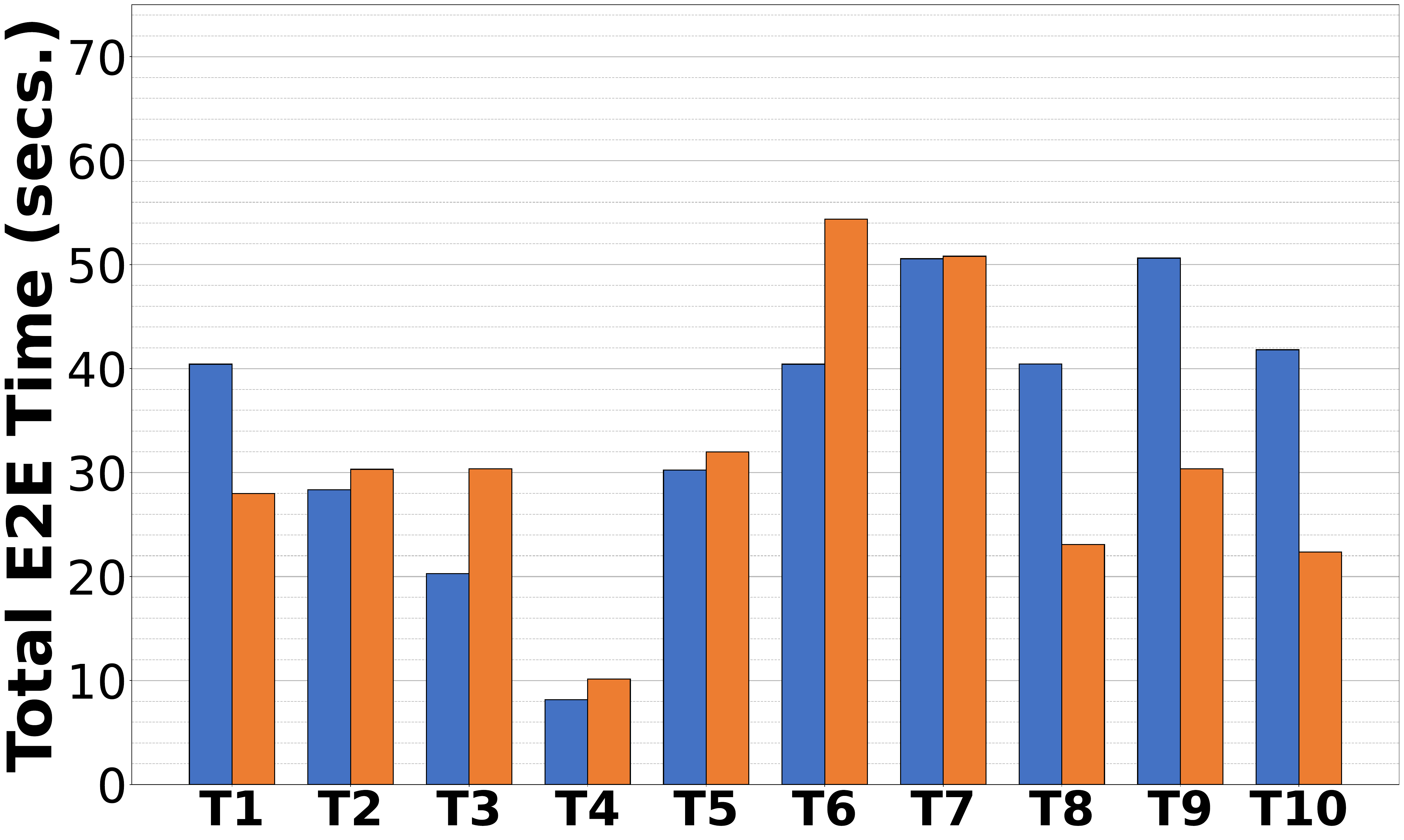}%
     \label{fig:tau-reflx}%
  }
\caption{Per-task latency on LiveMCPBench and TauBench for \reflect, comparing \textbf{N} and \textbf{MC}.}
\label{fig:other-benchmarks-reflect}
\end{figure}

Fig.~\ref{fig:other-benchmarks-reflect} extends the benchmark evaluation to \reflect. The latency behavior follows the same pattern as the main-text \react benchmark results: payload-heavy LiveMCPBench tasks benefit more from \textbf{MC}, while TauBench shows smaller and task-dependent changes because its tool outputs are compact.

\subsection{Per-Application Comparison with AgentCore} \label{app:results:cmp-fame}

\begin{figure}[t]
\centering
    \includegraphics[width=0.65\columnwidth]{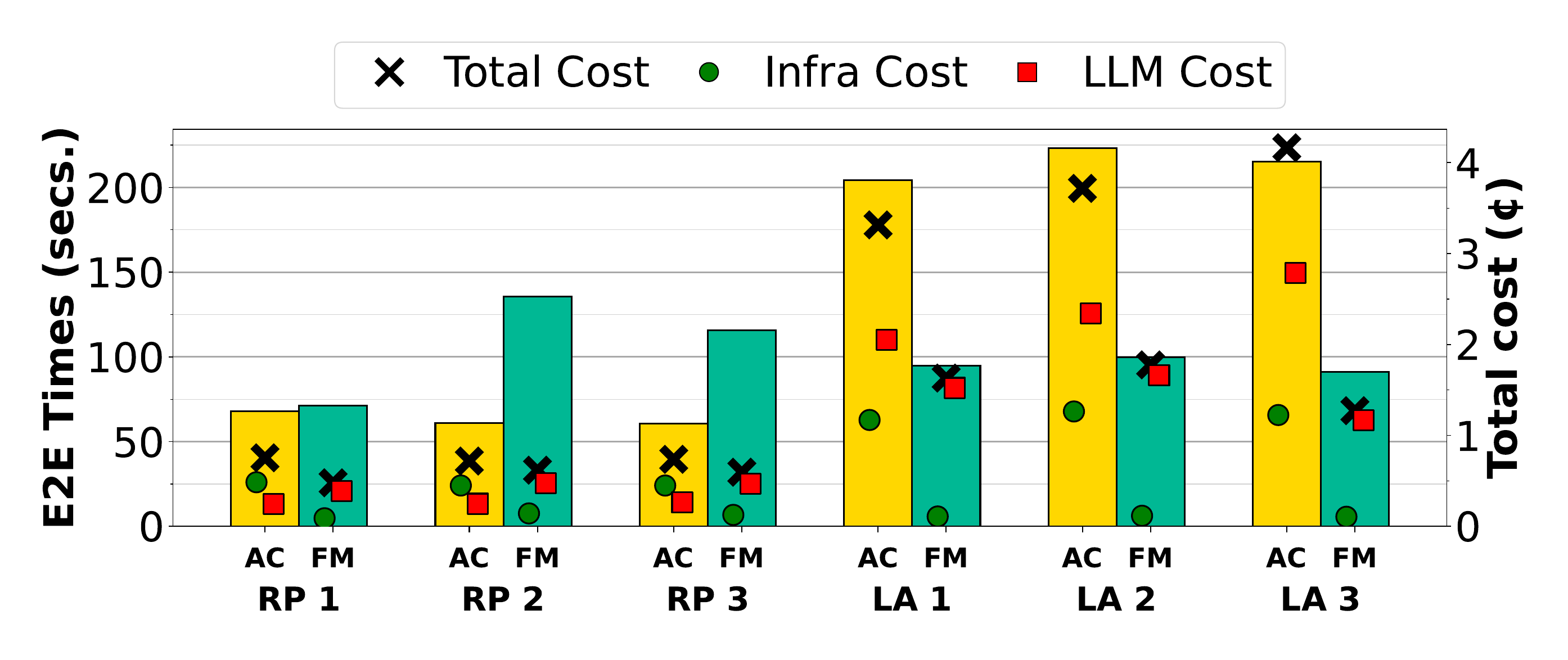}
\caption{Comparison of \fw and AgentCore on RS (P1--P3) and LA (F1--F3) under \textbf{MC}.}
    \label{fig:cmp-fame}
\end{figure}

Fig.~\ref{fig:cmp-fame} compares \fw and AgentCore under the strongest short-running configuration, \textbf{MC}. Latency is comparable across inputs, while \fw has lower total cost in this deployment. The difference reflects the use of FaaS-based role execution and MCP I/O optimizations in \fw, compared with the managed-runtime deployment used for AgentCore.

\begin{table*}[t]
\centering
\caption{Operations cost (\textcent) breakdown for Tabular workload between FAME (\textbf{F}) and Durable Functions (\textbf{DF}).}
\label{tab:tabular-concise-stacked}
\footnotesize
\setlength{\tabcolsep}{1.5pt} %
\begin{tabular}{l ccc ccc}
\toprule
\multicolumn{7}{c}{\textbf{(a) Unit Cost \& Base Operations}} \\
\cmidrule(l){1-7}
\textbf{Bridge} & $F_{\text{unit}}$ & $DF_{\text{unit}}$ & $\mathbf{R_u}$ ($F_{\text{unit}}/DF_{\text{unit}}$) & $F_{\text{ops}}$ & $DF_{\text{base}}$ & $\mathbf{R_b}$ ($F_{\text{ops}}/DF_{\text{base}}$) \\
\midrule
A-T & 0.00105 & 0.00258 & \textcolor{green!60!black}{$0.41\times$} & 0.04418 & 0.10320 & \textcolor{green!60!black}{$0.43\times$} \\
O-A & 0.00104 & 0.00086 & \textcolor{red!80!black}{$1.21\times$}   & 0.02284 & 0.01806 & \textcolor{red!80!black}{$1.26\times$} \\
\toprule
\multicolumn{7}{c}{\textbf{(b) DF Step Overhead \& Final Ratio}} \\
\cmidrule(l){1-7}
\textbf{Bridge} & \multicolumn{2}{c}{$DF_{\text{step}}$} & \multicolumn{2}{c}{$DF_{\text{ops}}$} & \multicolumn{2}{c}{$\mathbf{R_o}$ ($F_{\text{ops}}/DF_{\text{ops}}$)} \\
\midrule
A-T & \multicolumn{2}{c}{0.06020} & \multicolumn{2}{c}{0.16340} & \multicolumn{2}{c}{\textcolor{green!60!black}{0.27$\times$ ($\approx 3.7\times$ cheaper)}} \\
O-A & \multicolumn{2}{c}{0.03956} & \multicolumn{2}{c}{0.05762} & \multicolumn{2}{c}{\textcolor{green!60!black}{0.40$\times$ ($\approx 2.5\times$ cheaper)}} \\
\toprule
\multicolumn{7}{c}{\textbf{(c) Data Storage (KB) \& Cost}} \\
\cmidrule(l){1-7}
\textbf{Bridge} & \multicolumn{2}{c}{$DF_{\text{storage}}$ [Min--Max]} & \multicolumn{2}{c}{$F_{\text{storage}}$ [Min--Max]} & $\mathbf{R_{\text{data storage}}}$ ($F_s/DF_s$) & $\mathbf{R_{\text{data cost}}}$ ($F_{\text{cost}}/DF_{\text{cost}}$) \\
\midrule
A-T & \multicolumn{2}{c}{922.18 [640.50--922.18]} & \multicolumn{2}{c}{3138.90 [3138.90--4101.19]} & \textcolor{red!80!black}{3.40$\times$} & \textcolor{green!60!black}{0.0105$\times$ ($\approx 95.2\times$ cheaper)} \\
O-A & \multicolumn{2}{c}{1263.97 [1016.62--1263.97]} & \multicolumn{2}{c}{2881.08 [2881.08--4017.41]} & \textcolor{red!80!black}{2.28$\times$} & \textcolor{green!60!black}{0.0070$\times$ ($\approx 142.1\times$ cheaper)} \\
\bottomrule
\end{tabular}%
\end{table*}

\subsection{MCP Deployment Strategies} \label{sec:results:mcp-strategies}

\begin{figure}[t]
\centering%
  \subfloat[MCP Caching \textit{(Latency and Cache Hits)}]{%
     \includegraphics[width=0.45\columnwidth]{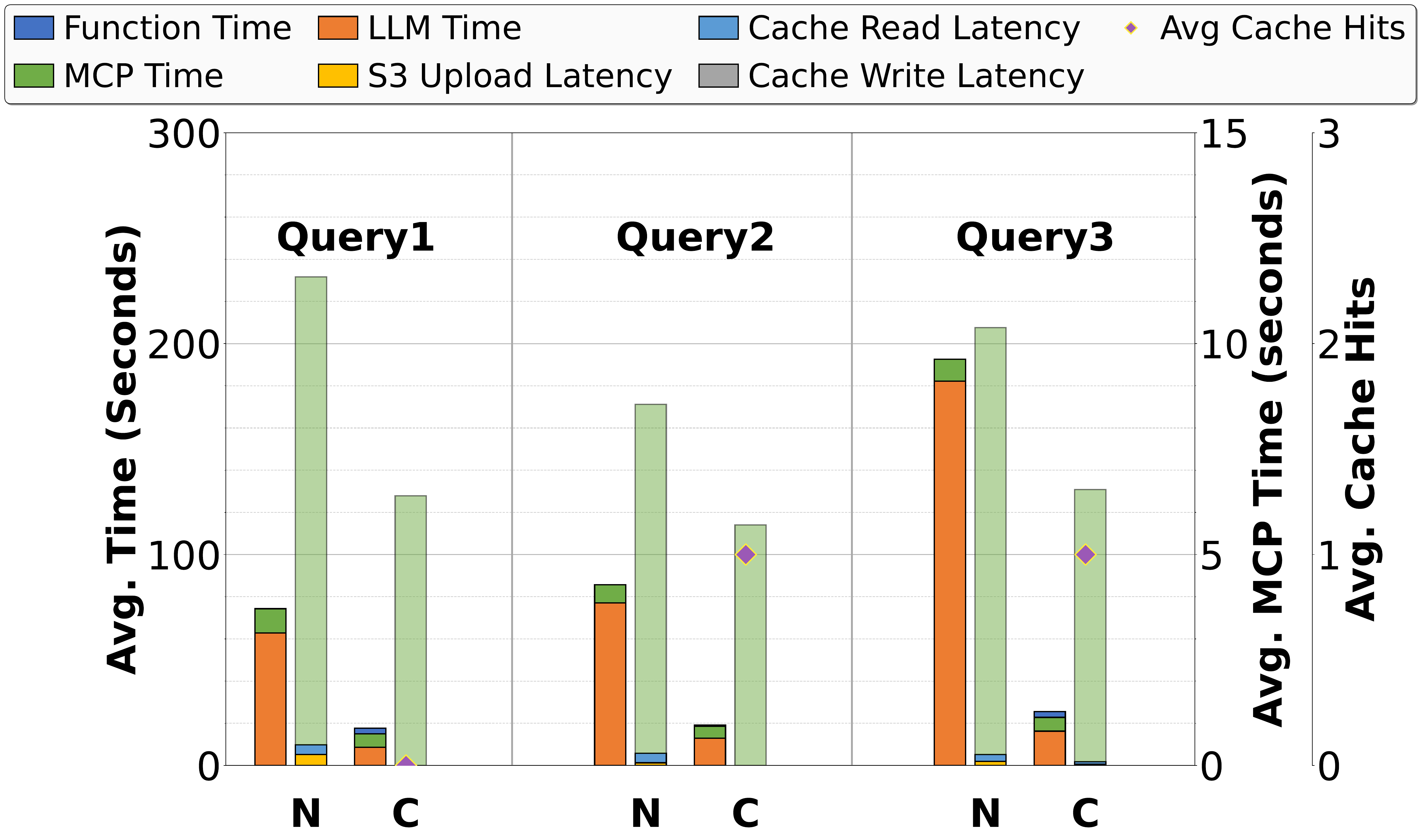}%
     \label{fig:mcp-cache}%
  }~~
  \subfloat[MCP Consolidation \textit{(Cold-start and sustained latency)}]{%
    \includegraphics[width=0.45\columnwidth]{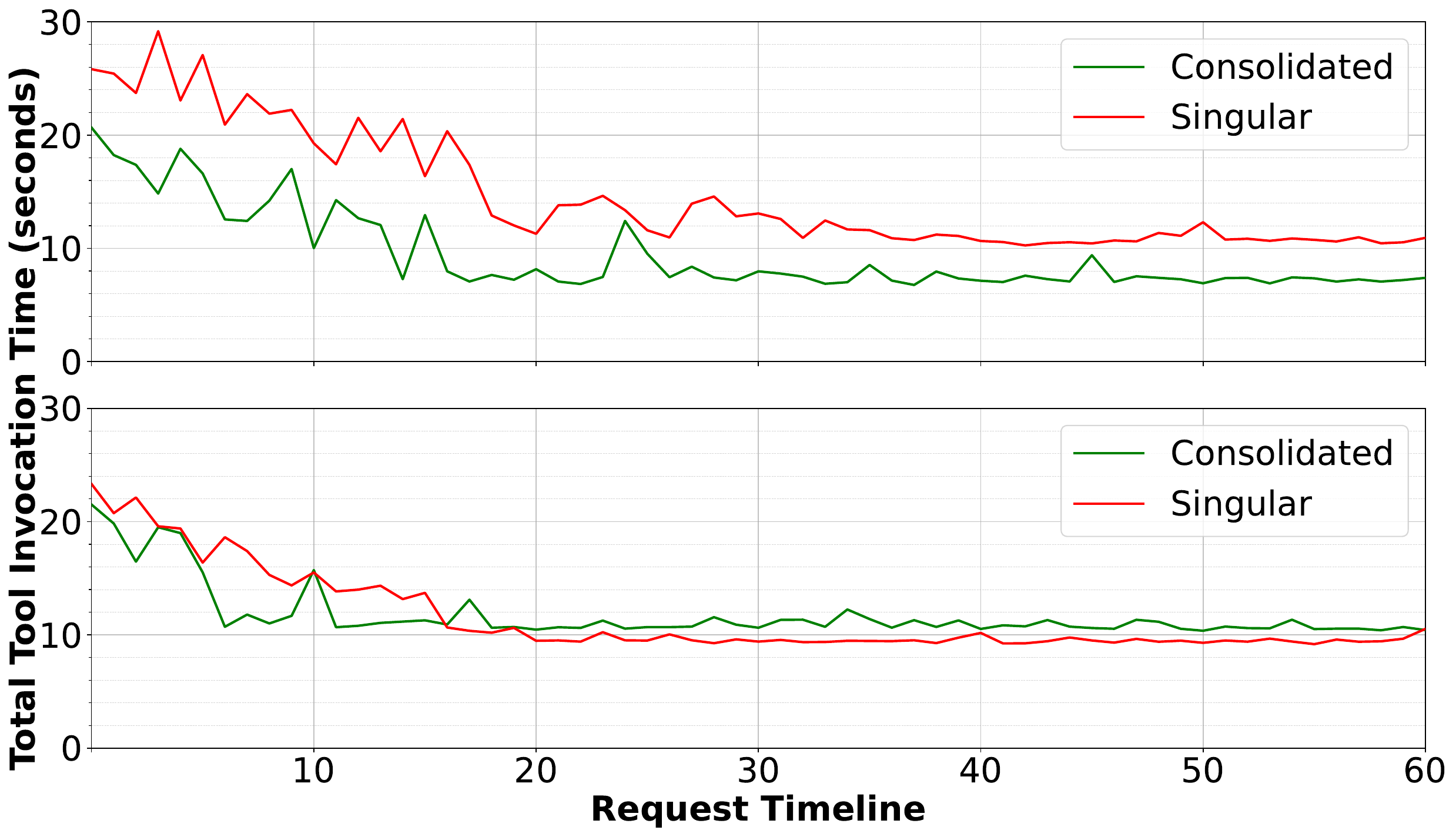}%
     \label{fig:mcp-cons}%
  }%
\caption{Benefits of MCP optimizations for \react on Research Paper Summarization (P1).}

\label{fig:mcp-opt}
\end{figure}

We compare \textit{singleton} deployments, with one Lambda per MCP server or tool, against \textit{consolidated} deployments that fuse frequently co-invoked tools. The experiment replays each application's tool-call trace at 1\,RPS for 120\,s. Consolidation amortizes cold starts and improves steady-state latency (Fig.~\ref{fig:mcp-cons}). For RS, total MCP latency drops from $10.2$\,s to $7.9$\,s and per-call cost falls from \textcent$2.0$ to \textcent$1.6$. Singleton deployments remain useful for rare tools, heterogeneous resource needs, and tighter fault isolation.

\section{Long-running Checkpoint Operation and Storage Breakdown} \label{app:results:long-cost}

Table~\ref{tab:tabular-concise-stacked} expands the Tabular-2022 checkpoint-cost comparison summarized in \S\ref{sec:results:failures}. We report the two async bridges separately: Agent--Tool (A--T) and Orchestrator--Agent (O--A). For \fw, an operation set consists of S3 \texttt{Put}, \texttt{List}, and \texttt{Get}: storing the serialized LangGraph checkpoint, finding the latest checkpoint for the \texttt{ThreadId}, and restoring the checkpoint on callback. For DF, the operation cost includes durable communication primitives and step overheads.

For the A--T bridge, \fw has lower base operation cost than DF even before step overheads are included. After adding DF step overheads, \fw is $\approx3.7\times$ cheaper on checkpoint operations. For the O--A bridge, \fw has a slightly higher base unit cost, but becomes $\approx2.5\times$ cheaper after DF step overheads are included.

The storage comparison shows the opposite size/cost trade-off. \fw stores larger explicit LangGraph checkpoints, but the S3-backed checkpoint store yields lower storage cost for this workload. Thus, the main advantage is not smaller checkpoint size, but fewer durable operations, lower storage cost, and no replay of intermediate LangGraph steps.

\begin{table*}[t]
\centering
\caption{Comparison of service runtimes used.}
\label{tab:runtime-comparison}
\footnotesize
\setlength{\tabcolsep}{1.5pt}
\renewcommand{\arraystretch}{1.12}
\begin{tabular}{L{2.2cm}||L{3.5cm} L{3.5cm} L{3.4cm} L{3.4cm}}
\hline
\textbf{Aspect} & \textbf{\fw} \textit{(our work)} & \textbf{AWS Bedrock AgentCore (AC)} & \textbf{AWS Lambda Durable Functions} & \textbf{AWS Lambda Step Functions} \\
\hline

\bf Role in evaluation &
Proposed QoS-aware FaaS middleware for MCP-enabled agentic workflows. &
Managed agent runtime baseline. &
Durable async FaaS baseline. &
Synchronous FaaS orchestration baseline. \\
\hline

\bf Orchestration model &
FaaS agent roles with callback-driven async orchestration for long EMCP waits. &
Managed distributed agent runtime. &
Durable workflow runtime with checkpoint and replay. &
Step Functions orchestrates Lambda roles synchronously. \\
\hline

\bf Long-running EMCP handling &
Dispatch, checkpoint, suspend, callback, restore, and resume across agent--tool and orchestrator--agent bridges. &
Long-running agent session with background task and polling in our evaluated baseline. &
Async execution through durable operations, callbacks, and replay. &
Coder Lambda waits synchronously for EMCP completion. \\
\hline

\bf State persistence &
Cross-turn AM in DynamoDB; within-turn LangGraph checkpoints in S3. &
Cross-turn AgentCore short term memory&
Durable execution state managed by Durable Functions(DF). &
Step Functions state plus Lambda-local execution state. \\
\hline

\bf Resume mechanism &
EMCP callback to agent \texttt{FunctionUrl}; agent callback to orchestrator; restore by \texttt{TaskId}/\texttt{ThreadId} and \texttt{NextNode}. &
Caller polls job status in the evaluated implementation. &
Durable callback and replay mechanism. &
No within-Lambda resume for long EMCP waits. \\
\hline

\bf Billing behavior during long waits &
FaaS agent and orchestrator are inactive during EMCP wait; cost shifts to checkpoint, callback, restore, and resume. &
Pure I/O wait does not bill idle CPU, but active background work, polling, retained session memory 
cost remain relevant. &
Avoids synchronous timeout, but durable steps, callbacks, data-written cost, and replay accumulate. &
Lambda remains active while waiting and can hit the function lifetime limit. \\
\hline

\bf Timeout behavior &
Timeout-safe for EMCP waits through callback-based resume. &
Times out when the Coder/Orchestrator background tasks extend beyond the runtime lifetime (8 hrs). &
Timeout-safe through durable async execution. &
Times out when the Coder Lambda waits beyond the FaaS lifetime. \\
\hline

\bf Main long-running cost drivers &
S3 \texttt{Put/List/Get}, callback, restore, and continuation; fewer checkpoints through selective checkpointing. &
Runtime/session resource use, background work or polling, retained memory
&
Durable operations (\texttt{invoke}, \texttt{callback}, \texttt{step}), data-written, storage, and replay. &
Lambda billed duration until timeout or completion. \\
\hline

\bf Main advantage &
Lowest measured async infrastructure cost; avoids replay and idle FaaS wait. &
Managed production runtime for agent deployment. &
Native durable async semantics. &
Simple orchestration for short-running FaaS workflows. \\
\hline

\bf Main limitation &
Requires \fw wrappers, checkpoint store, callbacks, and explicit async classification for EMCP services. &
Cost and session behavior depend on workload and implementation pattern. %
&
Higher durable-operation and replay overhead in evaluated workloads. &
Not suitable for long-running external tool calls. \\
\hline

\end{tabular}
\end{table*}

\end{document}